\documentclass{article}

\usepackage[final]{neurips_data_2022}

\usepackage[utf8]{inputenc} 
\usepackage[T1]{fontenc}    
\usepackage{hyperref}       
\usepackage{url}            
\usepackage{booktabs}       
\usepackage{amsfonts}       
\usepackage{nicefrac}       
\usepackage{microtype}      
\usepackage{xcolor,array}         
\usepackage{tabularx}
\usepackage{graphicx}
\usepackage{multirow,colortbl}
\usepackage{enumitem} 
\usepackage{amsmath}
\usepackage[frozencache,cachedir=.]{minted}
\usepackage{amssymb}
\usepackage{mathtools}
\usepackage{amsthm}
\usepackage{adjustbox}
\usepackage{subcaption}

\usepackage{booktabs}

\newcolumntype{Y}{>{\centering\arraybackslash}X}
\newcommand{\mname}{\texttt{PMO}} 
\newcommand{\nmethod}{25}
\newcommand{\noracle}{23}

\newcommand{\calO}{\mathcal{O}}

\newcommand{\calM}{\mathcal{M}}

\title{Sample Efficiency Matters: A Benchmark for Practical Molecular Optimization}

\author{Wenhao Gao$^{1*}$, Tianfan Fu$^{2*}$, Jimeng Sun$^3$, Connor W. Coley$^{1}$ \\
$^{1}$Massachusetts Institute of Technology,
$^{2}$Georgia Institute of Technology \\
$^3$University of Illinois at Urbana-Champaign\\
$^{*}$Equal Contributions \\[2ex]%
\texttt{\{whgao,ccoley\}@mit.edu},
\texttt{tfu42@gatech.edu},
\texttt{jimeng@illinois.edu} 
}

\begin{document}

\maketitle

\begin{abstract}
Molecular optimization is a fundamental goal in the chemical sciences and is of central interest to drug and material design. In recent years, significant progress has been made in solving challenging problems across various aspects of computational molecular optimizations, emphasizing high validity, diversity, and, most recently, synthesizability. Despite this progress, many papers report results on 
trivial or self-designed tasks, bringing additional challenges to directly assessing the performance of new methods. Moreover, the sample efficiency of the optimization---the number of molecules evaluated by the oracle---is rarely  discussed, despite being an essential consideration for realistic discovery applications.

To fill this gap, we have created an open-source benchmark for \textbf{p}ractical \textbf{m}olecular \textbf{o}ptimization, \mname, to facilitate the transparent and reproducible evaluation of algorithmic advances in molecular optimization. This paper thoroughly investigates the performance of \nmethod~molecular design algorithms on \noracle~single-objective (scalar) optimization tasks with a particular focus on sample efficiency. Our results show that most ``state-of-the-art'' methods fail to outperform  their predecessors under a limited oracle budget allowing 10K queries and that no existing algorithm can efficiently solve certain molecular optimization problems in this setting. We analyze the influence of the optimization algorithm choices, molecular assembly strategies, and oracle landscapes on the optimization performance to inform future algorithm development and benchmarking. \mname~provides a standardized experimental setup to comprehensively evaluate and compare new molecule optimization methods with existing ones. All code can be found at \url{https://github.com/wenhao-gao/mol\_opt}.
\end{abstract}

\section{Introduction}
\label{sec:intro}

Designing new functional molecules is a constrained multi-objective optimization problem that aims to find molecules with desired properties such as selective inhibition against a disease target, with additional desiderata and constraints to ensure the structures are stable and synthesizable. 
The importance of molecular design problems has attracted significant efforts to develop systematical molecular design methodologies instead of exhaustive searches, leveraging combinatorial optimization algorithms~\cite{jensen2019graph,xie2021mars}, predictive machine learning models~\cite{graff2021accelerating,gentile2022artificial}, and generative models~\cite{olivecrona2017molecular,gomez2018automatic}. 
Especially in recent years, we have witnessed significant progress in solving challenging problems across various aspects of computational molecular optimizations, such as achieving high validity~\cite{kusner2017grammar,jin2018junction,krenn2020self}, diversity~\cite{bengio2021gflownet}, and, most recently, synthesizability~\cite{bradshaw2020barking,gao2021amortized}.

Despite the exciting progress in the field and the abundance of new methods proposed, how these algorithms compare against each other remains unclear. Most method development papers and existing benchmarks such as Guacamol~\cite{brown2019guacamol}, Therapeutics Data Commons (TDC)~\cite{huang2021therapeutics} and Tripp et al.'s~\cite{tripp2021fresh} suffer from at least one of three problems:
(1) 
Lack of consideration of the oracle budget: 
Many papers~\cite{zhou2019optimization,nigam2019augmenting,gottipati2020learning} do not report how many times the oracle function is called to achieve the reported results (i.e., how many candidate molecules were evaluated), except in rare cases~\cite{korovina2020chembo,fu2021differentiable,bengio2021flow,maus2022local,grosnit2021high}, despite this range spanning orders of magnitude.
As most valuable oracles---experiments or high-accuracy simulations---require substantial costs, it is vital to identify the desired compound with as few oracle calls as possible. 
(2) 
Trivial oracles:
Some papers only report results on trivial oracles~\cite{nigam2019augmenting} like quantitative estimate of drug-likeness (QED)~\cite{bickerton2012quantifying}\footnote{
As demonstrated in this benchmark later, QED is likely to have a global maximum of 0.948 and even random sampling could reach that value. It is disabled to meaningfully distinguish different algorithms. 
} or penalized octanol-water partition coefficient (LogP)\footnote{
LogP is unbounded and the relationship between LogP values and molecular structures is fairly simple: adding carbons monotonically increases the estimated LogP value~\cite{you2018graph, fu2021differentiable}. This simple strategy makes the performance in LogP highly depend on the chemical space definition and the number of steps allowed, and provides no insights for distinguish algorithms' optimization ability. 
Besides, simply maximizing LogP is not a meaningful goal in drug design. Therefore, we exclude LogP in this benchmark.
}; other papers even introduce new self-designed tasks~\cite{gottipati2020learning,bengio2021flow}, which obfuscates a comparison to prior work. 
(3) Randomness: Another complication is that many algorithms are not deterministic and exhibit significant run-to-run variation, so reporting results from several independent trials is essential.
All of the existing benchmarks examined no more than five methods due to the significant variation between molecular optimization algorithms.
Thus we still lack a unified benchmark to assess which methods are beneficial in a realistic discovery scenario.

This paper presents a new reproducible large-scale experimental study with a sound experimental protocol for molecular design, \mname. 
We have benchmarked \nmethod~methods across \noracle~various widely-used oracle functions, with each of them tuned and run for multiple independent trials. 
To consider a combination of optimization ability and sample efficiency, we limit the number of maximum oracle calls up to 10,000 queries and measure model performance with the area under the curve (AUC) of the top-10 average performance versus oracle calls. 
Our results show that none of the existing molecular optimization algorithms are efficient enough to solve a \textit{de novo} molecular optimization problem within a realistic oracle budget of hundreds of experiments, and  ``state-of-the-art'' methods often fail to outperform their predecessors.
We analyze the algorithmic contribution and the influence of oracle landscapes on optimization performance to inform future algorithm development and benchmarking.
Our results highlight the necessity of standardized experimental reporting, including independent replicates and extensive hyperparameter tuning. 
We envision that the \mname~benchmark will make molecular optimizations more accessible and reproducible, thereby facilitating algorithmic advances and, ultimately, the broader adoption of molecular optimization techniques in experimental drug and materials discovery workflows.

\section{Algorithms}
\label{sec:algorithm}

A molecular optimization method has two major components: (1) a molecular assembly strategy that defines the chemical space by assembling a digital representation of compounds, and (2) an optimization algorithm that navigates this chemical space. This section will first introduce common strategies to assemble molecules, then introduce the benchmarked molecular optimization methods based on the core optimization algorithms. Table \ref{tab:method} summarizes current molecular design methods categorized based on assembly strategy and optimization method, including but not limited to the methods included in our baseline. We emphasize that our goal is not to make an exhaustive list but to include a group of methods that are representative enough to obtain meaningful conclusions.

\subsection{Preliminaries}
\label{sec:preliminaries}

In this paper, we limit our scope to general-purpose single-objective molecular optimization methods focusing on small organic molecules with scalar properties with some relevance to therapeutic design. Formally, we can formulate such a molecular design problem as an optimization problem:
\begin{equation}
    m^{*} = \arg\max_{m \in \calM}  \calO(m),
\end{equation}
where $m$ is a molecular structure, $\calM$ denotes the design space called chemical space that comprises all possible candidate molecules. The size of $\calM$ is impractically large, e.g., $10^{60}$~\cite{bohacek1996art}. We assume we have access to the ground truth value of a property of interest denoted by $\calO(m): \calM \rightarrow \mathcal{R}$, where an oracle, $\calO$, is a black-box function that evaluates certain chemical or biological properties of a molecule $m$ and returns the ground truth property $\calO(m)$ as a scalar. Note that neither the analytic form of oracles nor the derivatives of the properties are accessible. The most practical oracles---experiments or high-accuracy simulations--- typically require substantial costs. An algorithm able to optimize the oracle within a reasonable budget is thus necessary for automating the design of molecules to achieve high-level automated chemical design (ACD)~\cite{goldman2022defining} or function-oriented autonomous synthesis~\cite{gao2022autonomous}.

\begin{table*}[t!]
\centering
\caption{Representative molecule generation methods, categorised based on the molecular assembly strategies and the optimization algorithms. Columns are various molecular assembly strategies while rows are different optimization algorithms.
}
\label{tab:method}
\scriptsize
\begin{tabularx}{\textwidth}{c Y Y Y Y Y}
\toprule
& SMILES & SELFIES & Graph (atom) & Graph (fragment) & Synthesis \\ 
\midrule[0.1pt] 
GA & SMILES-GA~\cite{brown2019guacamol} & GA+D~\cite{nigam2019augmenting} STONED~\cite{nigam2021beyond} & - & Graph-GA~\cite{jensen2019graph} & SynNet~\cite{gao2021amortized} \\ \arrayrulecolor{gray}\midrule[0.7pt] 
MCTS & - & - & Graph-MCTS~\cite{jensen2019graph} & - & - \\ \midrule[0.01pt] 
BO & BOSS~\cite{moss2020boss} & - & - & GPBO~\cite{tripp2021fresh} & ChemBO~\cite{korovina2020chembo} \\ \midrule[0.1pt]
VAE & SMILES-VAE~\cite{gomez2018automatic} & SELFIES-VAE~\cite{maus2022local} & - & JTVAE~\cite{jin2018junction} & DoG-AE~\cite{bradshaw2020barking} \\ \midrule[0.1pt] 
GAN & ORGAN~\cite{sanchez2017optimizing} & - & MolGAN~\cite{de2018molgan} & - & - \\ \midrule[0.1pt]  
SBM & - & - & - & GFlowNet\cite{bengio2021gflownet}  MARS\cite{xie2021mars}  & - \\  \midrule[0.1pt] 
HC & SMILES LSTM~\cite{brown2019guacamol} & SELFIES LSTM & - & MIMOSA~\cite{fu2021mimosa} & DoG-Gen~\cite{bradshaw2020barking} \\ \midrule[0.1pt] 
RL & REINVENT~\cite{olivecrona2017molecular} & SELFIES-REINVENT & MolDQN~\cite{zhou2019optimization} GCPN~\cite{you2018graph} &  RationaleRL\cite{jin2020multi} FREED~\cite{yang2021hit} & PGFS~\cite{gottipati2020learning} REACTOR~\cite{horwood2020molecular} \\ \midrule[0.1pt] 
GRAD & - & Pasithea~\cite{shen2021deep} & - & DST~\cite{fu2021differentiable} & - \\ 
\arrayrulecolor{black}\bottomrule
\end{tabularx}
\end{table*}


\subsection{Molecular assembly strategies}
\label{sec:assemble}


\noindent\textbf{String-based}. 
String-based assembly strategies represent molecules as strings and explore chemical space by modifying strings directly: character-by-character, token-by-token, or through more complex transformations based on a specific grammar. We include two types of string representations:
(1) Simplified Molecular-Input Line-Entry System (SMILES)~\cite{weininger1988smiles}, a linear notation describing the molecular structure using short ASCII strings based on a graph traversal algorithm;
(2) SELF-referencIng Embedded Strings (SELFIES)~\cite{krenn2020self}, which avoids syntactical invalidity by enforcing the chemical validity rules in a formal grammar table.

\noindent\textbf{Graph-based}. 
Two-dimensional (2D) graphs can intuitively define molecular identities to a first approximation (ignoring stereochemistry\footnote{Incorporating certain stereochemical information in 2D molecular graphs is possible through various approaches~\cite{andersen2017chemical,pattanaik2020message,adams2021learning}.}): the nodes and edges represent the atoms and bonds. There are two main assembling strategies for molecular graphs: (1) an atom-based assembly strategy~\cite{zhou2019optimization} that adds or modifies atoms and bonds one at a time, which covers all valid chemical space;
(2) a fragment-based assembling strategy~\cite{jin2018junction} that summarizes common molecular fragments and operates one fragment at a time. Note that fragment-based strategy could also include atom-level operation.

\noindent\textbf{Synthesis-based}. 
Most of the above assembly strategies can cover a large chemical space, but an eventual goal of molecular design is to physically test the candidate; thus, a desideratum is to explore synthesizable candidates only. Designing molecules by assembling synthetic pathways from commercially-available starting materials and reliable chemical transformation adds a constraint of synthesizability to the search space. This class can be divided into template-free~\cite{bradshaw2020barking} and template-based~\cite{gao2021amortized} based on how to define reliable chemical transformations, but we will not distinguish between them in this paper as synthesis-based strategy is relatively less explored in general.

\subsection{Optimization algorithms}
\label{sec:optimization_algorithm}

\textbf{Screening} (a.k.a. virtual screening) involves searching over a pre-enumerated library of molecules. We include \underline{Screening} as a baseline, which randomly samples ZINC 250k~\cite{sterling2015zinc}. Model-based screening~\cite{svensson2017improving,hernandez2017parallel,ahmed2018efficient,gentile2020deep,graff2021accelerating,graff2022self} instead trains a surrogate model and prioritizes  molecules that are scored highly by the surrogate to accelerate screening. We adopt the implementation from the original paperof \underline{MolPAL}~\cite{graff2021accelerating} and treat it as a model-based version of screening.

\textbf{Genetic Algorithm (GA)} is a popular heuristic algorithm inspired by natural evolutionary processes. It combines \textit{mutation} and/or \textit{crossover} perturbing a \textit{mating pool} to enable exploration in the design space.
We include \underline{SMILES GA}~\cite{yoshikawa2018population} that defines actions based on SMILES context-free grammar and a modified version of \underline{STONED}~\cite{nigam2021beyond} that directly manipulates tokens in SELFIES strings. Unlike the string-based GAs that only have mutation steps, \underline{Graph GA}~\cite{jensen2019graph} derives crossover rules from graph matching and includes both atom- and fragment-level mutations. Finally, we include \underline{SynNet}~\cite{gao2021amortized} as a synthesis-based example that applies a genetic algorithm on binary fingerprints and decodes to synthetic pathways. 
We adopt the implementation of SMILES GA and Graph GA from Guacamol~\cite{brown2019guacamol}, STONED, and SynNet from the original paper.
We also include the original implementation of a deep learning enhanced version of SELFIES-based GA from~\cite{nigam2019augmenting} and label it as \underline{GA+D}.

\textbf{Monte-Carlo Tree Search (MCTS)} locally and randomly searches each branch of the current state (e.g., a molecule or partial molecule) and selects the most promising ones (those with highest property scores)  for the next iteration. \underline{Graph MCTS}~\cite{jensen2019graph} is an MCTS algorithm based on atom-level searching over molecular graphs. We adopt the implementation from Guacamol~\cite{brown2019guacamol}.

\textbf{Bayesian optimization (BO)}~\cite{shahriari2015taking} is a large class of method that builds a surrogate for the objective function using a Bayesian machine learning technique, such as Gaussian process (GP) regression, then uses an \textit{acquisition function} combining the surrogate and uncertainty to decide where to sample, which is naturally model-based.
However, as BO usually leverages a non-parametric model, it scales poorly with sample size and feature dimension~\cite{deisenroth2015distributed}.
We included a string-based model, BO over String Space (\underline{BOSS})~\cite{moss2020boss}, and a synthesis-based model, \underline{ChemBO}~\cite{korovina2020chembo}, but do not obtain meaningful results even with early stopping potentially due to the poor scaling of the string subsequence kernel (SSK) (see Section~\ref{sec:early_stop} for early stopping setting, and Section~\ref{sec:fail_method} for more analysis).
Finally, we adopt Gaussian process Bayesian optimization (\underline{GP BO})~\cite{tripp2021fresh} that optimizes the GP acquisition function with Graph GA methods in an inner loop. The implementation is from the original paper, and we treat it as a model-based version of Graph GA. Note that we categorize methods that apply BO to optimize molecules in latent space as a separate class below.

\textbf{Variational autoencoders (VAEs)}~\cite{kingma2013auto} are a class of generative method that maximize a lower bound of the likelihood (evidence lower bound (ELBO)) instead of estimating the likelihood directly. A VAE typically learns to map molecules to and from real space to enable the indirect optimization of molecules by numerically optimizing latent vectors, most commonly with BO~\cite{balandat2020botorch}. 
\underline{SMILES-VAE} \cite{gomez2018automatic} uses a VAE to model molecules represented as SMILES strings, and is implemented in MOSES~\cite{polykovskiy2020molecular}. 
We adopt the identical architecture to model SELFIES strings and denote it as \underline{SELFIES-VAE}. 
\underline{JT-VAE}~\cite{jin2018junction} abstracts a molecular graph into a junction tree (i.e., a cycle-free structure), and design message passing network as the encoder and tree-RNN as the decoder. 
\underline{DoG-AE}~\cite{bradshaw2020barking} uses Wasserstein autoencoder (WAE) to learn the distribution of synthetic pathways. 
Note that we include a set of vanilla methods for each kind while many variants 
have emerged, such as~\cite{grosnit2021high} and~\cite{maus2022local}. We leave the validation of variants for the future development of this benchmark.

\textbf{Score-based modeling (SBM)} formulates the problem of molecule design as a sampling problem where the target distribution is a function of the target property, featured by Markov-chain Monte Carlo (MCMC) methods that construct Markov chains with the desired distribution as their equilibrium distribution. MARkov molecular Sampling (\underline{MARS})~\cite{xie2021mars} is such an example that leverages a graph neural network to propose action steps adaptively in an MCMC with an annealing scheme. Generative Flow Network (\underline{GFlowNet})~\cite{bengio2021gflownet} views the generative process as a flow network and trains it with a temporal difference-like loss function based on the conservation of flow. By matching the property of interest with the volume of the flow, generation can sample a distribution proportional to the target distribution.

\textbf{Hill climbing (HC)} is an iterative learning method that incorporates the generated high-scored molecules into the training data and fine-tunes the generative model for each iteration. 
It is a variant of the cross-entropy method~\cite{de2005tutorial}, and can also be seen as a variant of REINFORCE~\cite{williams1992simple} with a particular reward shaping. 
We adopt \underline{SMILES-LSTM-HC} from Guacamol~\cite{brown2019guacamol} that leverages a LSTM to learn the molecular distribution represented in SMILES strings, and modifies it to a SELFIES version denoted as \underline{SELFIES-LSTM-HC}.
MultI-constraint MOlecule SAmpling (\underline{MIMOSA})~\cite{fu2021mimosa} leverages a graph neural network to predict the identity of a masked fragment node and trains it with a HC algorithm.
\underline{DoG-Gen}~\cite{bradshaw2020barking} instead learn the distribution of synthetic pathways as Directed Acyclic Graph (DAGs) with an RNN generator.


\textbf{Reinforcement Learning (RL)} learns how intelligent agents take actions in an environment to maximize the cumulative reward by transitioning through different states. 
In molecular design, a state is usually a partially generated molecule; actions are manipulations at the level of graphs or strings; rewards are defined as the generated molecules' property of interest. 
\underline{REINVENT}~\cite{olivecrona2017molecular} adopts a policy-based RL approach to tune RNNs to generate SMILES strings. We adopt the implementation from the original paper, and modify it to generate SELFIES strings, \underline{SELFIES-REINVENT}. 
\underline{MolDQN}~\cite{zhou2019optimization} uses a deep Q-network to generate molecular graph in an atom-wise manner. 


\textbf{Gradient ascent (GRAD)} methods learn to estimate the gradient direction based on the landscape of the molecular property over the chemical space, and back-propagate to optimize the molecules. 
\underline{Pasithea}~\cite{shen2021deep} exploits an MLP to predict properties from SELFIES strings, and back-propagate to modify tokens.
Differentiable scaffolding tree (\underline{DST})~\cite{fu2021differentiable} abstracts molecular graphs to scaffolding trees and leverages a graph neural network to estimate the gradient. We adopted the implementation from the original papers and modify them to update the surrogates online as data are acquired.

\section{Experiments}
\label{sec:experiments}

\subsection{Benchmark setup}
\label{sec:setup}

This section introduces the setup of \mname~benchmark. The main idea behind \mname~is the pursuit of an ideal \textit{de novo} molecular optimization algorithm that exhibits strong optimization ability, sample efficiency, generalizability to various optimization objectives, and robustness to hyperparameter selection and random seeds.


\noindent\textbf{Oracle}: To examine the generalizability of methods, we aim to include a broad range of pharmaceutically-relevant oracle functions. Systematic categorization of oracles based on their landscape is still challenging due to the complicated relationship between molecular structure and function. We have included the most commonly used oracles (see a recent discussion of commonly-used oracles in~\cite{tripp2022evaluation}). Several have been described as ``trivial'', but we assert this is only true when the number of oracle queries is not controlled. 
In total, \mname~ includes \noracle~oracle functions: QED~\cite{bickerton2012quantifying}, DRD2~\cite{olivecrona2017molecular}, GSK3$\beta$, JNK3~\cite{li2018multi}, and 19 oracles from Guacamol~\cite{brown2019guacamol}. QED is a relatively simple heuristic function that estimates if a molecule is likely to be a drug based on if it contains some ``red flags''. DRD2, GSK3$\beta$, and JNK3 are machine learning models (support vector machine (SVM), random forest (RF)) fit to experimental data to predict the bioactivities against their corresponding disease targets. Guacamol oracles are designed to mimic the drug discovery objectives based on multiple considerations, called multi-property objective (MPO), including similarity to target molecules, molecular weights, CLogP, etc. 
All oracle scores are normalized from 0 to 1, where 1 is optimal. 
Recently, docking scores that estimate the binding affinity between ligands and proteins have been adopted as oracles~\cite{cieplinski2020we, huang2021therapeutics, garcia2021dockstring}. However, as the simulations are more costly than above ones but are still coarse estimates that do not reflect true bioactivity, we leave it to future work.

\noindent\textbf{Metrics}: To consider the optimization ability and sample efficiency simultaneously, we report the area under the curve (AUC) of top-$K$ average property value versus the number of oracle calls (\textit{AUC top-$K$}) as the primary metric to measure the performance. Unlike using top-$K$ average property, AUC rewards methods that reach high values with fewer oracle calls. We use $K=10$ in this paper as it is useful to identify a small number of distinct molecular candidates to progress to later stages of  development.
We limit the number of oracle calls to 10000, though we expect methods to optimize well within hundreds of calls when using experimental evaluations.
The reported values of AUCs are min-max scaled to $[0, 1]$.

\noindent\textbf{Data}: We restrict all our methods to using the ZINC 250K dataset only whenever a database is required, which contains around 250K molecules sampled from the ZINC database~\cite{sterling2015zinc} for its pharmaceutical relevance, moderate size, and popularity. Screening and MolPAL search over this database; generative models such as VAEs, LSTMs are pretrained on this database; fragments required for JT-VAE, MIMOSA, DST are extracted from this database.

\noindent\textbf{Other details}: We tuned hyperparameters for most methods on the average AUC Top-10 from 3 independent runs of two Guacamol tasks: zaleplon\_mpo and perindopril\_mpo. Reported results are from 5 independent runs with various random seeds. All data, oracle functions, and metric evaluations are taken from the Therapeutic Data Commons (TDC)~\cite{huang2021therapeutics}~{(\url{https://tdcommons.ai})} and more details are described in Appendix. Note that the implementation of sitagliptin\_mpo and zaleplon\_mpo are different from the ones in Guacamol~\cite{brown2019guacamol}.

\begin{table*}[t!]
\centering
\caption{Performance of ten best performing molecular optimization methods based on mean AUC Top-10. We report the mean and standard deviation of \textbf{AUC Top-10} from 5 independent runs. 
The best model in each task is labeled bold. 
Full results are in the Appendix \ref{sec:main_results}.
}
\label{tab:task2}
\begin{adjustbox}{width=\textwidth}
\scriptsize
\begin{tabularx}{\textwidth}{c | Y | Y | Y | Y | Y}
\toprule
Method & REINVENT & Graph GA & REINVENT & GP BO & STONED \\
Assembly & SMILES & Fragments & SELFIES & Fragments & SELFIES \\
\midrule
albuterol\_similarity & 0.882$\pm$ 0.006 & 0.838$\pm$ 0.016 & 0.826$\pm$ 0.030 & \textbf{0.898$\pm$ 0.014} & 0.745$\pm$ 0.076 \\
amlodipine\_mpo & 0.635$\pm$ 0.035 & \textbf{0.661$\pm$ 0.020} & 0.607$\pm$ 0.014 & 0.583$\pm$ 0.044 & 0.608$\pm$ 0.046 \\
celecoxib\_rediscovery & 0.713$\pm$ 0.067 & 0.630$\pm$ 0.097 & 0.573$\pm$ 0.043 & \textbf{0.723$\pm$ 0.053} & 0.382$\pm$ 0.041 \\
deco\_hop & 0.666$\pm$ 0.044 & 0.619$\pm$ 0.004 & 0.631$\pm$ 0.012 & 0.629$\pm$ 0.018 & 0.611$\pm$ 0.008 \\
drd2 & 0.945$\pm$ 0.007 & 0.964$\pm$ 0.012 & 0.943$\pm$ 0.005 & 0.923$\pm$ 0.017 & 0.913$\pm$ 0.020 \\
fexofenadine\_mpo & 0.784$\pm$ 0.006 & 0.760$\pm$ 0.011 & 0.741$\pm$ 0.002 & 0.722$\pm$ 0.005 & \textbf{0.797$\pm$ 0.016} \\
gsk3b & \textbf{0.865$\pm$ 0.043} & 0.788$\pm$ 0.070 & 0.780$\pm$ 0.037 & 0.851$\pm$ 0.041 & 0.668$\pm$ 0.049 \\
isomers\_c7h8n2o2 & 0.852$\pm$ 0.036 & 0.862$\pm$ 0.065 & 0.849$\pm$ 0.034 & 0.680$\pm$ 0.117 & 0.899$\pm$ 0.011 \\
isomers\_c9h10n2o2pf2cl & 0.642$\pm$ 0.054 & 0.719$\pm$ 0.047 & 0.733$\pm$ 0.029 & 0.469$\pm$ 0.180 & 0.805$\pm$ 0.031 \\
jnk3 & \textbf{0.783$\pm$ 0.023} & 0.553$\pm$ 0.136 & 0.631$\pm$ 0.064 & 0.564$\pm$ 0.155 & 0.523$\pm$ 0.092 \\
median1 & \textbf{0.356$\pm$ 0.009} & 0.294$\pm$ 0.021 & 0.355$\pm$ 0.011 & 0.301$\pm$ 0.014 & 0.266$\pm$ 0.016 \\
median2 & 0.276$\pm$ 0.008 & 0.273$\pm$ 0.009 & 0.255$\pm$ 0.005 & \textbf{0.297$\pm$ 0.009} & 0.245$\pm$ 0.032 \\
mestranol\_similarity & 0.618$\pm$ 0.048 & 0.579$\pm$ 0.022 & 0.620$\pm$ 0.029 & \textbf{0.627$\pm$ 0.089} & 0.609$\pm$ 0.101 \\
osimertinib\_mpo & \textbf{0.837$\pm$ 0.009} & 0.831$\pm$ 0.005 & 0.820$\pm$ 0.003 & 0.787$\pm$ 0.006 & 0.822$\pm$ 0.012 \\
perindopril\_mpo & 0.537$\pm$ 0.016 & 0.538$\pm$ 0.009 & 0.517$\pm$ 0.021 & 0.493$\pm$ 0.011 & 0.488$\pm$ 0.011 \\
qed & 0.941$\pm$ 0.000 & 0.940$\pm$ 0.000 & 0.940$\pm$ 0.000 & 0.937$\pm$ 0.000 & 0.941$\pm$ 0.000 \\
ranolazine\_mpo & 0.760$\pm$ 0.009 & 0.728$\pm$ 0.012 & 0.748$\pm$ 0.018 & 0.735$\pm$ 0.013 & \textbf{0.765$\pm$ 0.029} \\
scaffold\_hop & \textbf{0.560$\pm$ 0.019} & 0.517$\pm$ 0.007 & 0.525$\pm$ 0.013 & 0.548$\pm$ 0.019 & 0.521$\pm$ 0.034 \\
sitagliptin\_mpo & 0.021$\pm$ 0.003 & \textbf{0.433$\pm$ 0.075} & 0.194$\pm$ 0.121 & 0.186$\pm$ 0.055 & 0.393$\pm$ 0.083 \\
thiothixene\_rediscovery & 0.534$\pm$ 0.013 & 0.479$\pm$ 0.025 & 0.495$\pm$ 0.040 & \textbf{0.559$\pm$ 0.027} & 0.367$\pm$ 0.027 \\
troglitazone\_rediscovery & \textbf{0.441$\pm$ 0.032} & 0.390$\pm$ 0.016 & 0.348$\pm$ 0.012 & 0.410$\pm$ 0.015 & 0.320$\pm$ 0.018 \\
valsartan\_smarts & \textbf{0.178$\pm$ 0.358} & 0.000$\pm$ 0.000 & 0.000$\pm$ 0.000 & 0.000$\pm$ 0.000 & 0.000$\pm$ 0.000 \\
zaleplon\_mpo & \textbf{0.358$\pm$ 0.062} & 0.346$\pm$ 0.032 & 0.333$\pm$ 0.026 & 0.221$\pm$ 0.072 & 0.325$\pm$ 0.027 \\
\midrule
Sum & 14.196 & 13.751 & 13.471 & 13.156 & 13.024 \\
Rank & 1 & 2 & 3 & 4 & 5 \\

\hline
\hline

Method & LSTM HC & SMILES GA & SynNet & DoG-Gen & DST \\
Assembly & SMILES & SMILES & Synthesis & Synthesis & Fragments \\
\midrule
albuterol\_similarity & 0.719$\pm$ 0.018 & 0.661$\pm$ 0.066 & 0.584$\pm$ 0.039 & 0.676$\pm$ 0.013 & 0.619$\pm$ 0.020 \\
amlodipine\_mpo & 0.593$\pm$ 0.016 & 0.549$\pm$ 0.009 & 0.565$\pm$ 0.007 & 0.536$\pm$ 0.003 & 0.516$\pm$ 0.007 \\
celecoxib\_rediscovery & 0.539$\pm$ 0.018 & 0.344$\pm$ 0.027 & 0.441$\pm$ 0.027 & 0.464$\pm$ 0.009 & 0.380$\pm$ 0.006 \\
deco\_hop & \textbf{0.826$\pm$ 0.017} & 0.611$\pm$ 0.006 & 0.613$\pm$ 0.009 & 0.800$\pm$ 0.007 & 0.608$\pm$ 0.008 \\
drd2 & 0.919$\pm$ 0.015 & 0.908$\pm$ 0.019 & \textbf{0.969$\pm$ 0.004} & 0.948$\pm$ 0.001 & 0.820$\pm$ 0.014 \\
fexofenadine\_mpo & 0.725$\pm$ 0.003 & 0.721$\pm$ 0.015 & 0.761$\pm$ 0.015 & 0.695$\pm$ 0.003 & 0.725$\pm$ 0.005 \\
gsk3b & 0.839$\pm$ 0.015 & 0.629$\pm$ 0.044 & 0.789$\pm$ 0.032 & 0.831$\pm$ 0.021 & 0.671$\pm$ 0.032 \\
isomers\_c7h8n2o2 & 0.485$\pm$ 0.045 & \textbf{0.913$\pm$ 0.021} & 0.455$\pm$ 0.031 & 0.465$\pm$ 0.018 & 0.548$\pm$ 0.069 \\
isomers\_c9h10n2o2pf2cl & 0.342$\pm$ 0.027 & \textbf{0.860$\pm$ 0.065} & 0.241$\pm$ 0.064 & 0.199$\pm$ 0.016 & 0.458$\pm$ 0.063 \\
jnk3 & 0.661$\pm$ 0.039 & 0.316$\pm$ 0.022 & 0.630$\pm$ 0.034 & 0.595$\pm$ 0.023 & 0.556$\pm$ 0.057 \\
median1 & 0.255$\pm$ 0.010 & 0.192$\pm$ 0.012 & 0.218$\pm$ 0.008 & 0.217$\pm$ 0.001 & 0.232$\pm$ 0.009 \\
median2 & 0.248$\pm$ 0.008 & 0.198$\pm$ 0.005 & 0.235$\pm$ 0.006 & 0.212$\pm$ 0.000 & 0.185$\pm$ 0.020 \\
mestranol\_similarity & 0.526$\pm$ 0.032 & 0.469$\pm$ 0.029 & 0.399$\pm$ 0.021 & 0.437$\pm$ 0.007 & 0.450$\pm$ 0.027 \\
osimertinib\_mpo & 0.796$\pm$ 0.002 & 0.817$\pm$ 0.011 & 0.796$\pm$ 0.003 & 0.774$\pm$ 0.002 & 0.785$\pm$ 0.004 \\
perindopril\_mpo & 0.489$\pm$ 0.007 & 0.447$\pm$ 0.013 & \textbf{0.557$\pm$ 0.011} & 0.474$\pm$ 0.002 & 0.462$\pm$ 0.008 \\
qed & 0.939$\pm$ 0.000 & 0.940$\pm$ 0.000 & \textbf{0.941$\pm$ 0.000} & 0.934$\pm$ 0.000 & 0.938$\pm$ 0.000 \\
ranolazine\_mpo & 0.714$\pm$ 0.008 & 0.699$\pm$ 0.026 & 0.741$\pm$ 0.010 & 0.711$\pm$ 0.006 & 0.632$\pm$ 0.054 \\
scaffold\_hop & 0.533$\pm$ 0.012 & 0.494$\pm$ 0.011 & 0.502$\pm$ 0.012 & 0.515$\pm$ 0.005 & 0.497$\pm$ 0.004 \\
sitagliptin\_mpo & 0.066$\pm$ 0.019 & 0.363$\pm$ 0.057 & 0.025$\pm$ 0.014 & 0.048$\pm$ 0.008 & 0.075$\pm$ 0.032 \\
thiothixene\_rediscovery & 0.438$\pm$ 0.008 & 0.315$\pm$ 0.017 & 0.401$\pm$ 0.019 & 0.375$\pm$ 0.004 & 0.366$\pm$ 0.006 \\
troglitazone\_rediscovery & 0.354$\pm$ 0.016 & 0.263$\pm$ 0.024 & 0.283$\pm$ 0.008 & 0.416$\pm$ 0.019 & 0.279$\pm$ 0.019 \\
valsartan\_smarts & 0.000$\pm$ 0.000 & 0.000$\pm$ 0.000 & 0.000$\pm$ 0.000 & 0.000$\pm$ 0.000 & 0.000$\pm$ 0.000 \\
zaleplon\_mpo & 0.206$\pm$ 0.006 & 0.334$\pm$ 0.041 & 0.341$\pm$ 0.011 & 0.123$\pm$ 0.016 & 0.176$\pm$ 0.045 \\
\midrule
Sum & 12.223 & 12.054 & 11.498 & 11.456 & 10.989 \\
Rank & 6 & 7 & 8 & 9 & 10 \\
\bottomrule
\end{tabularx}
\end{adjustbox}
\end{table*}

\begin{table*}[th!]
\centering
\caption{The ranking of each methods based on different metrics.
}
\label{tab:ranking}
\begin{adjustbox}{width=\textwidth}
\scriptsize
\begin{tabularx}{\textwidth}{c | Y | Y | Y | Y | Y | Y | Y}
\toprule
Method & AUC Top-1 & AUC Top-10 & AUC Top-100 & Top-1 & Top-10 & Top-100 & Mean \\
\midrule
REINVENT & 1 & 1 & 1 & 1 & 1 & 1 & 1 \\
Graph GA & 2 & 2 & 2 & 3 & 2 & 3 & 2.33 \\
SELFIES-REINVENT & 3 & 3 & 4 & 4 & 3 & 2 & 3.16 \\
SMILES-LSTM-HC & 5 & 6 & 7 & 2 & 4 & 4 & 4.66 \\
GP BO & 4 & 4 & 5 & 6 & 5 & 5 & 4.83 \\
STONED & 6 & 5 & 3 & 7 & 7 & 6 & 5.66 \\
DoG-GEN & 7 & 9 & 11 & 5 & 6 & 7 & 7.5 \\
SMILES GA & 9 & 7 & 6 & 10 & 8 & 8 & 8 \\
DST & 11 & 10 & 9 & 9 & 10 & 9 & 9.66 \\
SynNet & 8 & 8 & 8 & 11 & 11 & 14 & 10 \\
SELFIES-LSTM-HC & 13 & 14 & 13 & 8 & 9 & 11 & 11.33 \\
MIMOSA & 14 & 12 & 10 & 14 & 12 & 10 & 12 \\
MARS & 12 & 11 & 12 & 12 & 13 & 13 & 12.16 \\
MolPAL & 10 & 13 & 15 & 13 & 15 & 16 & 13.66 \\
GA+D & 23 & 17 & 14 & 15 & 14 & 12 & 15.83 \\
DoG-AE & 15 & 15 & 17 & 17 & 17 & 17 & 16.33 \\
GFlowNet & 20 & 16 & 16 & 19 & 16 & 15 & 17 \\
SELFIES-VAE & 16 & 18 & 21 & 16 & 18 & 21 & 18.33 \\
Screening & 17 & 19 & 19 & 18 & 19 & 19 & 18.5 \\
SMILES-VAE & 18 & 20 & 20 & 20 & 20 & 20 & 19.66 \\
GFlowNet-AL & 22 & 22 & 18 & 23 & 21 & 18 & 20.66 \\
Pasithea & 19 & 21 & 23 & 21 & 22 & 22 & 21.33 \\
JT-VAE & 21 & 23 & 22 & 22 & 23 & 23 & 22.33 \\
Graph MCTS & 24 & 24 & 24 & 24 & 24 & 24 & 24 \\
MolDQN & 25 & 25 & 25 & 25 & 25 & 25 & 25 \\
\bottomrule  
\end{tabularx}
\end{adjustbox}
\end{table*}

\subsection{Results \& Analysis}
\label{sec:results}

The primary results are summarized in Table \ref{tab:task2} and \ref{tab:ranking}. For clarity, we only show the ten best-performing models in the table. We show a selective set of optimization curves in Figure \ref{fig:curves}. The remaining results are in the Appendix \ref{sec:main_results} and \ref{sec:additional_results}.

\begin{figure*}[t!]
  \centering
  \includegraphics[width=\textwidth]{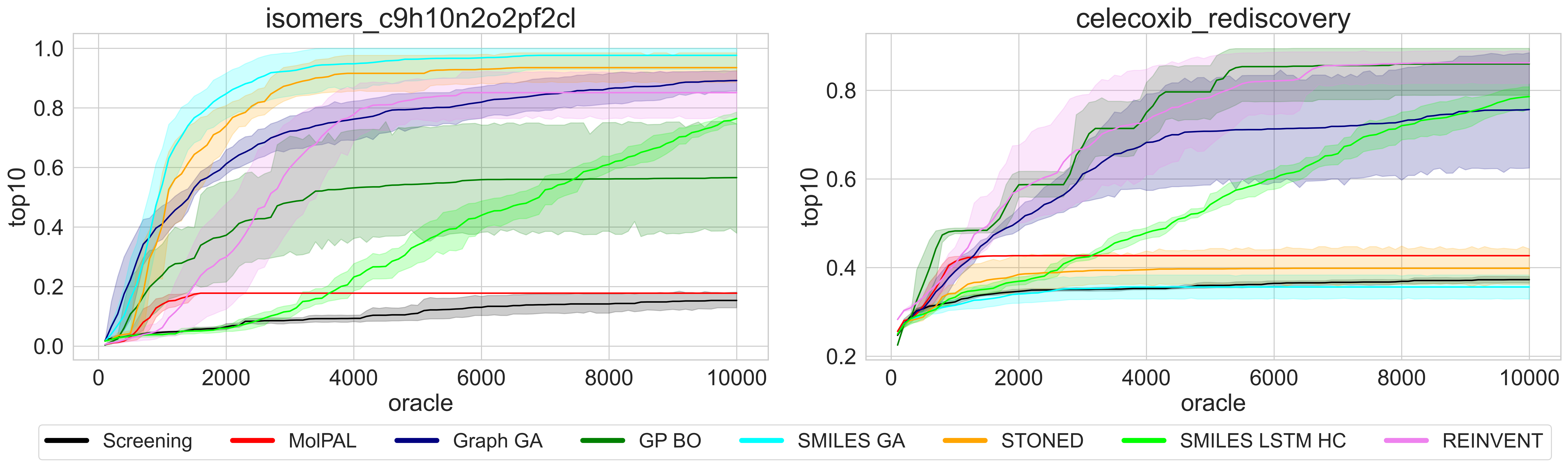}
  \caption{The optimization curves of top-10 average on optimizing isomer\_c9h10n2o2pf2cl and celecoxib\_rediscovery, as the representation of isomer-type and similarity-type oracles. Only 8 methods are displayed for clarity and full results are in the Appendix \ref{sec:main_results}.}
  \label{fig:curves}
\end{figure*}

\textbf{Sample efficiency matters.} A first observation from the results is that none of the methods we implemented can optimize the simple toy objectives within hundreds of oracle calls under our experimental settings, except some trivial ones like QED, DRD2, and osimertinib\_mpo, which emphasize the need for more efficient molecular optimization algorithms.
By comparing the ranking of AUC Top-10 and Top-10, we notice some methods have significantly different relative performances. For example, SMILES LSTM HC, which used to be seen as comparable to Graph GA, actually requires more oracle queries to achieve the same level of performance, while a related algorithm, REINVENT, requires far fewer (see Figure \ref{fig:curves}). These differences indicate the training algorithm of REINVENT is more efficient than HC, emphasizing the importance of AUC Top-10 as an evaluation metric. In addition, methods that assemble molecules either token-by-token or atom-by-atom from a single start point, such as GA+D, MolDQN, and Graph MCTS, are most data-inefficient. Those methods potentially cover broader chemical space and include many undesired candidates, such as unstable or unsynthesizable ones, which wastes a significant portion of the oracle budget and also imposes a strong requirement on the oracles' quality.




\textbf{Older algorithms are still powerful.} As shown in Table~\ref{tab:task2} and \ref{tab:ranking}, the best-performing algorithms are REINVENT and Graph GA among all the compared methods, despite both of them being released several years ago. 
However, we rarely see model development papers list these two methods as baselines.
The absence of a thorough benchmark has obfuscated the fact that  newer models  published in top AI conferences
do not seem to offer an improvement in performance by our metrics. Of course, we should acknowledge that some of the methods are developed to solve other problems in molecular optimization, such as strings' validity or synthesizability, and some might have opened new avenues to tackle the problem that could potentially be more efficient when mature. Still, some of the field's efforts and resources might be wasted due to a lack of a thorough and standardized benchmark. 

\begin{figure*}[t!]
  \centering
  \vspace{-20pt}
    \hspace{0cm}
    \begin{subfigure}{0.49\textwidth}
        \centering
        \includegraphics[width=\textwidth]{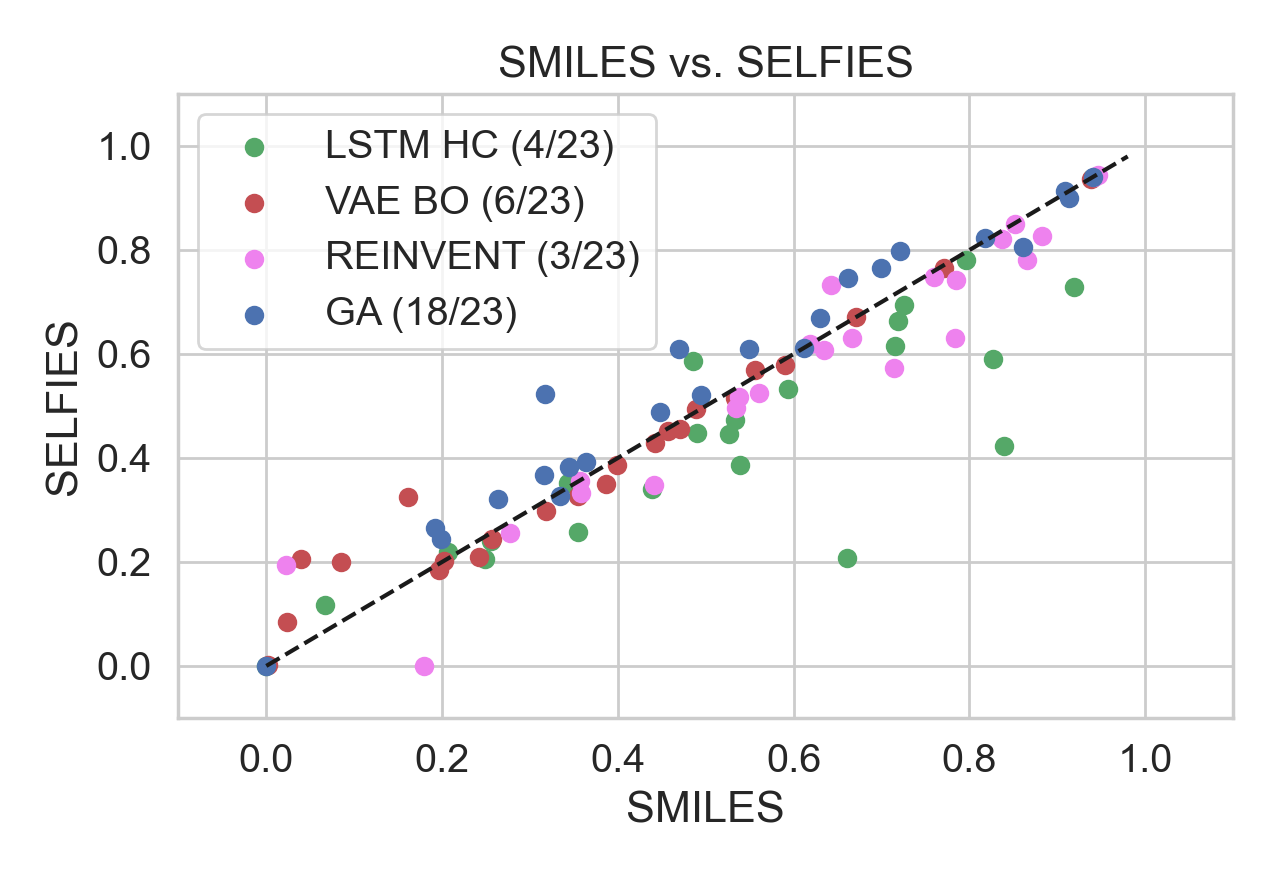}
        \caption{Comparison between SMILES- and SELFIES-based methods. Note GA is not a head-to-head comparison.}
        \label{fig:smiles_vs_selfies}
    \end{subfigure}
    \hspace{0cm}
    \begin{subfigure}{0.49\textwidth}
        \centering
        \includegraphics[width=\textwidth]{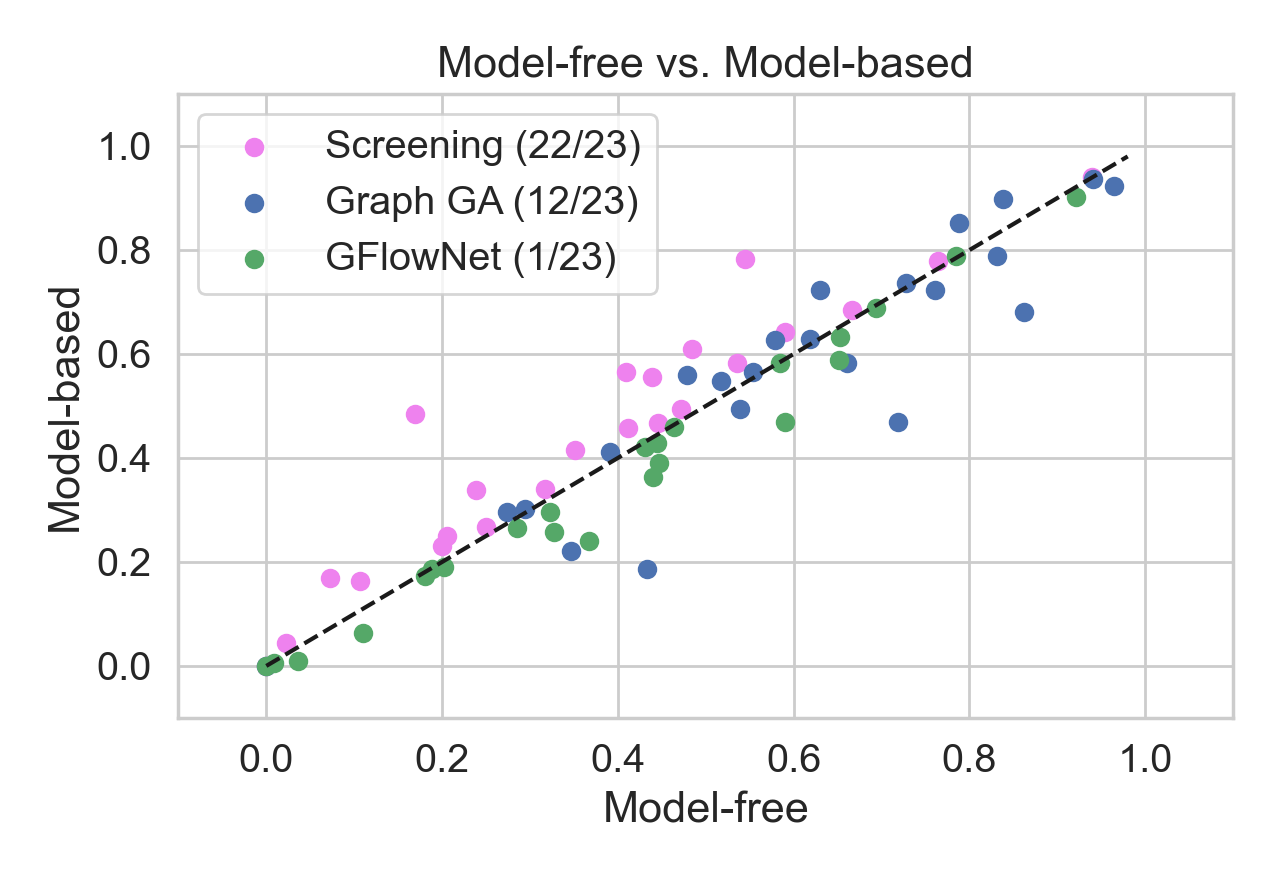}
        \caption{Comparison between model-free and corresponding model-based methods. }
        \label{fig:model_free_vs_based}
    \end{subfigure}
    \hspace{0cm}
  \caption{Each point represents the AUC Top-10 of one task, with x-axis the SMILES variant and y-axis the SELFIES variant of the same method. Colors are labeled by the optimization algorithms. The fractions of the tasks above the parity line are in parentheses.}
\end{figure*}

\textbf{There are no obvious shortcomings of SMILES.} 
SELFIES was designed as a substitute of SMILES to solve the syntactical invalidity problem met in SMILES representation and has been adopted by a number of recent studies. 
However, our head-to-head comparison of string-based methods, especially the ones leveraging language models, shows that most SELFIES variants cannot outperform their corresponding SMILES-based methods in terms of optimization ability and sample efficiency (Figure \ref{fig:smiles_vs_selfies}). 
We do observe some early methods like the initial version of SMILES VAE~\cite{gomez2018automatic} (2016) and ORGAN~\cite{sanchez2017optimizing} (2017) struggle to propose valid SMILES strings, but this is not an issue for more recent methods.
We believe this is partially because current language models are better able to learn the grammar of SMILES strings, which has flattened the advantage of SELFIES. Further, as shown in Appendix \ref{sec:bad_selfies}, more combinations of SELFIES tokens don't necessarily explore larger chemical space but might map to a small number of valid molecules that can be represented by truncated SELFIES strings, which implies that there are still syntax requirements in generating SELFIES strings to achieve effective exploration.

On the other hand, we observe a clear advantage of SELFIES-based GA compared to SMILES-based one, which indicates that SELFIES has an advantage over SMILES when we need to design the rules to manipulate the sequence.
However, we should note that the comparison is not head-to-head, as GAs' performances highly depend on the mutation and crossover rule design, but not the representation.
Graph GA's mutation rules are also encoded in SMARTS strings and operate on SMILES strings, which can also be seen as SMILES modification steps. Overall, when we need to design the generative action manually, the assembly strategy that could derive desired transformation more intuitively should be preferred.


\textbf{Model-based methods are potentially more efficient but need careful design.} 
It is widely recognized in the RL community that model-based optimization methods that explicitly leverage a predictive model (``world model'') are more sample efficient than the model-free ones~\cite{wang2019benchmarking}. Our results on MolPAL and screening verify the principle that training a predictive model is beneficial compared to random sampling (see Figure \ref{fig:model_free_vs_based}). However, the results of Graph GA (model-based variant: GP BO) and GFlowNet (model-based variant: GFlowNet-AL) indicate that simply adding a predictive model might not necessarily be helpful.
GP BO outperformed Graph GA in 12 tasks among 23, but Graph GA outperformed GP BO in the summation. 
GFlowNet outperformed GFlowNet-AL in almost every task. 
From the step-wise increment behavior (see Figure \ref{fig:curves}) and hyper-parameter tuning of GP BO (Appendix \ref{sec:hparams}), we conclude that the performance bottleneck is mainly the quality of the predictive model.
Further, GFlowNet-AL adopts a relatively naive model-based strategy that may suppress exploitation, especially when the model is not well-trained. Overall, we observe that model-based optimization algorithms have the potential to be more sample efficient but require careful design of the inner- and outer-loop optimization algorithms so the model does not lead the search astray.

\begin{figure*}[t!]
\parbox{.35\textwidth}{\caption{The heatmap and the clustering of oracles based on relative AUC Top-10. Relative AUC Top-10 is computed by normalizing AUC Top-10 values to a range from the lowest and the highest value within the task. The zaleplon\_mpo and sitagliptin\_mpo are multi-objective versions of isomer functions~\cite{brown2019guacamol}, while all other MPOs are based on similarity. Clear patterns emerge  between a large cluster of similarity-based oracles, four isomer-based oracles, and other non-clustered ones. Different types of landscape are more suitable for different kinds of methods to explore. The cluster tree was calculated with unweighted pair group method with arithmetic mean (UPGMA) using Euclidean distance.}
  \label{fig:cluster}
}
\parbox{.60\textwidth}{
  \includegraphics[width=0.75\textwidth]{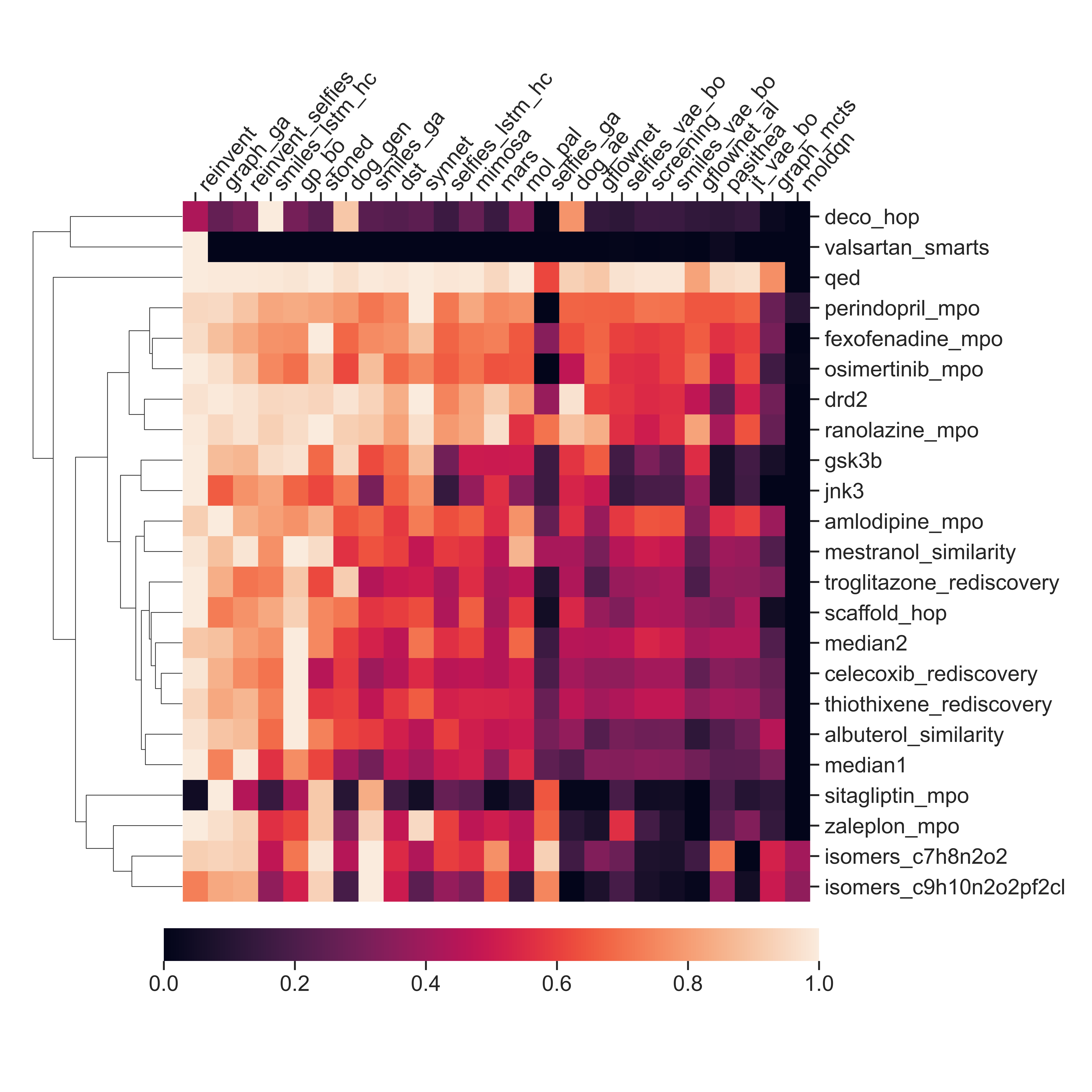}
} 
\vspace{-20pt}
\end{figure*}

\textbf{Different types of methods are more suitable for different kinds of landscapes.} As shown from Figure~\ref{fig:cluster} and Table~\ref{tab:task2}, we find that there are some clear clusters of oracles based on the relative performance of methods.
One clear pattern is that string-based GAs, such as SMILES GA and STONED, reach superior relative performance in tasks involving isomer functions, including isomer\_c7h8n2o2, isomer\_c9h10n2o2pf2cl, sitagliptin\_mpo, and zaleplon\_mpo. 
Isomer-type oracles are summations of atomic contribution, while all other MPOs are mainly based on similarity measured by fingerprints, and they generally have closer relative performance.
Among similarity-based oracles, the ones including logP and TPSA, such as fexofenadine\_mpo and osimertinib\_mpo, are clustered together against more naive similarities such as the rediscovery and median ones. 
The machine learning oracles predicting bioactivities belong to the same cluster of similarity-based oracles. While QED is too trivial that almost all methods reach very close values, deco\_hop, valsartan\_smarts, scaffold\_hop that are designed based on whether a molecule contains a substructure have varied performance. 
The results suggest that different types of landscape are better explored by different kinds of methods, such as string-based GA on isomer-type oracles. 
It is not evident which type of oracle is closest to a ``true'' pharmaceutical design objective, which is likely more complex and challenging to optimize; we leave further investigation on oracle landscapes and their influence on optimization to future work.


 
\textbf{Hyperparameter reoptimization and multiple runs are required when reporting results.} We also observed that the optimal set of hyper-parameters is always not the default ones suggested by a method's original paper (see Appendix \ref{sec:hparams}).
For example, REINVENT's performance is highly dependent on $\sigma$; we found the best-performing value to be much larger than the values suggested in the original paper (see Figure \ref{fig:tune_reinvent1} and \ref{fig:tune_reinvent2})~\cite{olivecrona2017molecular}. We conclude that this is due to unique demands of our setting of limited oracle budget, which was not a goal of the original study, and thus suggest reoptimizing the hyper-parameters whenever the testing environment is changed. 
Another challenge is the non-determinism of most algorithms. For example, Graph GA suffers from a relatively large variance due to its random-walk-like exploration, as does GP BO. 
If the oracle were a costly experimental evaluation, we might consider the worst-case performance as an endpoint to reduce the risk rather than the average performance, highlighting the importance of running multiple independent runs and reporting the distribution of outcomes.





\section{Conclusions}
\label{sec:conclusion}

This paper proposes \mname: a standardized molecular design benchmark focusing on sample efficiency as a key impediment to experimental adoption. 
We conduct a thorough investigation across \nmethod~methods and \noracle~objectives to determine the current state-of-the-art, investigate problems, and draw insights for future studies. Our primary observations are that (1) methods considered to be strong baselines, like LSTM HC, may be inefficient in data usage; (2) several older methods, like REINVENT and Graph GA, outperform more recent ones; (3) SELFIES does not seem to offer an immediate benefit in optimization performance compared to SMILES except in GA; (4) model-based methods have the potential to be more sample efficient but require careful design of the inner-loop, outer-loop, and the predictive model; and (5) different optimization algorithms may excel at different tasks, determined by the landscapes of oracle functions; which algorithm to select is still dependent on the use case and the type of tasks.

We acknowledge several limitations of the current study:  we cannot exhaustively explore every method and thoroughly tune every hyperparameter, the representative methods we implement might not be the best-in-class among all possible variants, our conclusion might be biased toward similarity-based oracles, and we are not thoroughly investigating other important quantities such as synthesizability~\cite{gao2020synthesizability} and diversity~\cite{huang2021therapeutics}. We also emphasize that our experiments consider the number of oracle calls from scratch, i.e., the data used to train the surrogate models in model-based methods are counted in the total budget. If a dataset has been collected previously, it may be prudent to train a surrogate model on this information and use a model-based method as illustrated by Tripp et al.~\cite{tripp2021fresh}. We will support the continued development of this benchmark to minimize the wasted effort caused by non-reproducibility and poor baselines to boost the field's growth toward solving practical molecular design problems.

We would like to conclude with recommendations for subsequent studies: (1) When comparing baselines, it is important to run algorithms under the same oracle budgets; (2) For general-purpose molecular design algorithms, one should test on multiple types of oracles;
(3) Conducting multiple independent runs and reporting the distribution of outcomes is critical for non-deterministic methods; (4) Whenever the tasks and testing environment are changed, hyperparameter tuning is necessary.


\begin{ack}
This research was supported by the Office of Naval Research under grant number N00014-21-1-2195 and the Machine Learning for Pharmaceutical Discovery and Synthesis consortium. Any opinions, findings, and conclusions or recommendations expressed in this material are those of the author(s) and do not necessarily reflect the views of the Office of Naval Research.
W.G. received additional funding from MIT-Takeda fellowship.
T.F. and J.S. were supported by NSF award SCH-2205289, SCH-2014438, IIS-1838042, NIH award R01 1R01NS107291-01.
We thank Samuel Goldman and John Bradshaw for commenting on the manuscript.
\end{ack}

\section*{Reproducibility Statement}

All code, parameters, and releasable data can be found at \url{https://github.com/wenhao-gao/mol_opt}, including instructions in a README file. All results generated in this experiment can be found at \url{https://figshare.com/articles/dataset/Results_for_practival_molecular_optimization_PMO_benchmark/20123453}.
Appendix \ref{sec:implement} describe the experimental setup, implementation details, datasets used, and hardware configuration. 


\bibliographystyle{unsrt}
\setcitestyle{numbers,open={[},close={]},citesep={,}}
\bibliography{references}

\section*{Checklist}

\begin{enumerate}

\item For all authors...
\begin{enumerate}
  \item Do the main claims made in the abstract and introduction accurately reflect the paper's contributions and scope?
    \answerYes{}
  \item Did you describe the limitations of your work?
    \answerYes{See Section~\ref{sec:conclusion}.} 
  \item Did you discuss any potential negative societal impacts of your work?
    \answerNA{}
  \item Have you read the ethics review guidelines and ensured that your paper conforms to them?
    \answerYes{}
\end{enumerate}

\item If you are including theoretical results...
\begin{enumerate}
  \item Did you state the full set of assumptions of all theoretical results?
    \answerNA{}
	\item Did you include complete proofs of all theoretical results?
    \answerNA{}
\end{enumerate}

\item If you ran experiments (e.g. for benchmarks)...
\begin{enumerate}
  \item Did you include the code, data, and instructions needed to reproduce the main experimental results (either in the supplemental material or as a URL)?
    \answerYes{All code, parameters, and releasable data can be found at \url{https://github.com/wenhao-gao/mol_opt}, including instructions in a README file.
Appendix \ref{sec:implement} describe the experimental setup, implementation details, datasets used, and hardware configuration. }
  \item Did you specify all the training details (e.g., data splits, hyperparameters, how they were chosen)?
    \answerYes{Please see Section~\ref{sec:implement}. }
	\item Did you report error bars (e.g., with respect to the random seed after running experiments multiple times)?
    \answerYes{Please see Table~\ref{tab:task2}. }
	\item Did you include the total amount of compute and the type of resources used (e.g., type of GPUs, internal cluster, or cloud provider)?
    \answerYes{Please see Section~\ref{sec:hardware}. }
\end{enumerate}

\item If you are using existing assets (e.g., code, data, models) or curating/releasing new assets...
\begin{enumerate}
  \item If your work uses existing assets, did you cite the creators?
    \answerYes{}
  \item Did you mention the license of the assets?
    \answerYes{Please see Section~\ref{sec:license} for details. }
  \item Did you include any new assets either in the supplemental material or as a URL?
    \answerYes{We release the code repository at \url{https://github.com/wenhao-gao/mol_opt}, including instructions in a README file.
Appendix \ref{sec:implement} and \ref{sec:configuration} describe the experimental setup, implementation details, datasets used, and hardware configuration. }
  \item Did you discuss whether and how consent was obtained from people whose data you're using/curating?
    \answerYes{All the data/codes we use are publicly available. Please see Section~\ref{sec:implement} for details. }
  \item Did you discuss whether the data you are using/curating contains personally identifiable information or offensive content?
    \answerYes{Our paper does not involve human subjects research. It also does not contain any personally identifiable information or offensive content.}
\end{enumerate}

\item If you used crowdsourcing or conducted research with human subjects...
\begin{enumerate}
  \item Did you include the full text of instructions given to participants and screenshots, if applicable?
    \answerNA{}
  \item Did you describe any potential participant risks, with links to Institutional Review Board (IRB) approvals, if applicable?
    \answerNA{}
  \item Did you include the estimated hourly wage paid to participants and the total amount spent on participant compensation?
    \answerNA{}
\end{enumerate}

\end{enumerate}


\appendix

\newpage

\section{Main Results}
\label{sec:main_results}

\subsection{AUC Top-10 Table}

\begin{table*}[h]
\centering
\caption{We report the mean and standard deviation of AUC Top-10 from 5 independent runs. We ranked the methods by the summation of mean AUC Top-10 of all tasks. 
(Continued)
}
\label{tab:full_1}
\begin{adjustbox}{width=\textwidth}
\scriptsize
\begin{tabularx}{\textwidth}{c | Y | Y | Y | Y | Y}
\toprule
Method & REINVENT & Graph GA & REINVENT SELFIES & GP BO & STONED \\
Assembly & SMILES & Fragments & SELFIES & Fragments & SELFIES \\
\midrule
albuterol\_similarity & 0.882$\pm$0.006 & 0.838$\pm$0.016 & 0.826$\pm$0.030 & \textbf{0.898$\pm$0.014} & 0.745$\pm$0.076 \\
amlodipine\_mpo & 0.635$\pm$0.035 & \textbf{0.661$\pm$0.020} & 0.607$\pm$0.014 & 0.583$\pm$0.044 & 0.608$\pm$0.046 \\
celecoxib\_rediscovery & 0.713$\pm$0.067 & 0.630$\pm$0.097 & 0.573$\pm$0.043 & \textbf{0.723$\pm$0.053} & 0.382$\pm$0.041 \\
deco\_hop & 0.666$\pm$0.044 & 0.619$\pm$0.004 & 0.631$\pm$0.012 & 0.629$\pm$0.018 & 0.611$\pm$0.008 \\
drd2 & 0.945$\pm$0.007 & 0.964$\pm$0.012 & 0.943$\pm$0.005 & 0.923$\pm$0.017 & 0.913$\pm$0.020 \\
fexofenadine\_mpo & 0.784$\pm$0.006 & 0.760$\pm$0.011 & 0.741$\pm$0.002 & 0.722$\pm$0.005 & \textbf{0.797$\pm$0.016} \\
gsk3b & \textbf{0.865$\pm$0.043} & 0.788$\pm$0.070 & 0.780$\pm$0.037 & 0.851$\pm$0.041 & 0.668$\pm$0.049 \\
isomers\_c7h8n2o2 & 0.852$\pm$0.036 & 0.862$\pm$0.065 & 0.849$\pm$0.034 & 0.680$\pm$0.117 & 0.899$\pm$0.011 \\
isomers\_c9h10n2o2pf2cl & 0.642$\pm$0.054 & 0.719$\pm$0.047 & 0.733$\pm$0.029 & 0.469$\pm$0.180 & 0.805$\pm$0.031 \\
jnk3 & \textbf{0.783$\pm$0.023} & 0.553$\pm$0.136 & 0.631$\pm$0.064 & 0.564$\pm$0.155 & 0.523$\pm$0.092 \\
median1 & \textbf{0.356$\pm$0.009} & 0.294$\pm$0.021 & 0.355$\pm$0.011 & 0.301$\pm$0.014 & 0.266$\pm$0.016 \\
median2 & 0.276$\pm$0.008 & 0.273$\pm$0.009 & 0.255$\pm$0.005 & \textbf{0.297$\pm$0.009} & 0.245$\pm$0.032 \\
mestranol\_similarity & 0.618$\pm$0.048 & 0.579$\pm$0.022 & 0.620$\pm$0.029 & \textbf{0.627$\pm$0.089} & 0.609$\pm$0.101 \\
osimertinib\_mpo & \textbf{0.837$\pm$0.009} & 0.831$\pm$0.005 & 0.820$\pm$0.003 & 0.787$\pm$0.006 & 0.822$\pm$0.012 \\
perindopril\_mpo & 0.537$\pm$0.016 & 0.538$\pm$0.009 & 0.517$\pm$0.021 & 0.493$\pm$0.011 & 0.488$\pm$0.011 \\
qed & 0.941$\pm$0.000 & 0.940$\pm$0.000 & 0.940$\pm$0.000 & 0.937$\pm$0.000 & 0.941$\pm$0.000 \\
ranolazine\_mpo & 0.760$\pm$0.009 & 0.728$\pm$0.012 & 0.748$\pm$0.018 & 0.735$\pm$0.013 & \textbf{0.765$\pm$0.029} \\
scaffold\_hop & \textbf{0.560$\pm$0.019} & 0.517$\pm$0.007 & 0.525$\pm$0.013 & 0.548$\pm$0.019 & 0.521$\pm$0.034 \\
sitagliptin\_mpo & 0.021$\pm$0.003 & \textbf{0.433$\pm$0.075} & 0.194$\pm$0.121 & 0.186$\pm$0.055 & 0.393$\pm$0.083 \\
thiothixene\_rediscovery & 0.534$\pm$0.013 & 0.479$\pm$0.025 & 0.495$\pm$0.040 & \textbf{0.559$\pm$0.027} & 0.367$\pm$0.027 \\
troglitazone\_rediscovery & \textbf{0.441$\pm$0.032} & 0.390$\pm$0.016 & 0.348$\pm$0.012 & 0.410$\pm$0.015 & 0.320$\pm$0.018 \\
valsartan\_smarts & \textbf{0.179$\pm$0.358} & 0.000$\pm$0.000 & 0.000$\pm$0.000 & 0.000$\pm$0.000 & 0.000$\pm$0.000 \\
zaleplon\_mpo & \textbf{0.358$\pm$0.062} & 0.346$\pm$0.032 & 0.333$\pm$0.026 & 0.221$\pm$0.072 & 0.325$\pm$0.027 \\
\midrule
Sum & 14.196 & 13.751 & 13.471 & 13.156 & 13.024 \\
Rank & 1 & 2 & 3 & 4 & 5 \\

\hline
\hline

Method & LSTM HC & SMILES GA & SynNet & DoG-Gen & DST \\
Assembly & SMILES & SMILES & Synthesis & Synthesis & Fragments \\
\midrule
albuterol\_similarity & 0.719$\pm$0.018 & 0.661$\pm$0.066 & 0.584$\pm$0.039 & 0.676$\pm$0.013 & 0.619$\pm$0.020 \\
amlodipine\_mpo & 0.593$\pm$0.016 & 0.549$\pm$0.009 & 0.565$\pm$0.007 & 0.536$\pm$0.003 & 0.516$\pm$0.007 \\
celecoxib\_rediscovery & 0.539$\pm$0.018 & 0.344$\pm$0.027 & 0.441$\pm$0.027 & 0.464$\pm$0.009 & 0.380$\pm$0.006 \\
deco\_hop & \textbf{0.826$\pm$0.017} & 0.611$\pm$0.006 & 0.613$\pm$0.009 & 0.800$\pm$0.007 & 0.608$\pm$0.008 \\
drd2 & 0.919$\pm$0.015 & 0.908$\pm$0.019 & \textbf{0.969$\pm$0.004} & 0.948$\pm$0.001 & 0.820$\pm$0.014 \\
fexofenadine\_mpo & 0.725$\pm$0.003 & 0.721$\pm$0.015 & 0.761$\pm$0.015 & 0.695$\pm$0.003 & 0.725$\pm$0.005 \\
gsk3b & 0.839$\pm$0.015 & 0.629$\pm$0.044 & 0.789$\pm$0.032 & 0.831$\pm$0.021 & 0.671$\pm$0.032 \\
isomers\_c7h8n2o2 & 0.485$\pm$0.045 & \textbf{0.913$\pm$0.021} & 0.455$\pm$0.031 & 0.465$\pm$0.018 & 0.548$\pm$0.069 \\
isomers\_c9h10n2o2pf2cl & 0.342$\pm$0.027 & \textbf{0.860$\pm$0.065} & 0.241$\pm$0.064 & 0.199$\pm$0.016 & 0.458$\pm$0.063 \\
jnk3 & 0.661$\pm$0.039 & 0.316$\pm$0.022 & 0.630$\pm$0.034 & 0.595$\pm$0.023 & 0.556$\pm$0.057 \\
median1 & 0.255$\pm$0.010 & 0.192$\pm$0.012 & 0.218$\pm$0.008 & 0.217$\pm$0.001 & 0.232$\pm$0.009 \\
median2 & 0.248$\pm$0.008 & 0.198$\pm$0.005 & 0.235$\pm$0.006 & 0.212$\pm$0.000 & 0.185$\pm$0.020 \\
mestranol\_similarity & 0.526$\pm$0.032 & 0.469$\pm$0.029 & 0.399$\pm$0.021 & 0.437$\pm$0.007 & 0.450$\pm$0.027 \\
osimertinib\_mpo & 0.796$\pm$0.002 & 0.817$\pm$0.011 & 0.796$\pm$0.003 & 0.774$\pm$0.002 & 0.785$\pm$0.004 \\
perindopril\_mpo & 0.489$\pm$0.007 & 0.447$\pm$0.013 & \textbf{0.557$\pm$0.011} & 0.474$\pm$0.002 & 0.462$\pm$0.008 \\
qed & 0.939$\pm$0.000 & 0.940$\pm$0.000 & \textbf{0.941$\pm$0.000} & 0.934$\pm$0.000 & 0.938$\pm$0.000 \\
ranolazine\_mpo & 0.714$\pm$0.008 & 0.699$\pm$0.026 & 0.741$\pm$0.010 & 0.711$\pm$0.006 & 0.632$\pm$0.054 \\
scaffold\_hop & 0.533$\pm$0.012 & 0.494$\pm$0.011 & 0.502$\pm$0.012 & 0.515$\pm$0.005 & 0.497$\pm$0.004 \\
sitagliptin\_mpo & 0.066$\pm$0.019 & 0.363$\pm$0.057 & 0.025$\pm$0.014 & 0.048$\pm$0.008 & 0.075$\pm$0.032 \\
thiothixene\_rediscovery & 0.438$\pm$0.008 & 0.315$\pm$0.017 & 0.401$\pm$0.019 & 0.375$\pm$0.004 & 0.366$\pm$0.006 \\
troglitazone\_rediscovery & 0.354$\pm$0.016 & 0.263$\pm$0.024 & 0.283$\pm$0.008 & 0.416$\pm$0.019 & 0.279$\pm$0.019 \\
valsartan\_smarts & 0.000$\pm$0.000 & 0.000$\pm$0.000 & 0.000$\pm$0.000 & 0.000$\pm$0.000 & 0.000$\pm$0.000 \\
zaleplon\_mpo & 0.206$\pm$0.006 & 0.334$\pm$0.041 & 0.341$\pm$0.011 & 0.123$\pm$0.016 & 0.176$\pm$0.045 \\
\midrule
Sum & 12.223 & 12.054 & 11.498 & 11.456 & 10.989 \\
Rank & 6 & 7 & 8 & 9 & 10 \\
\bottomrule
\end{tabularx}
\end{adjustbox}
\end{table*}

\newpage

\begin{table*}[h]
\centering
\caption{(Continued)
}
\label{tab:full_2}
\begin{adjustbox}{width=\textwidth}
\scriptsize
\begin{tabularx}{\textwidth}{c | Y | Y | Y | Y | Y}
\toprule
Method & MARS & MIMOSA & MolPal & LSTM HC SELFIES & DoG-AE \\
Assembly & Fragments & Fragments & - & SELFIES & Synthesis \\
\midrule
albuterol\_similarity & 0.597$\pm$0.124 & 0.618$\pm$0.017 & 0.609$\pm$0.002 & 0.664$\pm$0.030 & 0.533$\pm$0.034 \\
amlodipine\_mpo & 0.504$\pm$0.016 & 0.543$\pm$0.003 & 0.582$\pm$0.008 & 0.532$\pm$0.004 & 0.507$\pm$0.005 \\
celecoxib\_rediscovery & 0.379$\pm$0.060 & 0.393$\pm$0.010 & 0.415$\pm$0.001 & 0.385$\pm$0.008 & 0.355$\pm$0.012 \\
deco\_hop & 0.589$\pm$0.003 & 0.619$\pm$0.003 & 0.643$\pm$0.005 & 0.590$\pm$0.001 & 0.765$\pm$0.055 \\
drd2 & 0.891$\pm$0.020 & 0.799$\pm$0.017 & 0.783$\pm$0.009 & 0.729$\pm$0.034 & 0.943$\pm$0.009 \\
fexofenadine\_mpo & 0.711$\pm$0.006 & 0.706$\pm$0.011 & 0.685$\pm$0.000 & 0.693$\pm$0.004 & 0.679$\pm$0.017 \\
gsk3b & 0.552$\pm$0.037 & 0.554$\pm$0.042 & 0.555$\pm$0.011 & 0.423$\pm$0.018 & 0.601$\pm$0.091 \\
isomers\_c7h8n2o2 & 0.728$\pm$0.027 & 0.564$\pm$0.046 & 0.484$\pm$0.006 & 0.587$\pm$0.031 & 0.239$\pm$0.077 \\
isomers\_c9h10n2o2pf2cl & 0.581$\pm$0.013 & 0.303$\pm$0.046 & 0.164$\pm$0.003 & 0.352$\pm$0.019 & 0.049$\pm$0.015 \\
jnk3 & 0.489$\pm$0.095 & 0.360$\pm$0.063 & 0.339$\pm$0.009 & 0.207$\pm$0.013 & 0.469$\pm$0.138 \\
median1 & 0.207$\pm$0.011 & 0.243$\pm$0.005 & 0.249$\pm$0.001 & 0.239$\pm$0.009 & 0.171$\pm$0.009 \\
median2 & 0.181$\pm$0.011 & 0.214$\pm$0.002 & 0.230$\pm$0.000 & 0.205$\pm$0.005 & 0.182$\pm$0.006 \\
mestranol\_similarity & 0.388$\pm$0.026 & 0.438$\pm$0.015 & 0.564$\pm$0.004 & 0.446$\pm$0.009 & 0.370$\pm$0.014 \\
osimertinib\_mpo & 0.777$\pm$0.006 & 0.788$\pm$0.014 & 0.779$\pm$0.000 & 0.780$\pm$0.005 & 0.750$\pm$0.012 \\
perindopril\_mpo & 0.462$\pm$0.006 & 0.490$\pm$0.011 & 0.467$\pm$0.002 & 0.448$\pm$0.006 & 0.432$\pm$0.013 \\
qed & 0.930$\pm$0.003 & 0.939$\pm$0.000 & 0.940$\pm$0.000 & 0.938$\pm$0.000 & 0.926$\pm$0.003 \\
ranolazine\_mpo & 0.740$\pm$0.010 & 0.640$\pm$0.015 & 0.457$\pm$0.005 & 0.614$\pm$0.010 & 0.689$\pm$0.015 \\
scaffold\_hop & 0.469$\pm$0.004 & 0.507$\pm$0.015 & 0.494$\pm$0.000 & 0.472$\pm$0.002 & 0.489$\pm$0.010 \\
sitagliptin\_mpo & 0.016$\pm$0.003 & 0.102$\pm$0.023 & 0.043$\pm$0.001 & 0.116$\pm$0.012 & 0.009$\pm$0.005 \\
thiothixene\_rediscovery & 0.344$\pm$0.022 & 0.347$\pm$0.018 & 0.339$\pm$0.001 & 0.339$\pm$0.009 & 0.314$\pm$0.015 \\
troglitazone\_rediscovery & 0.256$\pm$0.016 & 0.299$\pm$0.009 & 0.268$\pm$0.000 & 0.257$\pm$0.002 & 0.259$\pm$0.016 \\
valsartan\_smarts & 0.000$\pm$0.000 & 0.000$\pm$0.000 & 0.000$\pm$0.000 & 0.000$\pm$0.000 & 0.000$\pm$0.000 \\
zaleplon\_mpo & 0.187$\pm$0.046 & 0.172$\pm$0.036 & 0.168$\pm$0.003 & 0.218$\pm$0.020 & 0.049$\pm$0.027 \\
\midrule
Sum & 10.989 & 10.651 & 10.268 & 10.246 & 9.790 \\
Rank & 11 & 12 & 13 & 14 & 15 \\

\hline
\hline

Method & GFlowNet & GA+D & VAE BO SELFIES & Screening & VAE BO SMILES \\
Assembly & Fragments & SELFIES & SELFIES & - & SMILES \\
\midrule
albuterol\_similarity & 0.447$\pm$0.012 & 0.495$\pm$0.025 & 0.494$\pm$0.012 & 0.483$\pm$0.006 & 0.489$\pm$0.007 \\
amlodipine\_mpo & 0.444$\pm$0.004 & 0.400$\pm$0.032 & 0.516$\pm$0.005 & 0.535$\pm$0.001 & 0.533$\pm$0.009 \\
celecoxib\_rediscovery & 0.327$\pm$0.004 & 0.223$\pm$0.025 & 0.326$\pm$0.007 & 0.351$\pm$0.005 & 0.354$\pm$0.002 \\
deco\_hop & 0.583$\pm$0.002 & 0.550$\pm$0.005 & 0.579$\pm$0.001 & 0.590$\pm$0.001 & 0.589$\pm$0.001 \\
drd2 & 0.590$\pm$0.070 & 0.382$\pm$0.205 & 0.569$\pm$0.039 & 0.545$\pm$0.015 & 0.555$\pm$0.043 \\
fexofenadine\_mpo & 0.693$\pm$0.006 & 0.587$\pm$0.007 & 0.670$\pm$0.004 & 0.666$\pm$0.004 & 0.671$\pm$0.003 \\
gsk3b & 0.651$\pm$0.026 & 0.342$\pm$0.019 & 0.350$\pm$0.034 & 0.438$\pm$0.034 & 0.386$\pm$0.006 \\
isomers\_c7h8n2o2 & 0.366$\pm$0.043 & 0.854$\pm$0.015 & 0.325$\pm$0.028 & 0.168$\pm$0.034 & 0.161$\pm$0.017 \\
isomers\_c9h10n2o2pf2cl & 0.110$\pm$0.031 & 0.657$\pm$0.020 & 0.200$\pm$0.030 & 0.106$\pm$0.021 & 0.084$\pm$0.009 \\
jnk3 & 0.440$\pm$0.022 & 0.219$\pm$0.021 & 0.208$\pm$0.022 & 0.238$\pm$0.024 & 0.241$\pm$0.026 \\
median1 & 0.202$\pm$0.004 & 0.180$\pm$0.009 & 0.201$\pm$0.003 & 0.205$\pm$0.005 & 0.202$\pm$0.006 \\
median2 & 0.180$\pm$0.000 & 0.121$\pm$0.005 & 0.185$\pm$0.001 & 0.200$\pm$0.004 & 0.195$\pm$0.001 \\
mestranol\_similarity & 0.322$\pm$0.007 & 0.371$\pm$0.016 & 0.386$\pm$0.009 & 0.409$\pm$0.019 & 0.399$\pm$0.005 \\
osimertinib\_mpo & 0.784$\pm$0.001 & 0.672$\pm$0.027 & 0.765$\pm$0.002 & 0.764$\pm$0.001 & 0.771$\pm$0.002 \\
perindopril\_mpo & 0.430$\pm$0.010 & 0.172$\pm$0.088 & 0.429$\pm$0.003 & 0.445$\pm$0.004 & 0.442$\pm$0.004 \\
qed & 0.921$\pm$0.004 & 0.860$\pm$0.014 & 0.936$\pm$0.001 & 0.938$\pm$0.000 & 0.938$\pm$0.000 \\
ranolazine\_mpo & 0.652$\pm$0.002 & 0.555$\pm$0.015 & 0.452$\pm$0.025 & 0.411$\pm$0.010 & 0.457$\pm$0.012 \\
scaffold\_hop & 0.463$\pm$0.002 & 0.413$\pm$0.009 & 0.455$\pm$0.004 & 0.471$\pm$0.002 & 0.470$\pm$0.003 \\
sitagliptin\_mpo & 0.008$\pm$0.003 & 0.281$\pm$0.022 & 0.084$\pm$0.015 & 0.022$\pm$0.003 & 0.023$\pm$0.004 \\
thiothixene\_rediscovery & 0.285$\pm$0.012 & 0.223$\pm$0.029 & 0.297$\pm$0.004 & 0.317$\pm$0.003 & 0.317$\pm$0.007 \\
troglitazone\_rediscovery & 0.188$\pm$0.001 & 0.152$\pm$0.013 & 0.243$\pm$0.004 & 0.249$\pm$0.003 & 0.257$\pm$0.003 \\
valsartan\_smarts & 0.000$\pm$0.000 & 0.000$\pm$0.000 & 0.002$\pm$0.003 & 0.000$\pm$0.000 & 0.002$\pm$0.004 \\
zaleplon\_mpo & 0.035$\pm$0.030 & 0.244$\pm$0.015 & 0.206$\pm$0.015 & 0.072$\pm$0.014 & 0.039$\pm$0.012 \\
\midrule
Sum & 9.131 & 8.964 & 8.887 & 8.635 & 8.587 \\
Rank & 16 & 17 & 18 & 19 & 20 \\
\bottomrule
\end{tabularx}
\end{adjustbox}
\end{table*}

\newpage

\begin{table*}[h]
\centering
\caption{(Continued)
}
\label{tab:full_3}
\begin{adjustbox}{width=\textwidth}
\scriptsize
\begin{tabularx}{\textwidth}{c | Y | Y | Y | Y | Y}
\toprule
Method & Pasithea & GFlowNet-AL & JT-VAE BO & Graph MCTS & MolDQN \\
Assembly & SELFIES & Fragments & Fragments & Atoms & Atoms \\
\midrule
albuterol\_similarity & 0.447$\pm$0.007 & 0.390$\pm$0.008 & 0.485$\pm$0.029 & 0.580$\pm$0.023 & 0.320$\pm$0.015 \\
amlodipine\_mpo & 0.504$\pm$0.003 & 0.428$\pm$0.002 & 0.519$\pm$0.009 & 0.447$\pm$0.008 & 0.311$\pm$0.008 \\
celecoxib\_rediscovery & 0.312$\pm$0.007 & 0.257$\pm$0.003 & 0.299$\pm$0.009 & 0.264$\pm$0.013 & 0.099$\pm$0.005 \\
deco\_hop & 0.579$\pm$0.001 & 0.583$\pm$0.001 & 0.585$\pm$0.002 & 0.554$\pm$0.002 & 0.546$\pm$0.001 \\
drd2 & 0.255$\pm$0.040 & 0.468$\pm$0.046 & 0.506$\pm$0.136 & 0.300$\pm$0.050 & 0.025$\pm$0.001 \\
fexofenadine\_mpo & 0.660$\pm$0.015 & 0.688$\pm$0.002 & 0.667$\pm$0.010 & 0.574$\pm$0.009 & 0.478$\pm$0.012 \\
gsk3b & 0.281$\pm$0.038 & 0.588$\pm$0.015 & 0.350$\pm$0.051 & 0.281$\pm$0.022 & 0.241$\pm$0.008 \\
isomers\_c7h8n2o2 & 0.673$\pm$0.030 & 0.241$\pm$0.055 & 0.103$\pm$0.016 & 0.530$\pm$0.035 & 0.431$\pm$0.035 \\
isomers\_c9h10n2o2pf2cl & 0.345$\pm$0.145 & 0.064$\pm$0.012 & 0.090$\pm$0.035 & 0.454$\pm$0.067 & 0.342$\pm$0.026 \\
jnk3 & 0.154$\pm$0.018 & 0.362$\pm$0.021 & 0.222$\pm$0.009 & 0.110$\pm$0.019 & 0.111$\pm$0.008 \\
median1 & 0.178$\pm$0.009 & 0.190$\pm$0.002 & 0.179$\pm$0.003 & 0.195$\pm$0.005 & 0.122$\pm$0.007 \\
median2 & 0.179$\pm$0.004 & 0.173$\pm$0.001 & 0.180$\pm$0.003 & 0.132$\pm$0.002 & 0.088$\pm$0.003 \\
mestranol\_similarity & 0.361$\pm$0.016 & 0.295$\pm$0.004 & 0.356$\pm$0.013 & 0.281$\pm$0.008 & 0.188$\pm$0.007 \\
osimertinib\_mpo & 0.749$\pm$0.007 & 0.787$\pm$0.003 & 0.775$\pm$0.004 & 0.700$\pm$0.004 & 0.674$\pm$0.006 \\
perindopril\_mpo & 0.421$\pm$0.008 & 0.421$\pm$0.002 & 0.430$\pm$0.009 & 0.277$\pm$0.013 & 0.213$\pm$0.043 \\
qed & 0.931$\pm$0.002 & 0.902$\pm$0.005 & 0.934$\pm$0.002 & 0.892$\pm$0.006 & 0.731$\pm$0.018 \\
ranolazine\_mpo & 0.347$\pm$0.012 & 0.632$\pm$0.007 & 0.508$\pm$0.055 & 0.239$\pm$0.027 & 0.051$\pm$0.020 \\
scaffold\_hop & 0.456$\pm$0.003 & 0.460$\pm$0.002 & 0.470$\pm$0.005 & 0.412$\pm$0.003 & 0.405$\pm$0.004 \\
sitagliptin\_mpo & 0.088$\pm$0.013 & 0.006$\pm$0.001 & 0.046$\pm$0.027 & 0.056$\pm$0.012 & 0.003$\pm$0.002 \\
thiothixene\_rediscovery & 0.288$\pm$0.006 & 0.266$\pm$0.005 & 0.282$\pm$0.008 & 0.231$\pm$0.004 & 0.099$\pm$0.007 \\
troglitazone\_rediscovery & 0.240$\pm$0.002 & 0.186$\pm$0.003 & 0.237$\pm$0.005 & 0.224$\pm$0.009 & 0.122$\pm$0.004 \\
valsartan\_smarts & 0.006$\pm$0.012 & 0.000$\pm$0.000 & 0.000$\pm$0.000 & 0.000$\pm$0.000 & 0.000$\pm$0.000 \\
zaleplon\_mpo & 0.091$\pm$0.013 & 0.010$\pm$0.001 & 0.125$\pm$0.038 & 0.058$\pm$0.019 & 0.010$\pm$0.005 \\
\midrule
Sum & 8.556 & 8.406 & 8.358 & 7.803 & 5.620 \\
Rank & 21 & 22 & 23 & 24 & 25 \\
\bottomrule
\end{tabularx}
\end{adjustbox}
\end{table*}

\newpage

\subsection{Optimization Curves}

\begin{figure*}[h!]
  \centering
  \includegraphics[width=\textwidth]{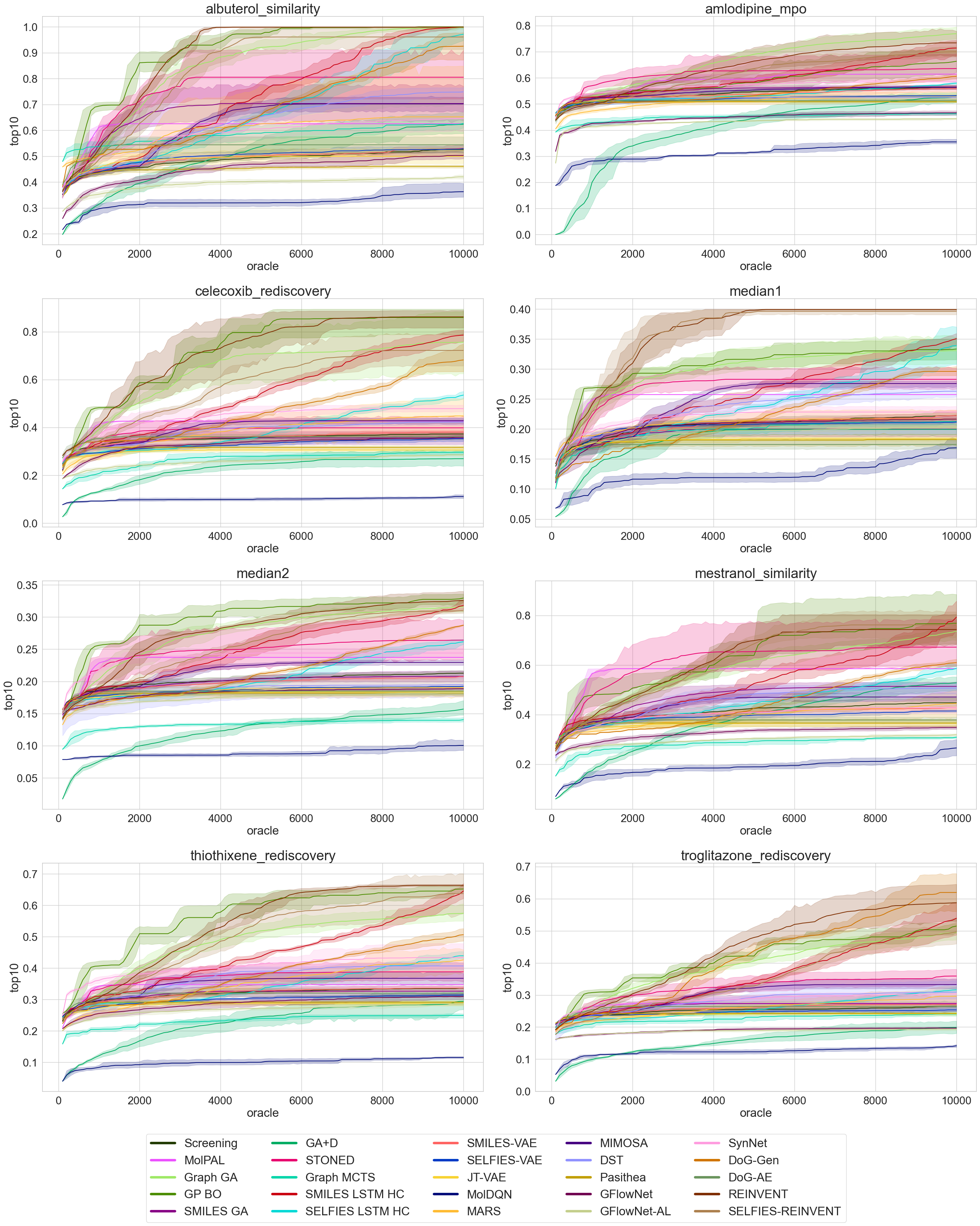}
  \caption{The optimization curves of top-10 average on optimizing similarity-based oracles.}
  \label{fig:curves_similarity}
\end{figure*}

\newpage

\begin{figure*}[h!]
  \centering
  \includegraphics[width=\textwidth]{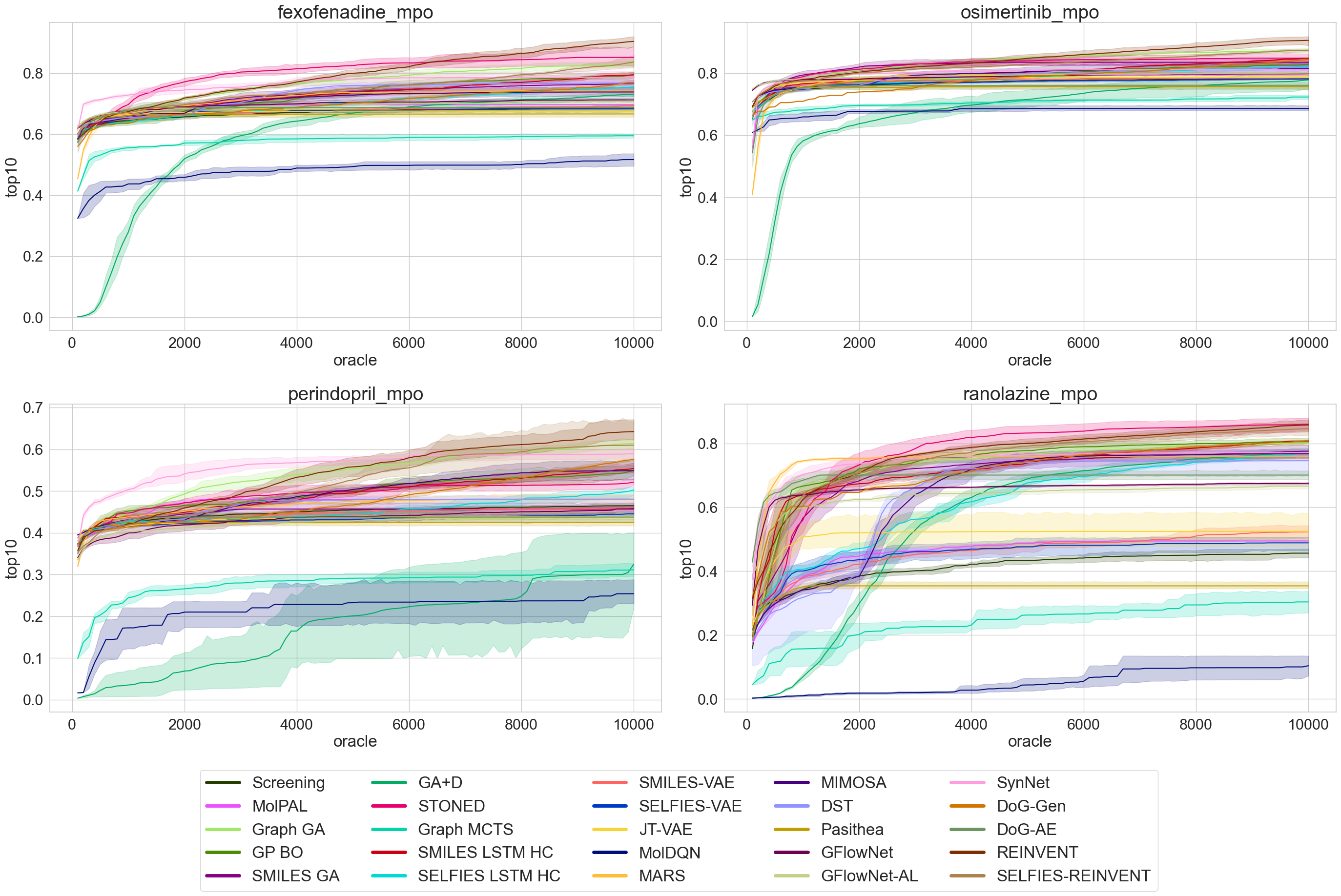}
  \caption{The optimization curves of top-10 average on optimizing similarity-based MPO oracles.}
  \label{fig:curves_similarity_mpo}
\end{figure*}

\begin{figure*}[h!]
  \centering
  \includegraphics[width=\textwidth]{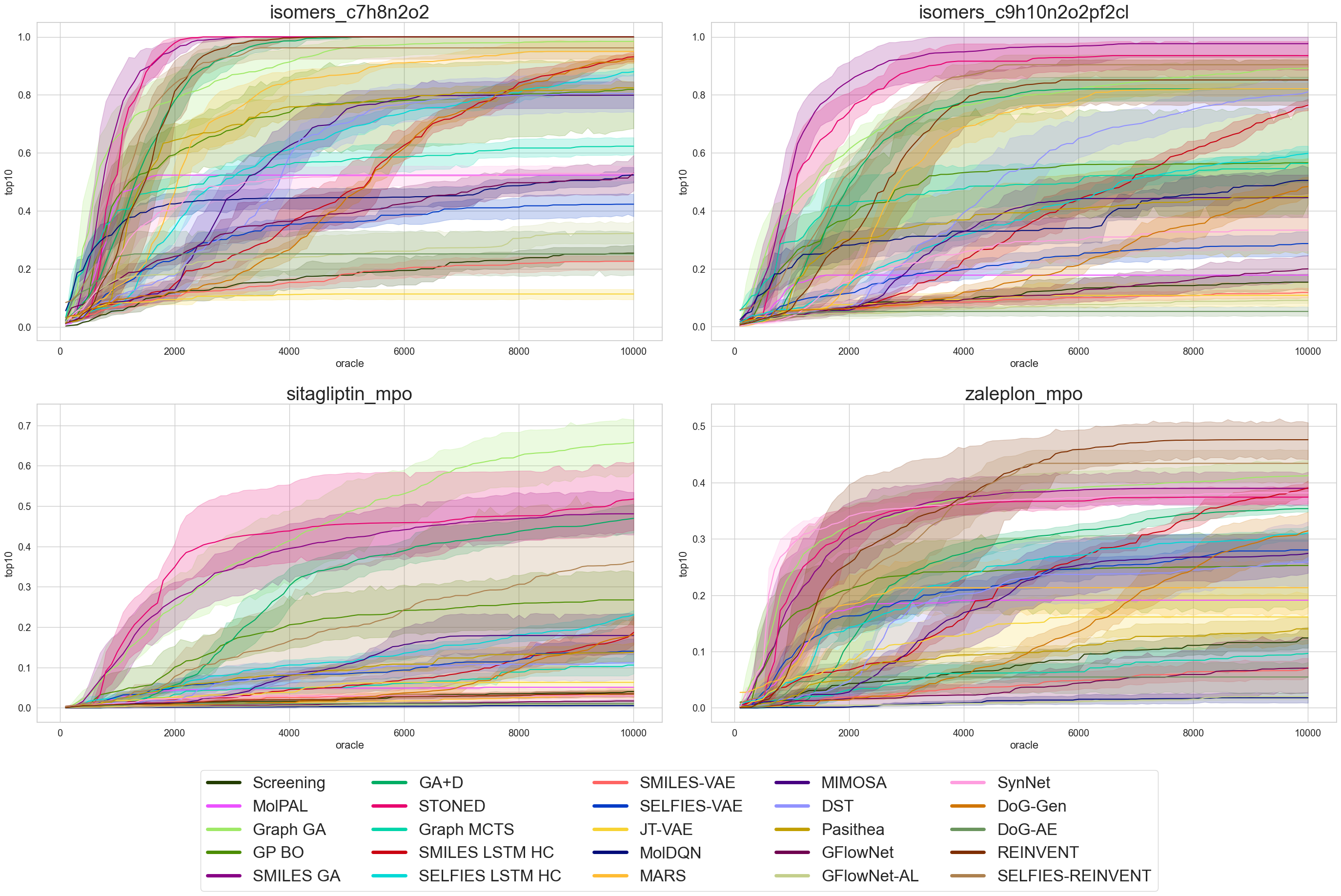}
  \caption{The optimization curves of top-10 average on optimizing isomer-based oracles.}
  \label{fig:curves_isomer}
\end{figure*}

\newpage

\begin{figure*}[h!]
  \centering
  \includegraphics[width=\textwidth]{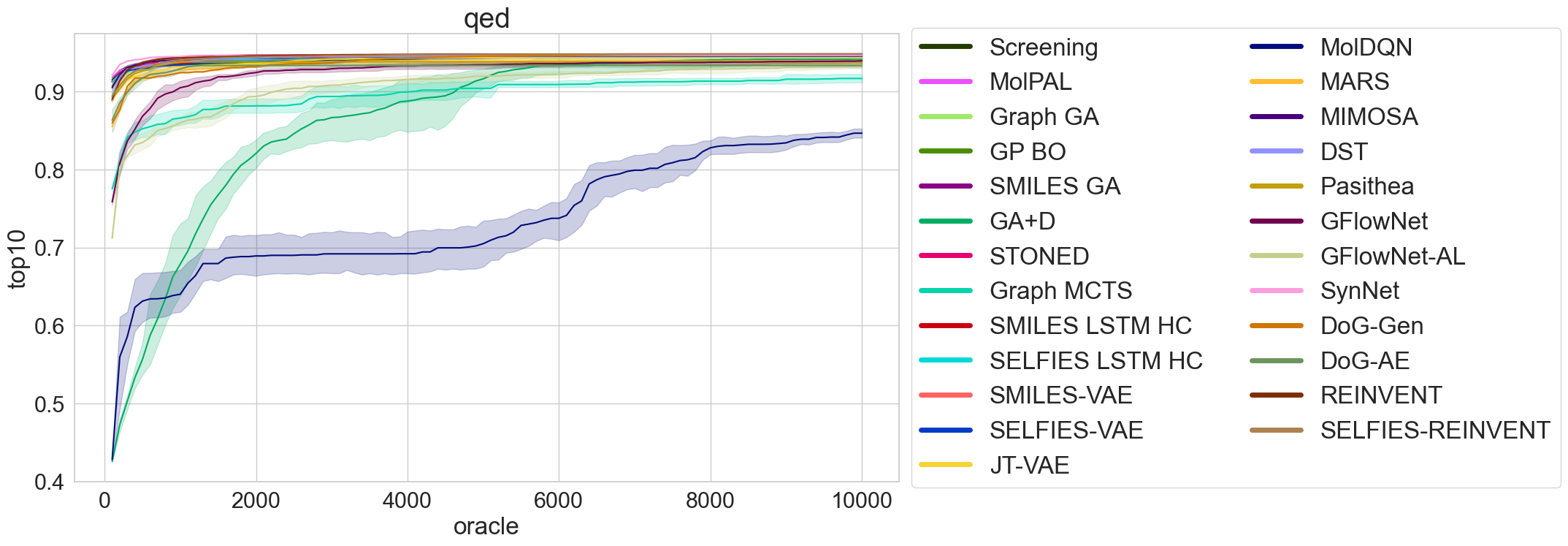}
  \caption{The optimization curves of top-10 average on optimizing QED.}
  \label{fig:curves_qed}
  \vspace{-5mm}
\end{figure*}

\begin{figure*}[h!]
  \centering
  \includegraphics[width=\textwidth]{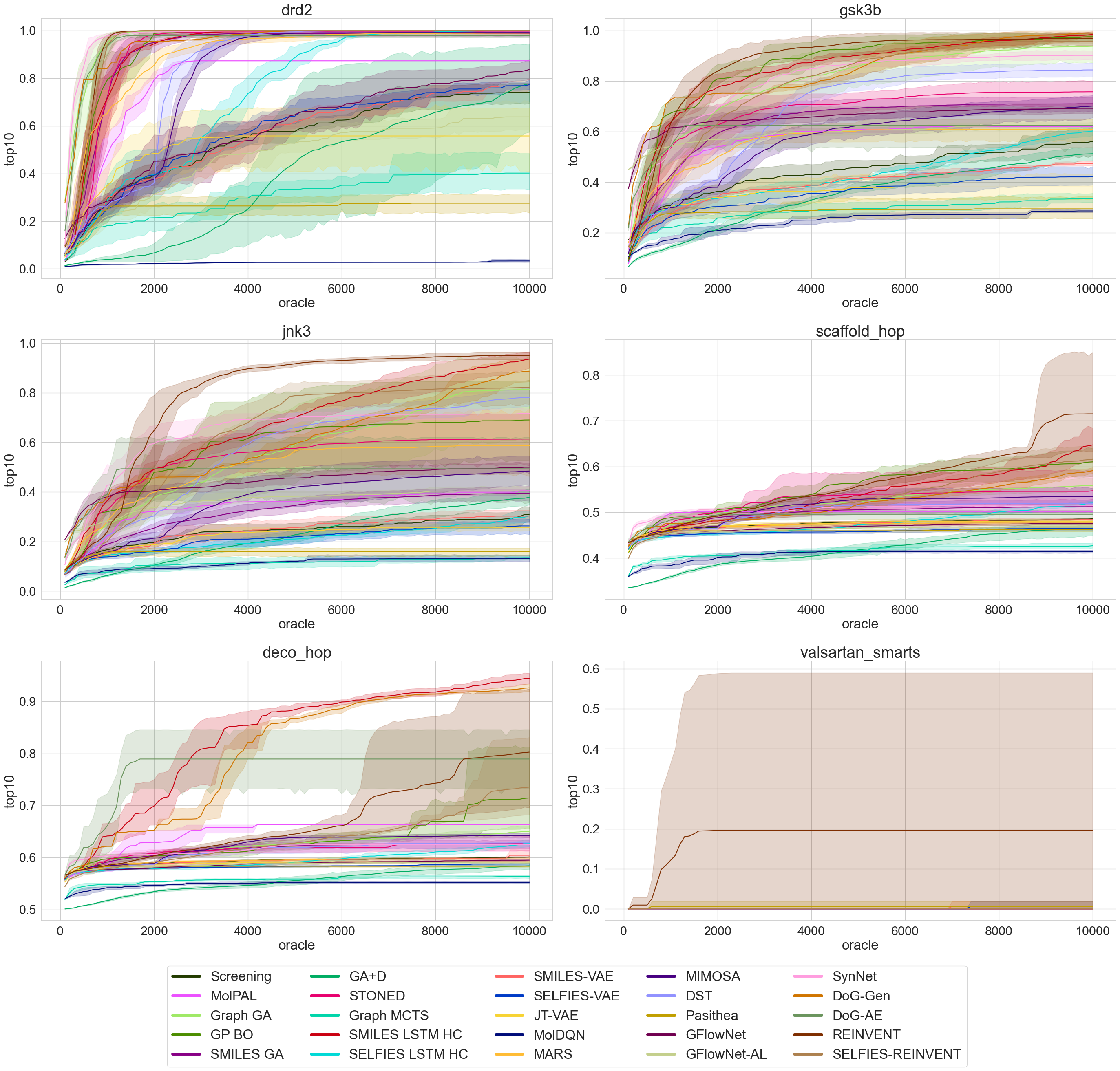}
  \caption{The optimization curves of top-10 average on optimizing SMARTS-based oracles and machine learning oracle.}
  \label{fig:curves_smarts_ml}
\end{figure*}

\newpage
\subsection{Synthesizability}
\begin{figure*}[h!]
  \centering
  \includegraphics[width=0.9\textwidth]{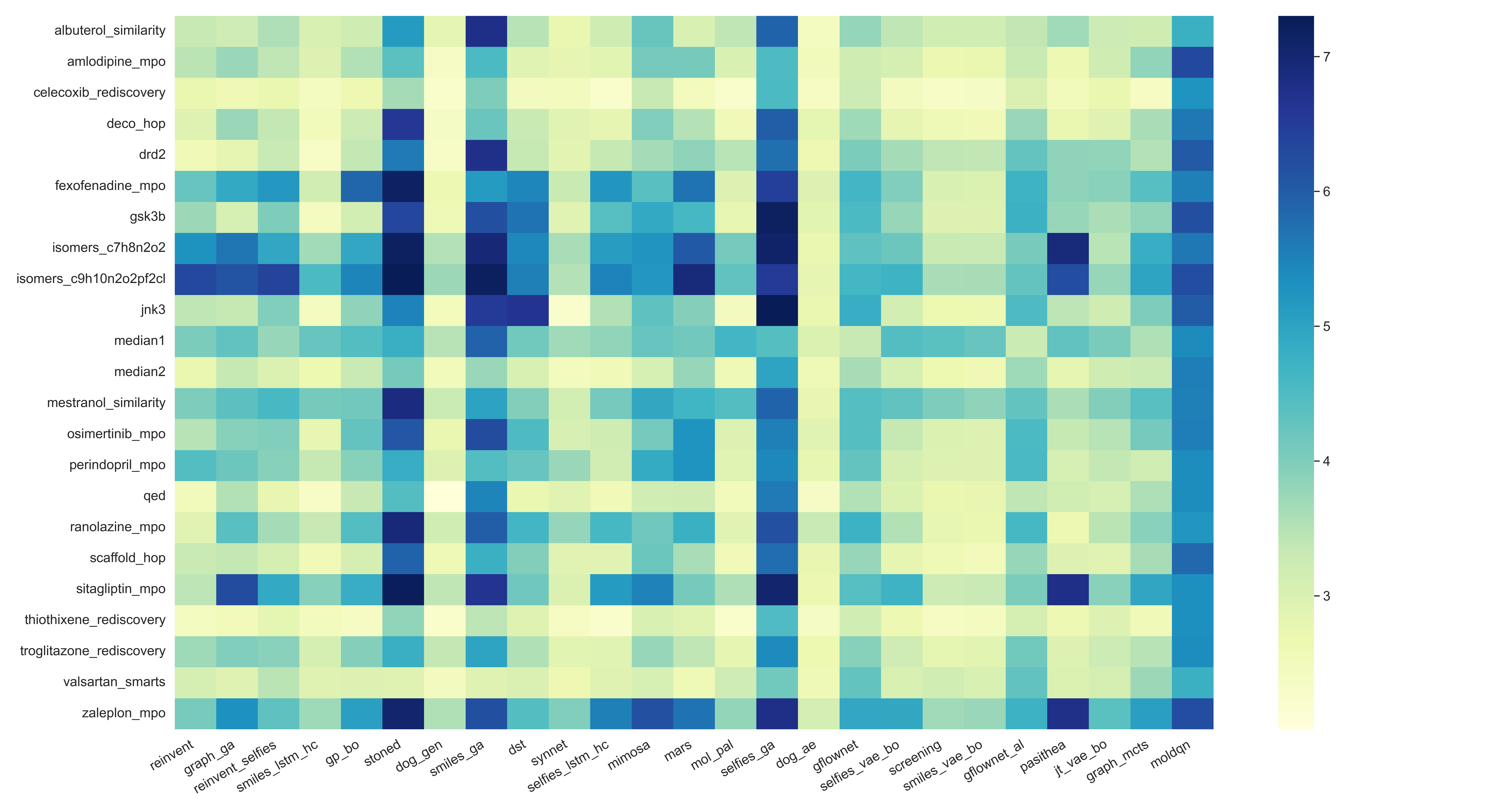}
  \caption{The heat map of SA\_Score (the lower the better) calculated from the Top-100 molecules from each method, averaged from all runs.}
  \label{fig:heatmap_sa}
\end{figure*}

We computed the SA\_Score of Top-100 molecules from each run and visualized the values in the Figure \ref{fig:heatmap_sa}. Though SA\_Score is not a great metric, we could see that synthesis-based methods have consistently lower SA\_Score in all tasks.

\newpage
\subsection{Diversity}
\begin{figure*}[h!]
  \centering
  \includegraphics[width=0.9\textwidth]{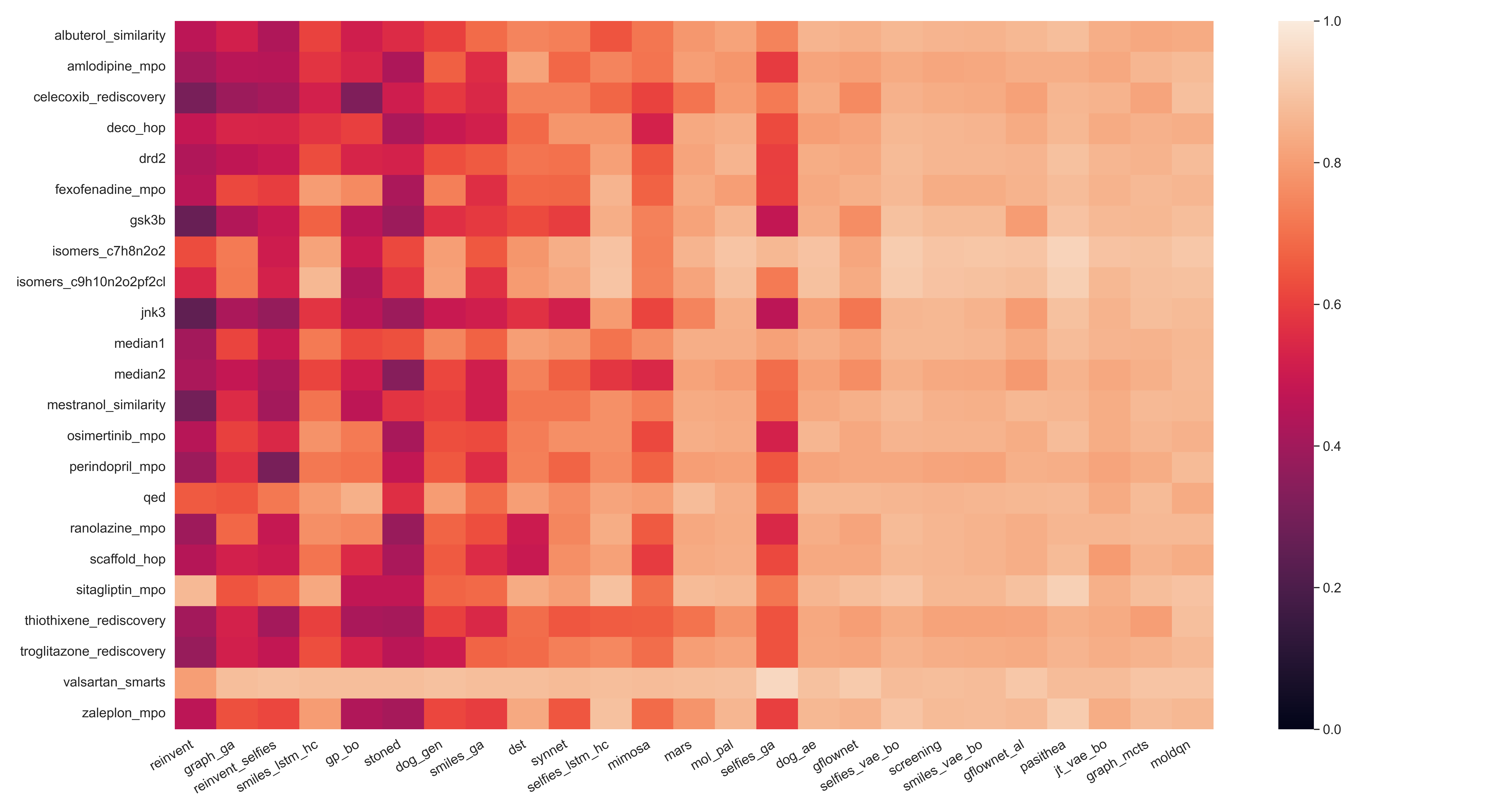}
  \caption{The heat map of diversity (the higher the better) calculated from the Top-100 molecules from each method, averaged from all runs.}
  \label{fig:heatmap_div}
\end{figure*}

We computed the diversity of the Top-100 molecules from each run and visualized the values in the Figure \ref{fig:heatmap_div}. The diversity is defined as the averaged internal distance within a batch of molecules, measured by Tanimoto similarity. We could see a general trend that the stronger a model is in optimization, the less diverse the results are. The methods with higher diversity would have an advantage, especially when the oracles have non-ignorable noise.

\newpage
\section{Implementation Details}
\label{sec:implement}

In this section, we elaborate the implementation details for each method. 
We summarize some shared properties of all the methods in Table~\ref{tab:method_property}.

\begin{table*}[h!]
\centering
\caption{The comparison of all the methods. AR represents auto-regressive model. 
Bayesian optimization usually leverages non-parametric (``non-param'' in the column ``model size'') model, the model size will increase as more training data come in. 
Run time is the average rough clock time for a single run in our benchmark and do not involve the time for pretraining and data preprocessing. 
}
\label{tab:method_property}
\footnotesize
\begin{adjustbox}{max width=\textwidth}
\begin{tabularx}{\textwidth}{Y c Y  Y Y Y Y Y Y }

\toprule
Categ. & Method & Assemb. & Model & Pretrain & Model Size (M) & Action Space & Type & Run Time (min) \\ \midrule 
\multirow{3}{*}{VS} & Screening & -  & - & N & 0 & - & model-free & 2 \\ 
 & MolPal & - & -  & Y & 3.2 & - & model-based & 60 \\ 
\hline 
\multirow{9}{*}{GA} 
 & SELFIES-GA & SELFIES  & - & N & 0 & cross & model-free & 20 \\
 & SMILES-GA & SMILES  & - & N & 0 & cross & model-free & 2 \\ 
 & STONED & SELFIES  & - & N & 0 & mutate & model-free & 3 \\ 
 & Graph-GA & fragment  & - & N & 0 & cross & model-free & 3 \\ 
 & SynNet & synthesis  & MLP & Y & 2,158 & cross & model-free & 300 \\ 
\hline 
\multirow{1}{*}{BO}
 & GPBO & fragment  & GP & N & non-param & one-hot & model-based & 15 \\ 
\hline 
\multirow{8}{*}{VAE+BO} 
 & SMILES-VAE & SMILES & RNN & Y & 17.9 & one-hot & model-based & 17 \\ 
 & SELFIES-VAE & SELFIES  & RNN & Y & 18.7 & one-hot & model-based & 21 \\ 
 & JTVAE & fragment & GNN\& treeRNN & Y & 21.8 &  one-hot & model-based & 17 \\ 
 & DoG-AE & synthesis & RNN & Y & 8.9 & one-hot & model-based & 47 \\ 
\hline 
\multirow{7}{*}{HC} 
 & SMILES-LSTM-HC & SMILES & RNN & Y & 98.9 & AR & model-free & 3 \\ 
 & SELFIES-LSTM-HC & SELFIES & RNN & Y & 30.4 & AR & model-free & 4 \\ 
 & MIMOSA & fragment  & GNN & Y & 0.25 &  one-hot & model-free & 10 \\ 
 & DoG-Gen & synthesis & RNN & Y & 51.0 & AR & model-free & 120 \\ 
\hline 
\multirow{5}{*}{RL} 
 & REINVENT & SMILES &  RNN & Y & 16.3 & AR & model-free & 2 \\ 
 & SELFIES-REINVENT & SELFIES &  RNN & Y & 16.5 & AR & model-free & 3 \\ 
 & MolDQN & atom &  MLP & N & 6.4 & AR & model-free & 52 \\ 
\hline 
\multirow{5}{*}{SBM} 
 & MARS & fragment  & GNN & N & 16.5 & one-hot & model-free & 21 \\ 
 & GFlowNet & fragment  & GNN & Y & 5.7 & one-hot & model-free & 28 \\ 
 & GFlowNet-AL & fragment  & GNN & Y & 5.7 & one-hot & model-based & 29 \\ 
\hline 
\multirow{3}{*}{Gradient} 
 & Pasithea & SELFIES & MLP & Y & 2.2 & one-hot & model-based & 46 \\ 
 & DST & fragment  & GNN & Y & 0.23 & one-hot & model-based & 300 \\ 
\hline 
\multirow{1}{*}{MCTS} 
 & Graph MCTS & atom & - & N & 0 & one-hot & model-free & 2 \\ 
\bottomrule
\end{tabularx}
\end{adjustbox}
\end{table*}

\subsection{Hyperparameter setup}  We have ``hparams\_default.yaml'' and ``hparams\_tune.yaml'' file for each method in their folders, where ``hparams\_default.yaml'' specify the default setup and ``hparams\_tune.yaml'' specify several possible choices of hyperparameter for tuning.

\subsection{Shared Setup: dataset} 
\label{sec:shared_data}
To avoid the bias introduced by different dataset, e.g., ZINC, ChemBL, for all the methods, we use ZINC to (i) train/pretrain the model; (ii) provide initial molecule set and (iii) extract vocabulary set.

\subsection{Shared Setup: Early Stop}
\label{sec:early_stop}
We utilize early stop strategies to save computational cost for iterative learning methods, e.g., BO, HC, GA based methods. The default patience is set to 5. That is, when the performance does not improve for 5 iteration (generation), we would terminate the process earlier. 
The methods that uses early stop strategy are ChemBO, DoG-AE, DST, Graph-GA, JTVAE, MIMOSA, RationaleRL, REINVENT, REINVENT-SELFIES, Screening, SELFIES-LSTM-HC, SELFIES-VAE, SMILES-LSTM-HC, SMILES-GA, SMILES-VAE.

\subsection{Shared Setup: Bayesian optimization for all VAEs}
\label{sec:bo_appendix}
We unify the implementation of Bayesian optimization for all VAE based methods for fair comparison, including JTVAE, SElFIES-VAE, SMILES-VAE, DoG-AE. Specifically, it is implemented by the python package ``BoTorch''~\cite{balandat2020botorch} with an exact Gaussian process model. Note that changing the exact GPs to sparse GPs would help increase the scalability~\cite{titsias2009variational}, but we found the VAE+BO methods reached at least several thousands oracle calls in most tasks, which is enough for a meaningful comparison. 
For all the VAE+BO methods, we pretrain the VAE model, provided the pretrained model so that users can start from BO process.

\subsection{Shared Setup: Pretraining}
\label{sec:pretraining}

Pretraining strategy has demonstrated its effectiveness in enhancing the optimization in many approaches, including VAE and HC methods. 
We pretrain the models on ZINC database, and the pretrained models are available in our repository. Worth to mention that the pretraining process does not require oracle calls.

\subsection{Shared Setup: PyTorch based}
We want to build a unified software environment to standardize the molecule optimization process and all the methods uses PyTorch~\cite{paszke2019pytorch} to build neural network models and Adam~\cite{kingma2014adam} to train them. 

\subsection{Action space manipulation}
\label{sec:action}
There are several types of state-action space: (1) \textit{auto-regressive} (AR): growing the molecule via adding a building bloack each step, conditioned on the partially generated one, e.g, RL method; (2) \textit{one-hot}: constructing or modifying the  molecule as a whole, e.g., VAE based method, gradient ascent method (DST, Pasithea); (3) \textit{cross}: maintain a population of molecules and exchange the structural information between molecules. This kind of action space is only used by GA based methods. 
The action space for all methods are available in Table~\ref{tab:method_property}. 

\subsection{Screening}
\label{sec:screening_appendix}
Screening searches over the molecule database (ZINC in this paper) sequentially via randomly selecting molecules and evaluating their properties. It does not involve a learning process. 

\subsection{MolPAL}
\label{sec:molpal}
MolPal~\cite{graff2021accelerating} is a machine learning enhanced version of screening (Section~\ref{sec:screening_appendix}). 
Specifically, it train a machine learning model to predict the molecular property and prioritize the molecule with higher predicted scores in ZINC to replace the random search in screening. 
Concretely, it firstly trains a molecular property predictor, which is a two-layer message-passing neural network, the hidden dimension is 300, activation function is ReLU. 
When training the message passing network, the initial training size is set to 500. Then during screening process, it updates the message passing network in online manner with batch size 100. 
It uses an Adam optimizer with an initial learning rate of 1e-4.

\subsection{SMILES-VAE}
\label{sec:smilesvae}
SMILES-VAE~\cite{gomez2018automatic} first trains string based VAE model on ZINC database. Both the encoder and decoder use single-directional GRU as neural architecture. For encoder GRU, the hidden dimension is 256, number of layers is 1, dropout rate is set to 0.5. The VAE latent variable's dimension is 128. The decoder GRU has three layers, dropout rate is 0, hidden dimension is 512. Optimizer is Adam with initial learning rate 3e-4. Gradient is clipped to 50 during training. The batch size is 512. The training and validation data is all the molecules in ZINC database. After training the VAE, it uses Bayesian optimization (BO) to explore the continuous latent variable space, the BO setup has been described in Section~\ref{sec:bo_appendix}. The pretrained SMILES-VAE model is available in the repository. 

\subsection{SELFIES-VAE}
\label{sec:selfiesvae}
It shares the same setup (neural architecture and learning process) with SMILES-VAE~\cite{gomez2018automatic} (Section~\ref{sec:smilesvae}), except the vocabulary. 

\subsection{DST}
\label{sec:dst}
Differentiable Scaffolding Tree (DST)~\cite{fu2021differentiable} utilize graph convolutional network (GCN) as property predictor. 
In GCN, the number of layer is 3, the hidden dimension and input embedding dimension are both 100. ReLU is used as activation function in hidden layers. DST leverages GA-like process, generate offspring based on a population of molecule candidates in each iteration, and select the promising ones from the offspring set and save them in the population for the next iteration.   
In each iteration, the population size is set to 50. When generating the offspring pool, it used determinantal point process (DPP) to enhance the diversity of the population, where $\lambda$ controls the weight of diversity compared with fitness. It is set to 2. The pool size is set to 500, which means in each iteration, we generate at most 500 offspring. $\epsilon=0.7$ controls the probability threshold to add a substructure from the current now. $k=5$ represent the maximal number of substructures that are sampled from a single branch during expansion. 
The substructure can be either a single ring or an atom. The vocabulary set contains 82 most frequent substructures in ZINC databases, whose frequencies are greater than 1,000. 
In the inner loop, when optimize DST for each single molecule, we use Adam optimizer with initial learning rate 1e-3 and the maximal iteration number is set to 5K, with early stop strategy. 
During the optimization process, we use the new labelled molecules to update the GCN in online manner.

\subsection{Pasithea}
\label{sec:pasithea}
Pasithea~\cite{shen2021deep} is also a gradient ascent method like DST and utilize SELFIES as representation. 
It differentiate the molecule and back-propagate the gradient of the neural network to update the molecule iteratively. 
It uses four layer multiple layer perceptron (MLP) as neural model with ReLU function as activation to provide nonlinearity. 
SELFIES strings are converted into multi-hot vector as the input of the MLP. 
The hidden dimensions are all set to 500. The output layer is to predict the property, so the output-layer dimension is 1. It first use 800 molecules to train the neural network as predictor and then online update it during the optimization process. 
The training epoch is set to 5, the optimizer is Adam with initial learning rate 1e-3. 
During inference, i.e., updating differentiable molecule, Pasithea uses Adam as optimizer with initial learning rate 5e-3, epoch number is 50.

\subsection{MolDQN}
\label{sec:moldqn}
Molecule Deep Q-Network (MolDQN)~\cite{zhou2019optimization} formulates molecule optimization as a Markov Decision Process. In each step of a single episode, it add an atom from vocabulary (C, N, O) to any eligible position of the current molecular graph and choose one molecule with highest estimated Q-value for the next step. Q-value is the estimated by deep Q-network. 
The maximal number of steps in each episode is 40. Each step calls oracle once. 
The discount factor is 0.9. $\epsilon$ controls the weight of exploration and exploitation, we tune the $\epsilon$ to make it more exploration at the beginning of learning process and more exploitation at the end (i.e., use up oracle calls).
Deep Q-network is a multilayer perceptron (MLP) whose hidden dimensions are 1024, 512, 128, 32, respectively, the output dimension is 1. The input of the Q-network is a 1025-dimensional vector, which is the concatenation of the molecule feature (1024-bit Morgan fingerprint, with a radius of 2) and the number of left steps. 
Adam is used as an optimizer with 1e-4 as the initial learning rate.
Only rings with a size of 5 and 6 are allowed.

\subsection{MIMOSA}
\label{sec:mimosa}
Multi-constraint Molecule Sampling ({MIMOSA})~\cite{fu2021mimosa} reformulate molecule optimization as a MCMC sampling problem and the property oracles are encoded in the target distribution. We use Adam optimizer with a learning rate of 0.001. In pretraining phase, MIMOSA to set GNNs with 5 layers and 300-dimensional hidden units. 
MIMOSA randomly masks a single node (a substructure) for each molecule and predict its substructure category based on other feature. 
The substructure can be either a single ring or an atom. The vocabulary set contains 82 most frequent substructures in ZINC databases, whose frequencies are greater than 1,000, same as DST (Section~\ref{sec:dst}). 
Then during inference phase, in each iteration, it samples new molecules via masking one random-selected node (i.e., substructure), and use GNN to predict the substructure's categorical distribution, and flip the node to a new substructures with highest probability. 
It samples at most 500 molecules and online updates the GNN using the top-300 scored molecules.

\subsection{MARS}
\label{sec:mars}
Markov Molecular Sampling (MARS)~\cite{xie2021mars} is based on MCMC sampling. 
It uses a graph neural network to imitate the MCMC proposal distribution. The GNN is three layer, the dimension of node embedding is 64, the dimension of edge embedding is 128.  
It uses simulated annealing to sampling and adaptive proposal (online updated) from the target distribution. It collects 1000 frequent fragments as vocabulary. The batch size is set to 128 during training. 

\subsection{GFlowNet}
\label{sec:gflownet}
Generative Flow Network (GFlowNet) is a MCMC sampling method~\cite{bengio2021gflownet}. 
It predefine 72 basic building blocks as vocabulary set, which are selected from ZINC database. 
It uses message passing neural network (MPNN) to estimate the flow and takes the atom graph as the input feature. 
The hidden state dimension and embedding dimension are both set to 256. 
The number of layer is set to 3. LeadyReLU is used as activation function. $\epsilon$ is set to 2e-8, which is defined in Equation 12 in original paper and is used to avoid taking the logarithm of a tiny number. 
It uses Adam as optimizer with initial learning rate 5e-4, where $\beta_1=0.9, \beta_2=0.999$. The batch size is set to 4. 

\subsection{GFlowNet-AL}
\label{sec:gflownetal}

GFlowNet-AL is a model-based version of GFlowNet that uses predictive model to enhance GFlowNet. GFlowNet-AL share the same setup (neural architecture, learning process) with GFlowNet.

\subsection{JTVAE}
\label{sec:jtvae}
Junction Tree VAE (JTVAE)~\cite{jin2018junction} represent the molecule graph into junction tree, which is cycle-free and easier to generation. JTVAE leverage design message passing network as encoder and tree RNN as decoder. 
Encoder represent both molecular graph and junction tree into latent variable, decoder first generate junction tree and then reconstruct molecular graph conditioned on the junction tree. 
The hidden size of message passing network and tree RNN is 450. The dimension of latent variable is 56, where the dimensions of latent variable for both molecular graph and junction tree are 28. The depth of junction tree level message passing network is 20 and the depth of molecular graph-level message passing network is 3. After training the VAE, it uses Bayesian optimization (BO) to explore the continuous latent variable space, the BO setup has been described in Section~\ref{sec:bo_appendix}. 
The original implementation was based on Python 2, we adapt it to Python 3. Also, we re-implement BO process using BoTorch (Section~\ref{sec:bo_appendix}).

\subsection{GP BO}
\label{sec:gpbo}
Gaussian Process BO (GP BO)~\cite{tripp2021fresh} utilizes Gaussian process as the surrogate model and optimize the acquisition function with Graph GA methods internally. We treat it as a model-based version of Graph GA, where we adopt 2-radius 2048 bit molecular fingerprint as molecular feature. It should be noted that there are other types of fingerprints, such as fragprints~\cite{thawani2020photoswitch} and MAP4~\cite{capecchi2020one}, and the choice of that could be a major performance determinant. In GA, the initial population size is 340; 
the maximal BO iteration is 10000;
BO's batch size is 1180; 
maximal generations is 60; 
Size of offspring set is 150; 
the mutation rate is 0.01; 
population size is 820. 
We adopt the implementation from the original paper~\cite{tripp2021fresh}. 




\subsection{DoG-AE}
\label{sec:dog_ae}
The autoencoder version of DAGs of molecular graphs (DoG)~\cite{bradshaw2020barking} uses autoencoder (AE) to learn the distribution of synthesizable molecules. 
The dimension of latent variable of autoencoder is 25, for the molecular graph embedder (encoder), the hidden layer size is 80, embedding dimension is 50, number of layer is 4. 
for DAG embedder, the hidden layer size is 50, number of layer is 7. 
Decoder is a GRU, whose input size is 50, hidden size 200, num of layers is 3, dropout rate is 0.1. 
Bayesian optimization is utilized to optimize the continuous latent space. 
DoG is a basic generator that constructs synthesizable molecules from building blocks via virtual chemical reactions.

\subsection{DoG-Gen}
\label{sec:dog_gen}
DoG-Gen is the hill climbing\footnote{Section~\ref{sec:optimization_algorithm}} version of DoG~\cite{bradshaw2020barking}. In each iteration, it samples 3,000 molecules and keep 1,000 ones with the best fitness scores for the next iterations. 
It uses the Molecular Transformer~\cite{schwaller2019molecular} as a black box oracle for reaction prediction. The molecular transformer is pretrained on USPTO dataset. It uses gated graph neural network (GGNN)~\cite{li2015gated} to learn molecular embedding and GRU to generate the molecule. 

\subsection{SynNet} \label{sec:synnet} SynNet~\cite{gao2021amortized} use GA to manipulate binary molecular fingerprint. It uses MLP as the neural architecture and molecular fingerprint as the input feature of the neural network. It uses 2-radius 4096 bit fingerprint as the input of MLP. During GA-process, the population size is 128, offspring size is 512. mutation probability is set to 0.5. For each element, the number of mutation is set to 24. 
SynNet consists of four modules, each containing a multi-layer perceptron (MLP), 
(1.) An Action Type selection neural network that classifies action types among the four possible actions (“Add”, “Expand”, “Merge”, and “End”) in building the synthetic tree. The input dimension is 3*4096, the hidden dimension is set to 500, output dimension is 4.  
(2). A First Reactant selection neural network that predicts an embedding for the first reactant. A candidate molecule is identified for the first reactant through a k-nearest neighbors (k-NN) search from the list of potential building blocks. The input dimension is 3*4096, the hidden dimension is set to 1,200, output dimension is 1.  
(3). A Reaction selection neural network whose output is a probability distribution over available reaction templates, from which inapplicable reactions are masked (based on reactant 1) and a suitable template is then sampled using a greedy search. The input dimension is 4*4096, the hidden dimension is set to 3000, output dimension is 91.
(4). A Second Reactant selection neural network that identifies the second reactant if the sampled template is bi-molecular. The model predicts an embedding for the second reactant, and a candidate is then sampled via a k-NN search from the masked set of building blocks. The input dimension is 4*4096+91, the hidden dimension is set to 3000, output dimension is 1.
All the 4 MLP has 5 layers. Adam optimizer is used with initial learning rate 1e-4. 

\subsection{REINVENT}
\label{sec:reinvent}

\begin{figure*}[h]
  \centering
  \includegraphics[width=\textwidth]{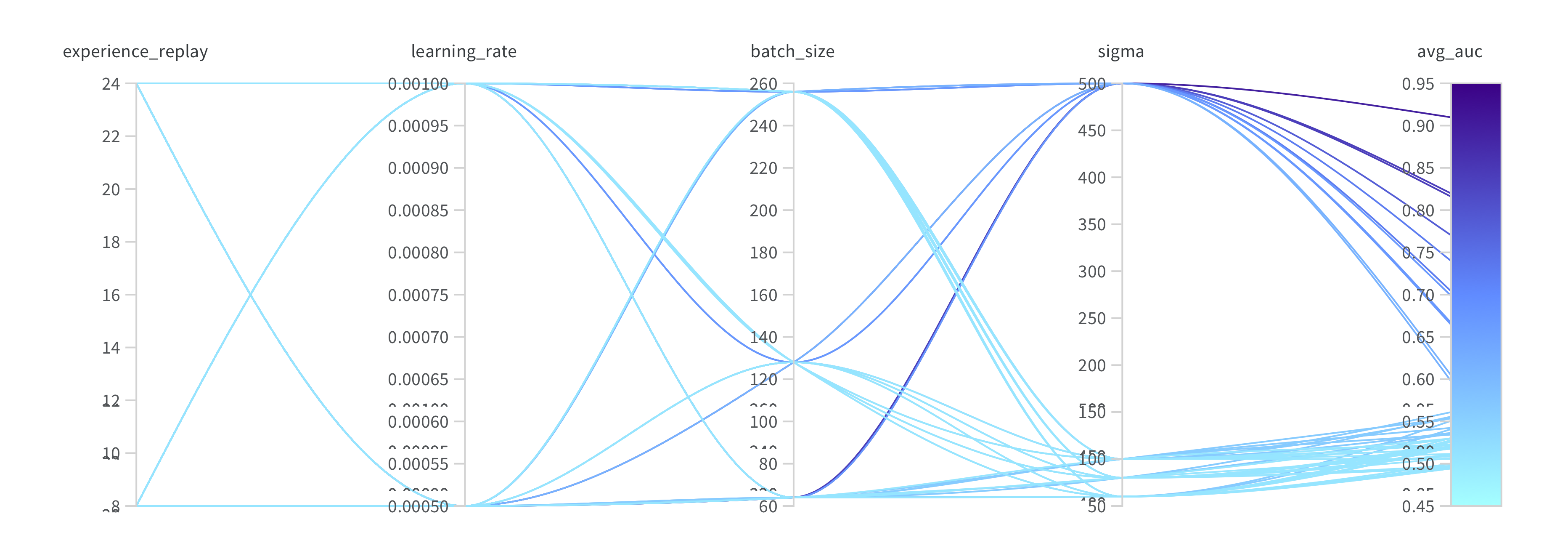}
  \caption{The hyper-parameter tuning result of REINVENT (part 1). We can see that sigma ($\sigma$) has large impact on optimization performance, and the optimal value is much larger than the default setting in the original paper~\cite{olivecrona2017molecular}.}
  \label{fig:tune_reinvent2}
\end{figure*}

\begin{figure*}[h]
  \centering
  \includegraphics[width=\textwidth]{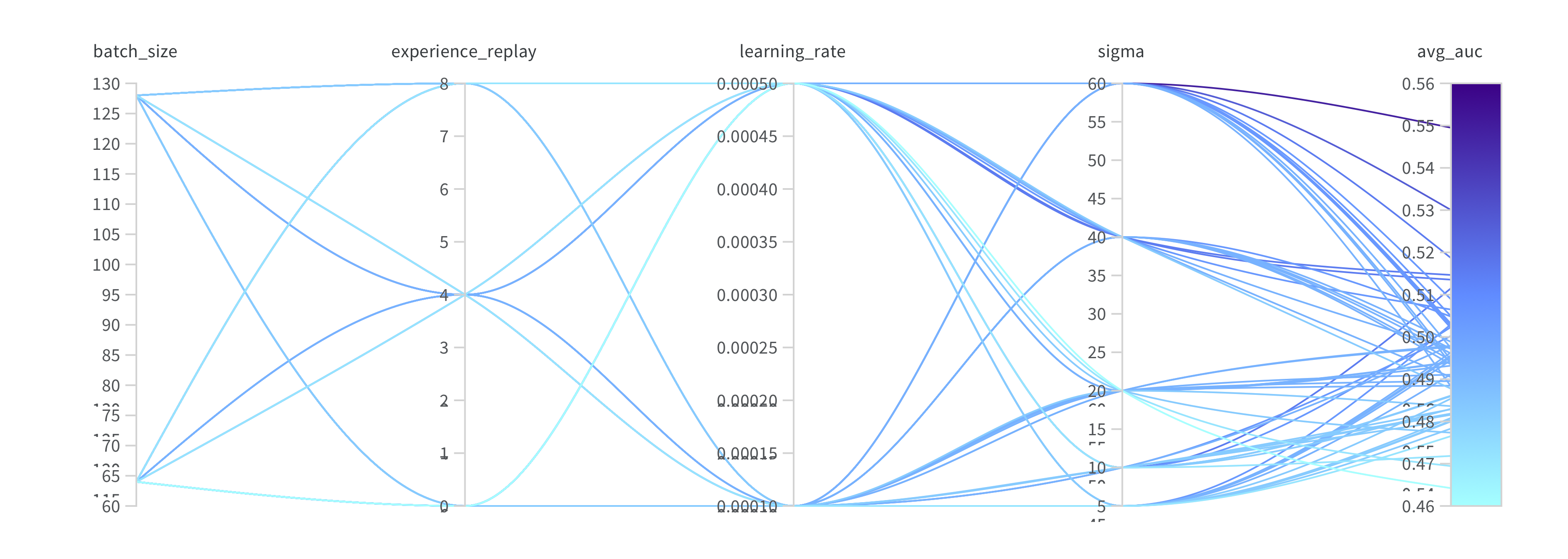}
  \caption{The hyper-parameter tuning result of REINVENT (part 2).}
  \label{fig:tune_reinvent1}
\end{figure*}

REINVENT~\cite{olivecrona2017molecular} is the top-1 method as shown in Table~\ref{tab:task2}. 
REINVENT uses SMILES string as representation and recurrent neural network (RNN) as neural model, which contains multiple GRU cells. The embedding dimension of input token is set to 128, the hidden dimensions of all GRU are set to 512. 
In REINVENT, the whole objective contains (i) prior likelihood to encourage the generated SMILES to be close to training SMILES string and (ii) a reward function for optimization. 
The $\sigma$ control the importance of reward function in the whole objective and plays a critical role in optimization performance, as shown in Figure~\ref{fig:tune_reinvent2} and~\ref{fig:tune_reinvent1} ($\sigma$ is sigma). 
After intensive tuning, $\sigma$ is set to $500$. It is even not found by the original paper, where $\sigma$ is set to 60. 
Based on our empirical studies, the selection of $\sigma$ is vital to the optimization performance. Also, the batch size during the training is set to 64. Adam is used as optimizer with initial learning rate 5e-4. REINVENT is pretrained on ZINC data, the pretrained model is used in two ways: (1) provide a warm start and are finetuned during optimization; (2) evaluate the prior likelihood of the generated SMILES string to measure their SMILES likeness. 

\subsection{SELFIES-REINVENT}

It uses SELFIES string as molecular representation and shares the same setup (neural architecture, learning process) with REINVENT~\cite{olivecrona2017molecular} (Section~\ref{sec:reinvent}), except the vocabulary.

\subsection{SMILES-LSTM Hill Climbing (SMILES-LSTM HC)}
\label{sec:smileslstm}
SMILES-LSTM Hill Climbing~\cite{brown2019guacamol} uses three-layer LSTM as neural model, the hidden size is 512. It pretrains the LSTM using ZINC data. 
It use Adam as optimizer with initial learning rate 1e-3. During hill climbing, the population size is 100; the epoch is set to 10; batch size is 256; each epoch sample 1024 molecule and keep the best 512 molecules (highest scores) for the next epoch. The maximal length of SMILES is 100. 

\subsection{SELFIES-LSTM Hill Climbing (SELFIES-LSTM HC)}
It uses SELFIES string to represent molecule and shares the same setup (neural architecture, learning process) with SMILES-LSTM Hill Climbing~\cite{brown2019guacamol} (Section~\ref{sec:smileslstm}), except the vocabulary.

\subsection{GA+D (SELFIES-GA)}
\label{sec:selfiesga}
Genetic Algorithm with Discriminator (GA+D)~\cite{nigam2019augmenting} utilizes SELFIES string to represent molecule and apply genetic algorithm. It is enhanced by a discriminator neural network. 
The discriminator neural network is a fully connected neural network with ReLU activation and sigmoid output layer.  the number of molecules in the generation (i.e., population)  is 300. 
The patience value is set to 5. 
beta ($\beta$) is the weight of discriminator neural network's score in fitness evaluation, which is used to select most promising molecules in each generation. After empirical studies, we do not find $\beta$ has positive contribution to the performance. Thus, the default value is set to 0.
The maximal generation number is 1000.

\subsection{STONED}
\label{sec:stoned}
Superfast Traversal, Optimization, Novelty, Exploration and Discovery (STONED)~\cite{nigam2021beyond} implements genetic algorithm (only mutation operator, without crossover) on SELFIES string. After tuning, we find when the generation size is set to 500, STONED reached best optimization performance. Like other genetic algorithm, it does not need any learnable parameter, is super-fast and easy to implement. 

\subsection{SMILES-GA}
SMILES-GA~\cite{brown2019guacamol} manipulates SMILES string with only mutation operation. The crossover operation is not conducted because it would lead to poor chemical validity. 
The population size is set to 100. In each generation, the number of mutated molecule is set to 300. The maximal length of SMILES string is set to 200. Mutation randomly flips a randomly-selected bit in the current SMILES string. 
The initial population is randomly selected from ZINC. It uses early stop strategy and the patience is set to 5. 

\subsection{Graph-GA}
Graph-GA~\cite{jensen2019graph} manipulates molecular graph with crossover and mutation operators successively. 
The population size is set to 120. offspring size is set to 70. 
The mutation rate is set to 0.067. That is, the new child molecule will be mutated with probability 6.7\%. 
The mutation operations includes (1) insert an atom; (2) change bond's order; (3) delete cyclic bond; (4) add ring; (5) delete an atom; (6) change an atom and (7) append an atom. 

\subsection{Graph-MCTS}
Graph level Monte Carlo Tree Search (Graph-MCTS)~\cite{jensen2019graph} manipulate molecular graph using MCTS. 
Like GA algorithms, it does not involve any learnable parameters. 
It start from Ethane, whole SMILES string is ``CC''. 
During the searching process, it constrains the maximal number of atoms to 60. 
For each state (molecular graph), the maximal number of children is 5. 
The root node simulates 22 times. 
Exploration coefficient balances the weight of exploitation and exploration and is set to 4.3. Larger exploration coefficient indicates more exploration instead of exploitation. 

\subsection{Methods Not Included}
\label{sec:fail_method}

In this section, we describe some other methods that are representative but not included in our benchmark. 
We also analyze the reasons. 
These methods contain Bayesian Optimization over String Space (BOSS)~\cite{moss2020boss}, synthesis-based Bayesian optimization (ChemBO)~\cite{korovina2020chembo}, Objective-Reinforced Generative Adversarial Network (ORGAN)~\cite{guimaraes2017objective}, 
Generative Adversarial Network (MolGAN)~\cite{de2018molgan}, rationaleRL~\cite{jin2020multi}. 
BO based methods (BOSS and ChemBO) are non-parametric methods and use the combination of training data to approximate the landscape. The evaluation of the approximate function relies on the number of training data and the evaluation of kernel function relies on the data's dimension. The optimization process requires intensive evaluation of both approximate function and kernel function, thus BO scales poorly with both data dimension and number and is computationally prohibitive~\cite{snoek2012practical}. In our experiment, BOSS and ChemBO are only available to generate less than 200 molecules and stop early, which is not comparable with other methods in our benchmark. One potential reason is the poor scaling of string subsequence kernels~\cite{teo2006fast}. Thus we decided not to incorporate them. 
ORGAN uses SMILES as molecular representation and the generated molecules has lower validity ($<$1\%). 
MolGAN does not achieve comparable optimization performance. 
RationaleRL requires extracting property-aware rationale as the basic building block, the process relies on Monte Carlo Tree Search and requires intensive oracle calls (more than 10K). 

\newpage


\section{Configuration}
\label{sec:configuration}

\subsection{Software}
\label{sec:software}

We build a unifying conda environment for most of the methods, which relies on the following python packages. 
\begin{itemize}
 \item \textbf{Python}. We use Python 3.7. 
 \item \textbf{PyTorch} is used to build neural network. We recommend to install PyTorch 1.10.2. 
 \item \textbf{PyTDC} (Therapeutic Data Commons)~\cite{huang2021therapeutics}. TDC provides dataloader for ZINC, evaluator (diversity, novelty, etc) and oracle scoring (all the oracles in this paper). 
 \item \textbf{RDKit} is an open-Source cheminformatics software and is used for molecule manipulation. We use RDKit 2020.09.1.0. It can be installed using conda via ``conda install -c rdkit rdkit ''.
 \item \textbf{wandb}~\cite{wandb} is used to record the learning process. It can be installed using pip. And users need to register a wandb account. It also supports automatic hyperparameter tuning and visualize the results in an intuitive manner. 
 \item \textbf{YaML} is used to setup configuration file. It can be installed using pip. We have ``hparams\_default.yaml'' and ``hparams\_tune.yaml'' file. 
 \item \textbf{SELFIES (optional)}~\cite{krenn2020self} is only used for SELFIES related methods. It can be installed using pip.
 \item \textbf{BoTorch (optional)}~\cite{balandat2020botorch}  is a library for Bayesian Optimization built on PyTorch and is only used for BO related methods. It can be installed using pip.
\end{itemize}

\noindent\textbf{Individual conda environment}. 
The following methods need an individual conda environment. 
\begin{itemize}
 \item \textbf{ChemBO} require install our modified dragonfly package and TensorFlow. The modified dragonfly is already in our repository.  
 \item \textbf{DoG-AE and DoG-Gen} required installing two individual conda environment following their original instruction in \url{https://github.com/john-bradshaw/synthesis-dags}. 
\end{itemize}

\subsection{Hardware}
\label{sec:hardware}

We use (i) Intel Xeon E5-2690 machine with 256G RAM and 8 NVIDIA Pascal Titan X GPUs and (ii)
Most of the NN based methods require GPU to acclerate learning process. 

\subsection{License}
\label{sec:license}

Our package uses the MIT license. We use ZINC database for all the methods, ZINC is free to use for everyone~\cite{sterling2015zinc}. All the 25 methods' implementation are publicly available at GitHub.

\subsection{Run with one-line code}
\label{sec:run}

All the methods can be run in one line of code after the setup of conda environment. 
We provide the pretrained model (if needed) and other necessary data/configuration files. 

\begin{minted}[frame=lines,
framesep=2mm,
baselinestretch=1,
fontsize=\footnotesize
]{bash}
cd mol_opt 
python run.py graph_ga
\end{minted}

\begin{minted}[frame=lines,
framesep=2mm,
baselinestretch=1,
fontsize=\footnotesize
]{bash}
python run.py dst --task production --n_runs 5 --oracles qed jnk3 drd2 
\end{minted}

\begin{minted}[frame=lines,
framesep=2mm,
baselinestretch=1,
fontsize=\footnotesize
]{bash}
python run.py graph_ga --task tune --n_runs 30 --smi_file ./data/zinc.txt \ 
                       --wandb offline --max_oracle_calls 10000 --patience 5 
\end{minted}

\newpage
\section{Additional Results}
\label{sec:additional_results}

\subsection{SELFIES strings collapse}
\label{sec:bad_selfies}

\begin{figure*}[h]
  \centering
  \includegraphics[width=\textwidth]{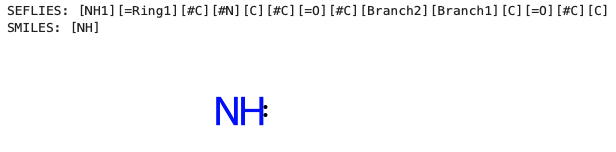}
  \caption{An example of SELFIES string that is valid but doesn't provide meaningful exploration of chemical space.}
  \label{fig:bad_selfies1}
\end{figure*}

\begin{figure*}[h]
  \centering
  \includegraphics[width=\textwidth]{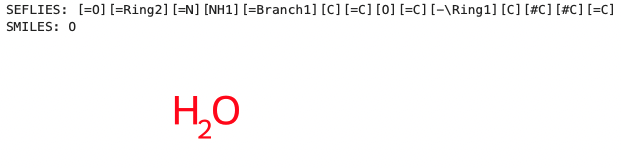}
  \caption{An example of SELFIES string that is valid but doesn't provide meaningful exploration of chemical space.}
  \label{fig:bad_selfies2}
\end{figure*}

\begin{figure*}[h]
  \centering
  \includegraphics[width=\textwidth]{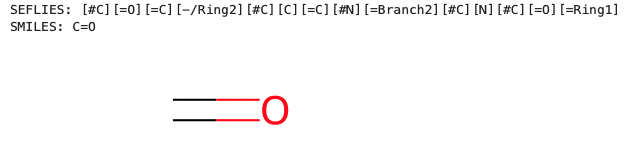}
  \caption{An example of SELFIES string that is valid but doesn't provide meaningful exploration of chemical space.}
  \label{fig:bad_selfies3}
\end{figure*}

Though most SELFIES strings represent valid molecules, replacing SMILES with SELFIES doesn't necessarily lead to an immediate advantage in molecular optimization. One reason is that different combinations of SELFIES strings could collapse to a single truncated SELFIES strings and don't provide an additional exploration of chemical space. See Figure \ref{fig:bad_selfies1},  \ref{fig:bad_selfies2}, and \ref{fig:bad_selfies3} for examples. These SELFIES strings were constructed with tokens from the vocabulary of ZINC 250k and converted to SMILES strings using the \verb+decoder+ provided in the official Github repository.

\newpage
\subsection{Hyper-parameter Tuning}
\label{sec:hparams}

Most algorithms are sensitive to the choice of hyper-parameters. We tried to tune most algorithms within a reasonably large hyper-parameter space and visualize some of the results here to show how hyper-parameters affect the performance. For each algorithm, the endpoint is a summation of AUC Top-10 of zaleplon\_mpo (an isomer-based oracle) and perindopril\_mpo (a similarity-based oracle), averaged from 3 runs for each task. We tuned and visualized them with the wandb~\cite{wandb}. The oracles are chosen to discriminate most algorithms and be representative based on preliminary results.

\begin{figure*}[h]
  \centering
  \includegraphics[width=\textwidth]{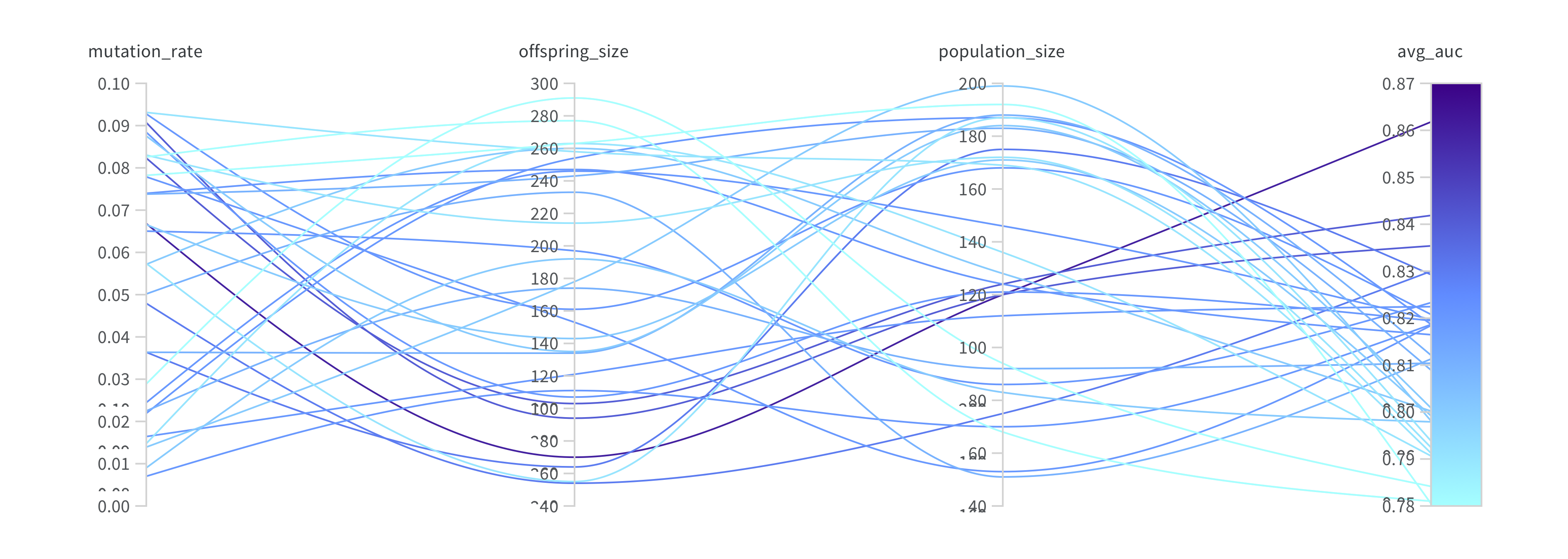}
  \caption{The hyper-parameter tuning result of Gaph GA.}
  \label{fig:tune_graph_ga}
\end{figure*} 

\begin{figure*}[h]
  \centering
  \includegraphics[width=\textwidth]{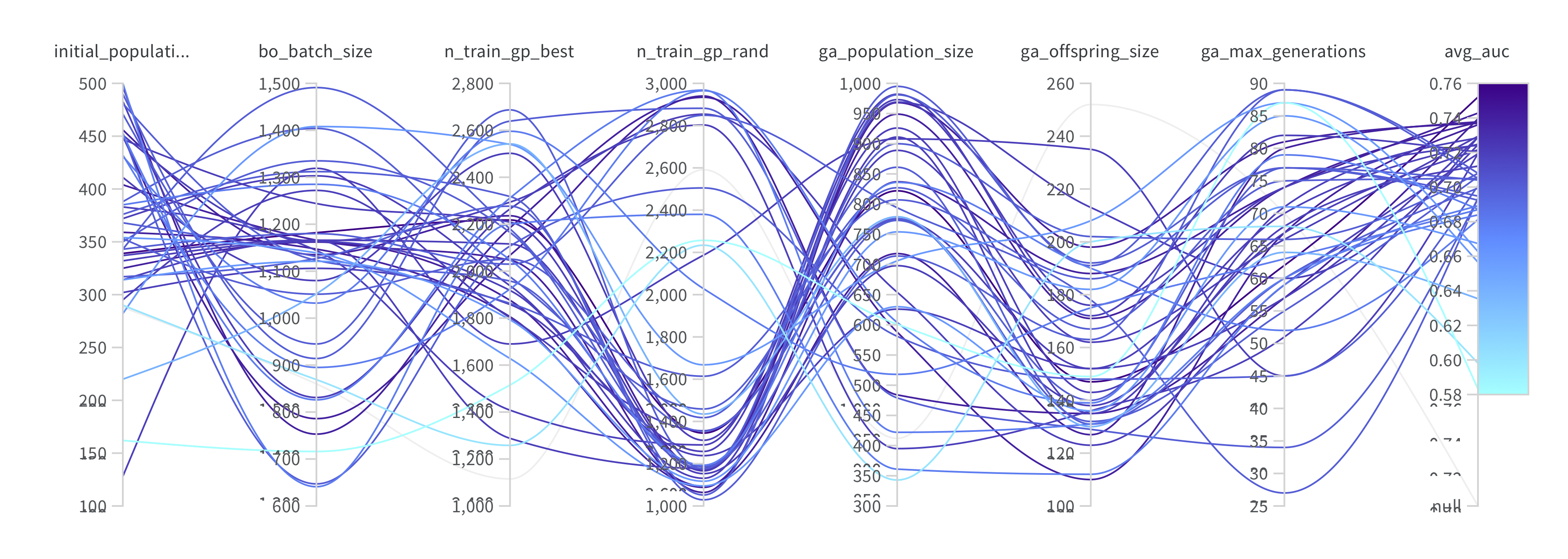}
  \caption{The hyper-parameter tuning result of GP BO.}
  \label{fig:tune_gp_bo}
\end{figure*} 

\begin{figure*}[h]
  \centering
  \includegraphics[width=\textwidth]{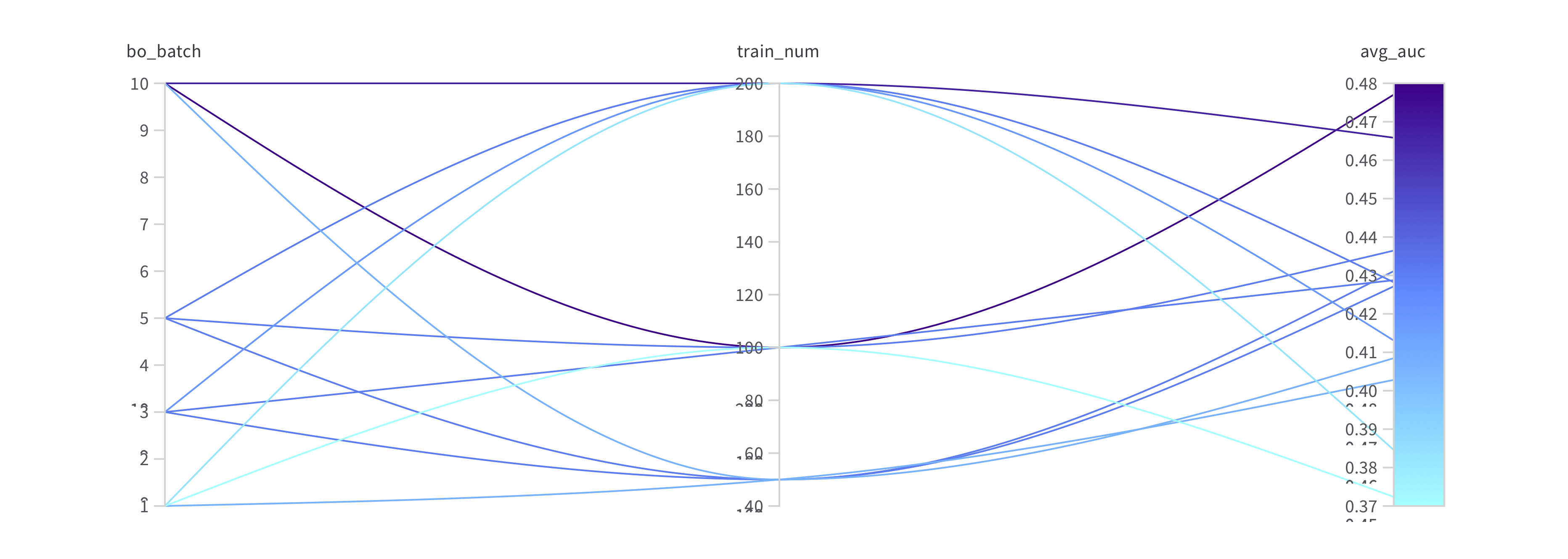}
  \caption{The hyper-parameter tuning result of DoG-AE.}
  \label{fig:tune_dog_ae}
\end{figure*} 

\begin{figure*}[h]
  \centering
  \includegraphics[width=\textwidth]{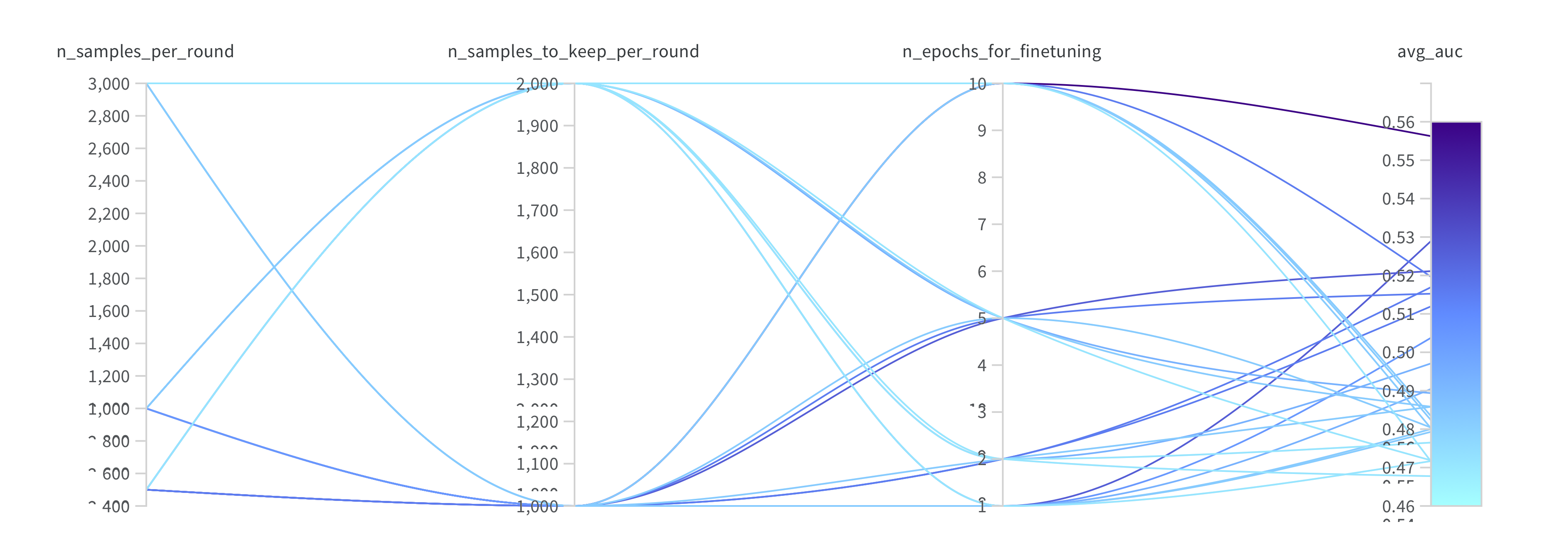}
  \caption{The hyper-parameter tuning result of DoG-Gen.}
  \label{fig:tune_dog_gen}
\end{figure*} 

\begin{figure*}[h]
  \centering
  \includegraphics[width=\textwidth]{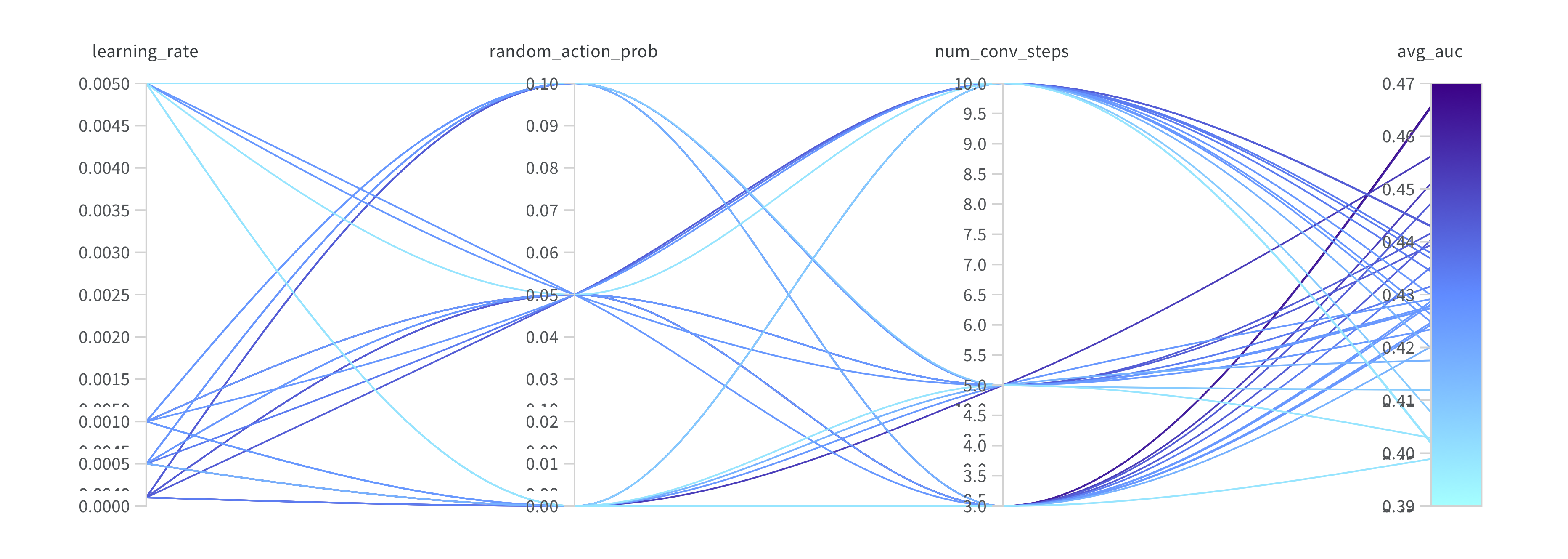}
  \caption{The hyper-parameter tuning result of GFlowNet.}
  \label{fig:tune_gflownet}
\end{figure*} 

\begin{figure*}[h]
  \centering
  \includegraphics[width=\textwidth]{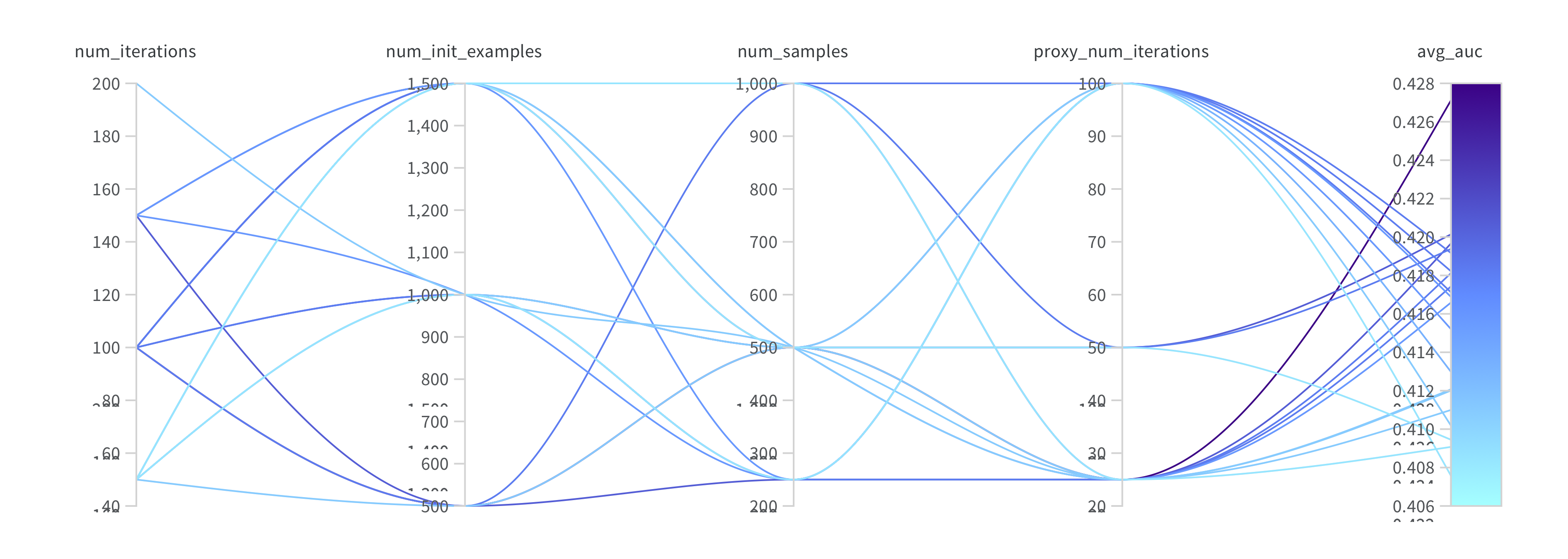}
  \caption{The hyper-parameter tuning result of GFlowNet-AL.}
  \label{fig:tune_gflownet_al}
\end{figure*} 

\begin{figure*}[h]
  \centering
  \includegraphics[width=\textwidth]{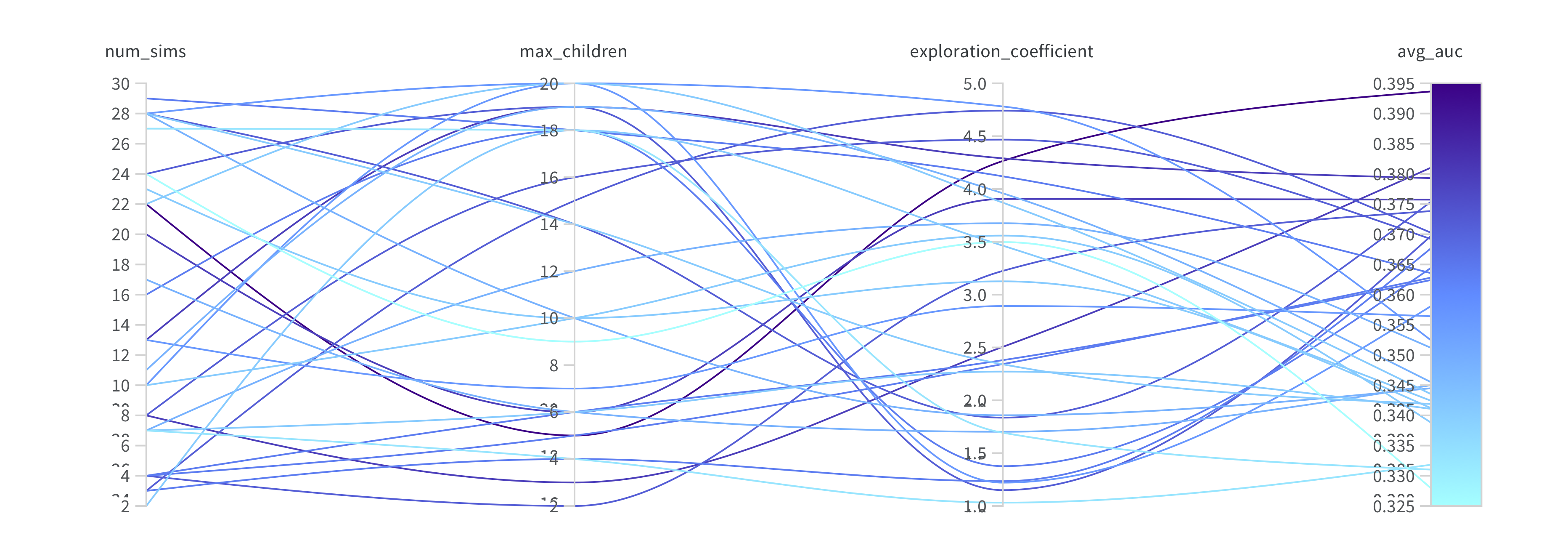}
  \caption{The hyper-parameter tuning result of Graph MCTS.}
  \label{fig:tune_graph_mcts}
\end{figure*} 

\begin{figure*}[h]
  \centering
  \includegraphics[width=\textwidth]{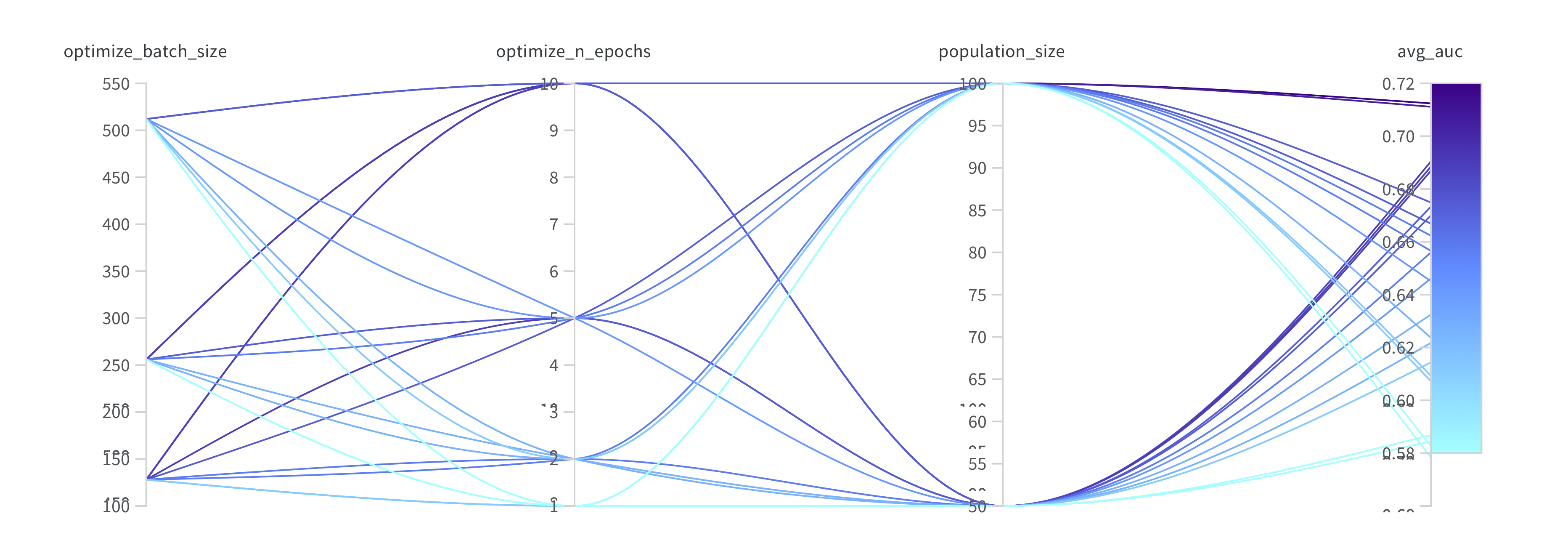}
  \caption{The hyper-parameter tuning result of SMILES LSTM HC.}
  \label{fig:tune_lstm_hc}
\end{figure*}

\begin{figure*}[h]
  \centering
  \includegraphics[width=\textwidth]{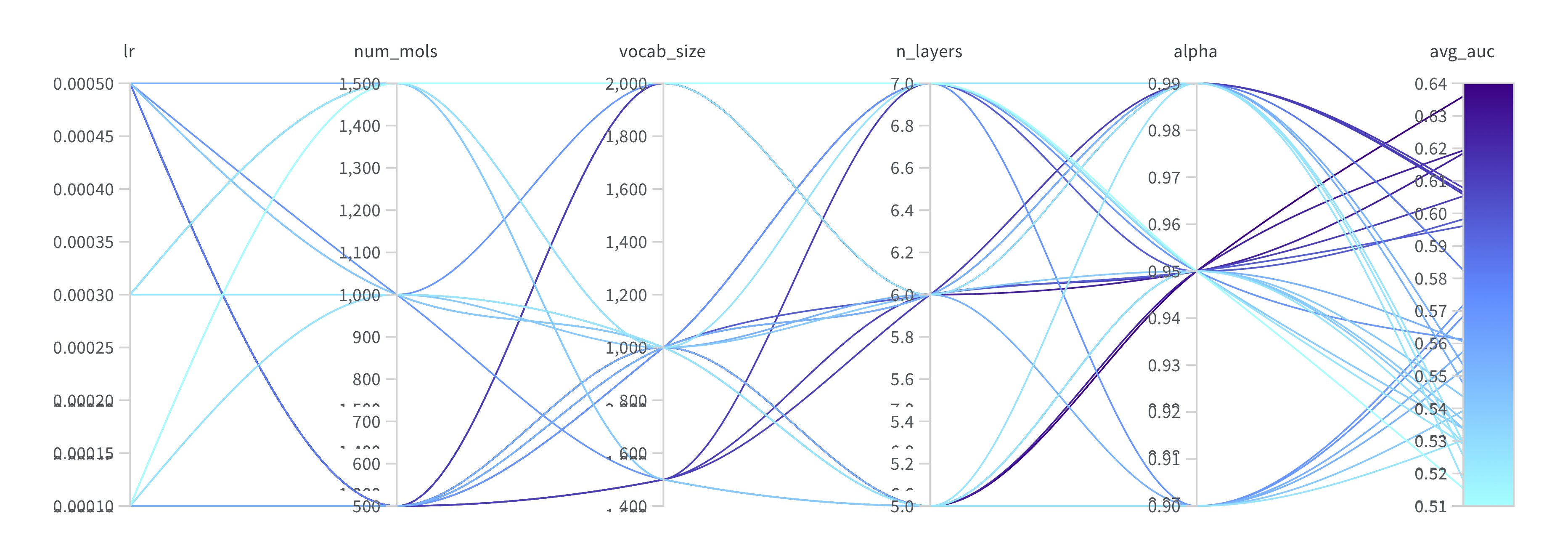}
  \caption{The hyper-parameter tuning result of MARS (part 1).}
  \label{fig:tune_mars1}
\end{figure*}

\begin{figure*}[h]
  \centering
  \includegraphics[width=\textwidth]{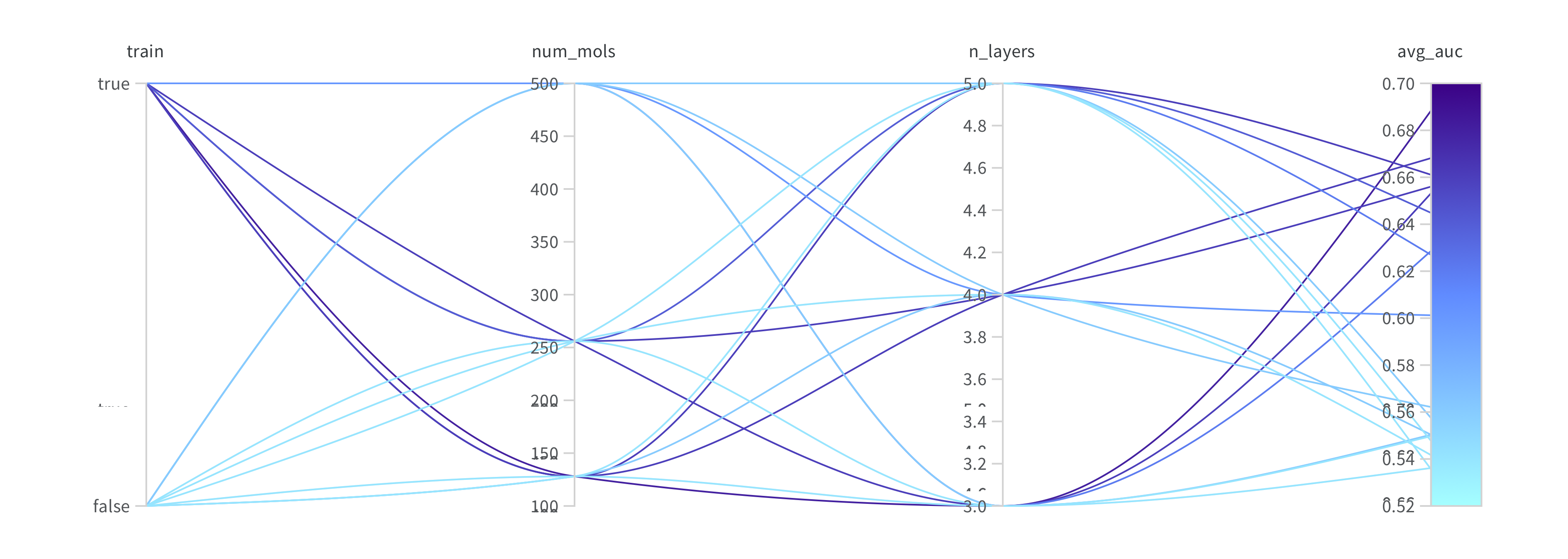}
  \caption{The hyper-parameter tuning result of MARS (part 2).}
  \label{fig:tune_mars2}
\end{figure*}

\begin{figure*}[h]
  \centering
  \includegraphics[width=\textwidth]{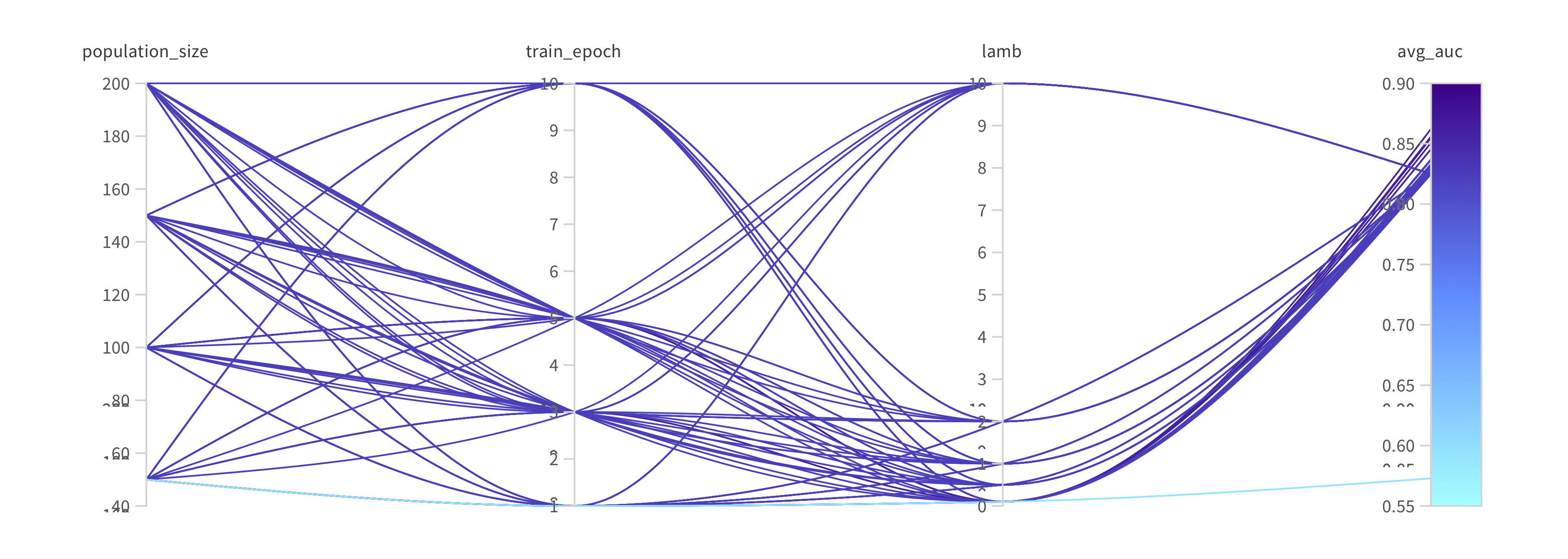}
  \caption{The hyper-parameter tuning result of MIMOSA.}
  \label{fig:tune_mimosa}
\end{figure*}

\begin{figure*}[h]
  \centering
  \includegraphics[width=\textwidth]{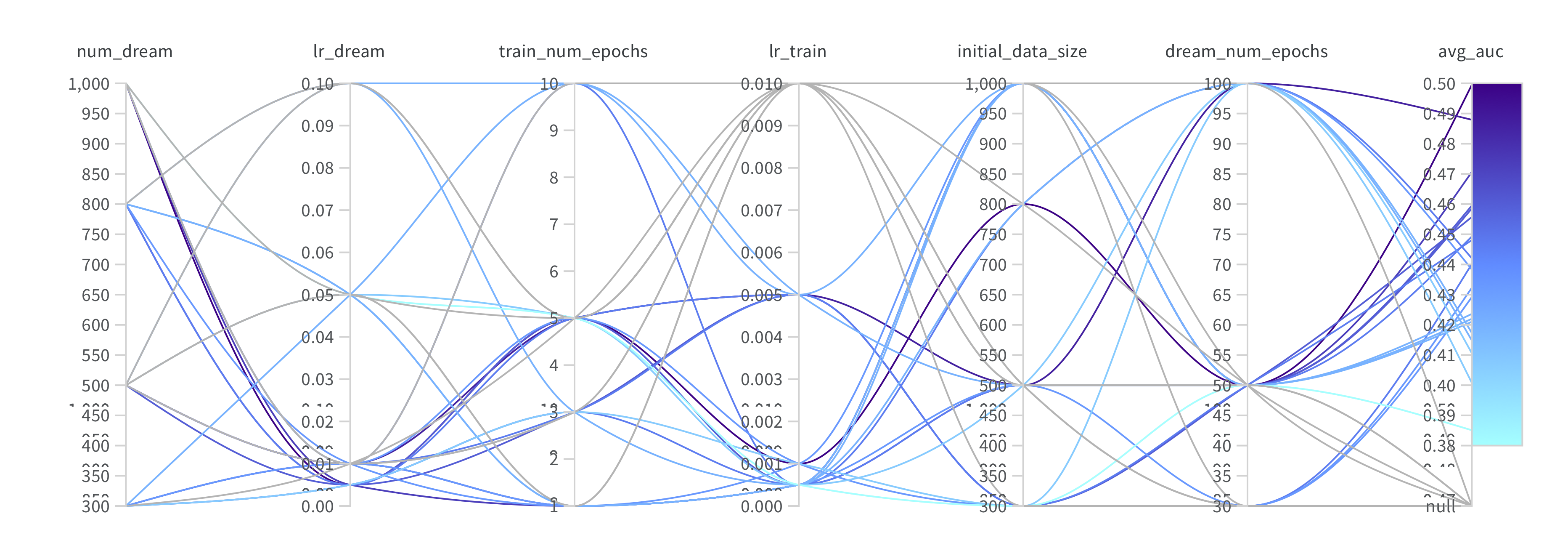}
  \caption{The hyper-parameter tuning result of Pasithea.}
  \label{fig:tune_pasithea}
\end{figure*} 

\begin{figure*}[h]
  \centering
  \includegraphics[width=\textwidth]{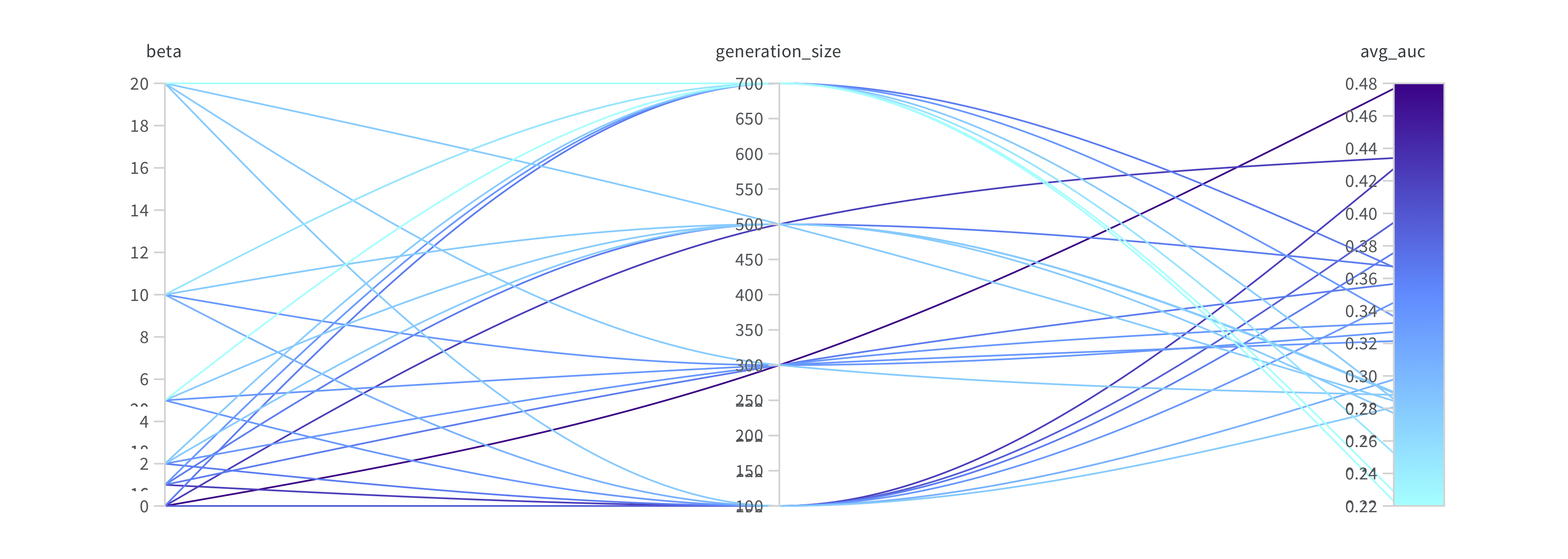}
  \caption{The hyper-parameter tuning result of GA+D.}
  \label{fig:tune_selfies_ga}
\end{figure*} 

\begin{figure*}[h]
  \centering
  \includegraphics[width=\textwidth]{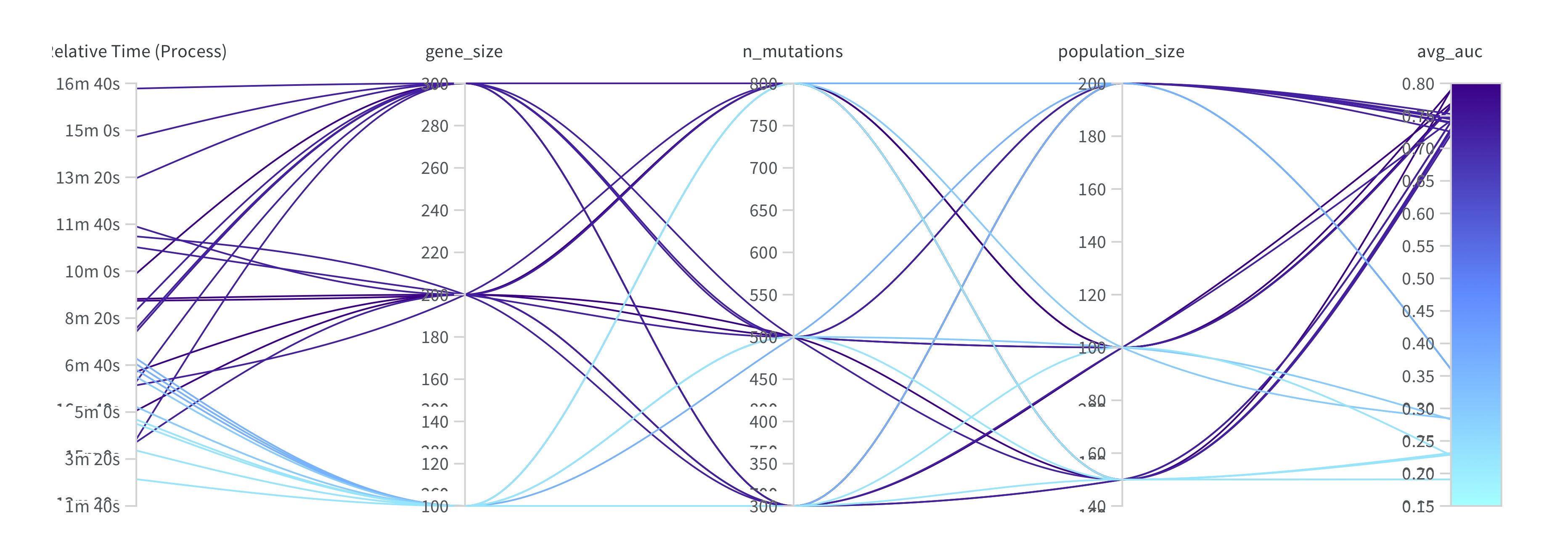}
  \caption{The hyper-parameter tuning result of SMILES GA.}
  \label{fig:tune_smiles_ga}
\end{figure*} 

\begin{figure*}[h]
  \centering
  \includegraphics[width=\textwidth]{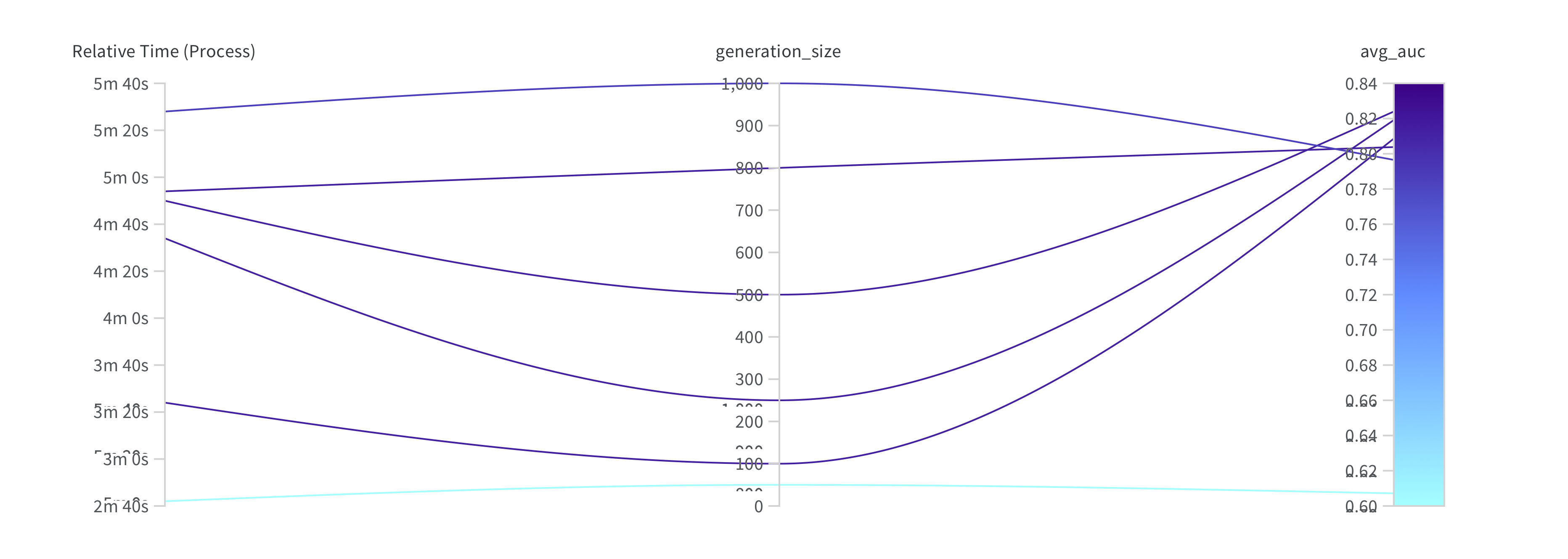}
  \caption{The hyper-parameter tuning result of STONED.}
  \label{fig:tune_stoned}
\end{figure*} 

\newpage
\subsection{Additional Tables}

\begin{table*}[h]
\centering
\caption{The mean and standard deviation of \textbf{AUC Top-1} from 5 independent runs. We ranked the methods by the summation of mean \textbf{AUC Top-1} of all tasks. 
(Continued)
}
\label{tab:auctop1_1}
\begin{adjustbox}{width=\textwidth}
\scriptsize
\begin{tabularx}{\textwidth}{c | Y | Y | Y | Y | Y}
\toprule
Method & REINVENT & Graph GA & REINVENT SELFIES & GP BO & LSTM HC \\
Assembly & SMILES & Fragments & SELFIES & Fragments & SMILES \\
\midrule
albuterol\_similarity & 0.903$\pm$0.003 & 0.875$\pm$0.022 & 0.853$\pm$0.032 & \textbf{0.922$\pm$0.011} & 0.798$\pm$0.030 \\
amlodipine\_mpo & 0.652$\pm$0.037 & \textbf{0.685$\pm$0.021} & 0.626$\pm$0.020 & 0.607$\pm$0.044 & 0.636$\pm$0.020 \\
celecoxib\_rediscovery & 0.801$\pm$0.098 & 0.683$\pm$0.122 & 0.616$\pm$0.039 & \textbf{0.808$\pm$0.075} & 0.619$\pm$0.030 \\
deco\_hop & 0.679$\pm$0.047 & 0.624$\pm$0.005 & 0.645$\pm$0.022 & 0.645$\pm$0.026 & \textbf{0.888$\pm$0.008} \\
drd2 & 0.969$\pm$0.007 & 0.992$\pm$0.001 & 0.980$\pm$0.003 & 0.957$\pm$0.007 & 0.957$\pm$0.012 \\
fexofenadine\_mpo & 0.801$\pm$0.007 & 0.774$\pm$0.011 & 0.762$\pm$0.004 & 0.740$\pm$0.007 & 0.753$\pm$0.010 \\
gsk3b & 0.893$\pm$0.044 & 0.826$\pm$0.069 & 0.823$\pm$0.035 & 0.877$\pm$0.040 & 0.935$\pm$0.014 \\
isomers\_c7h8n2o2 & 0.882$\pm$0.029 & 0.899$\pm$0.060 & 0.888$\pm$0.033 & 0.747$\pm$0.112 & 0.615$\pm$0.058 \\
isomers\_c9h10n2o2pf2cl & 0.673$\pm$0.059 & 0.765$\pm$0.046 & 0.780$\pm$0.024 & 0.513$\pm$0.172 & 0.465$\pm$0.034 \\
jnk3 & \textbf{0.813$\pm$0.024} & 0.597$\pm$0.141 & 0.670$\pm$0.069 & 0.592$\pm$0.159 & 0.787$\pm$0.057 \\
median1 & 0.367$\pm$0.009 & 0.319$\pm$0.027 & \textbf{0.367$\pm$0.012} & 0.315$\pm$0.017 & 0.298$\pm$0.019 \\
median2 & 0.289$\pm$0.009 & 0.288$\pm$0.008 & 0.269$\pm$0.006 & \textbf{0.309$\pm$0.009} & 0.276$\pm$0.014 \\
mestranol\_similarity & 0.637$\pm$0.048 & 0.615$\pm$0.027 & 0.646$\pm$0.033 & 0.665$\pm$0.082 & 0.613$\pm$0.054 \\
osimertinib\_mpo & \textbf{0.849$\pm$0.010} & 0.845$\pm$0.006 & 0.831$\pm$0.002 & 0.803$\pm$0.004 & 0.815$\pm$0.003 \\
perindopril\_mpo & 0.553$\pm$0.017 & 0.559$\pm$0.010 & 0.533$\pm$0.022 & 0.511$\pm$0.013 & 0.514$\pm$0.010 \\
qed & \textbf{0.943$\pm$0.000} & 0.942$\pm$0.000 & 0.942$\pm$0.000 & 0.941$\pm$0.000 & 0.942$\pm$0.000 \\
ranolazine\_mpo & \textbf{0.786$\pm$0.009} & 0.758$\pm$0.013 & 0.777$\pm$0.018 & 0.762$\pm$0.013 & 0.756$\pm$0.011 \\
scaffold\_hop & 0.572$\pm$0.021 & 0.526$\pm$0.008 & 0.540$\pm$0.015 & 0.562$\pm$0.023 & \textbf{0.628$\pm$0.058} \\
sitagliptin\_mpo & 0.055$\pm$0.015 & \textbf{0.492$\pm$0.068} & 0.257$\pm$0.116 & 0.237$\pm$0.061 & 0.128$\pm$0.030 \\
thiothixene\_rediscovery & 0.557$\pm$0.013 & 0.506$\pm$0.026 & 0.517$\pm$0.046 & \textbf{0.591$\pm$0.026} & 0.485$\pm$0.015 \\
troglitazone\_rediscovery & 0.458$\pm$0.034 & 0.410$\pm$0.016 & 0.371$\pm$0.014 & 0.431$\pm$0.015 & 0.405$\pm$0.025 \\
valsartan\_smarts & \textbf{0.187$\pm$0.374} & 0.000$\pm$0.000 & 0.000$\pm$0.000 & 0.000$\pm$0.000 & 0.000$\pm$0.000 \\
zaleplon\_mpo & \textbf{0.383$\pm$0.062} & 0.366$\pm$0.033 & 0.369$\pm$0.020 & 0.252$\pm$0.071 & 0.286$\pm$0.021 \\
\midrule
Sum & 14.711 & 14.356 & 14.077 & 13.798 & 13.611 \\
Rank & 1 & 2 & 3 & 4 & 5 \\

\hline
\hline

Method & STONED & DoG-Gen & SynNet & SMILES GA & MolPal \\
Assembly & SELFIES & Synthesis & Synthesis & SMILES & - \\
\midrule
albuterol\_similarity & 0.755$\pm$0.078 & 0.747$\pm$0.014 & 0.645$\pm$0.052 & 0.679$\pm$0.056 & 0.694$\pm$0.003 \\
amlodipine\_mpo & 0.616$\pm$0.048 & 0.555$\pm$0.004 & 0.580$\pm$0.006 & 0.564$\pm$0.004 & 0.621$\pm$0.010 \\
celecoxib\_rediscovery & 0.388$\pm$0.044 & 0.525$\pm$0.012 & 0.485$\pm$0.032 & 0.350$\pm$0.026 & 0.496$\pm$0.002 \\
deco\_hop & 0.612$\pm$0.009 & 0.874$\pm$0.003 & 0.626$\pm$0.011 & 0.613$\pm$0.007 & 0.804$\pm$0.019 \\
drd2 & 0.933$\pm$0.019 & \textbf{0.992$\pm$0.000} & 0.983$\pm$0.002 & 0.930$\pm$0.017 & 0.902$\pm$0.007 \\
fexofenadine\_mpo & \textbf{0.803$\pm$0.018} & 0.730$\pm$0.007 & 0.778$\pm$0.017 & 0.729$\pm$0.016 & 0.704$\pm$0.001 \\
gsk3b & 0.702$\pm$0.055 & \textbf{0.958$\pm$0.007} & 0.854$\pm$0.044 & 0.667$\pm$0.039 & 0.776$\pm$0.002 \\
isomers\_c7h8n2o2 & 0.913$\pm$0.010 & 0.580$\pm$0.034 & 0.607$\pm$0.050 & \textbf{0.930$\pm$0.022} & 0.832$\pm$0.005 \\
isomers\_c9h10n2o2pf2cl & 0.822$\pm$0.028 & 0.365$\pm$0.031 & 0.433$\pm$0.084 & \textbf{0.881$\pm$0.062} & 0.361$\pm$0.009 \\
jnk3 & 0.543$\pm$0.093 & 0.707$\pm$0.022 & 0.722$\pm$0.042 & 0.339$\pm$0.025 & 0.457$\pm$0.024 \\
median1 & 0.281$\pm$0.020 & 0.242$\pm$0.003 & 0.235$\pm$0.010 & 0.204$\pm$0.011 & 0.301$\pm$0.000 \\
median2 & 0.249$\pm$0.033 & 0.229$\pm$0.003 & 0.251$\pm$0.007 & 0.203$\pm$0.006 & 0.266$\pm$0.000 \\
mestranol\_similarity & 0.621$\pm$0.103 & 0.487$\pm$0.010 & 0.424$\pm$0.020 & 0.480$\pm$0.029 & \textbf{0.708$\pm$0.006} \\
osimertinib\_mpo & 0.827$\pm$0.012 & 0.800$\pm$0.004 & 0.810$\pm$0.004 & 0.823$\pm$0.011 & 0.803$\pm$0.001 \\
perindopril\_mpo & 0.493$\pm$0.012 & 0.505$\pm$0.003 & \textbf{0.579$\pm$0.014} & 0.453$\pm$0.011 & 0.495$\pm$0.003 \\
qed & 0.942$\pm$0.000 & 0.939$\pm$0.000 & 0.943$\pm$0.000 & 0.942$\pm$0.000 & 0.942$\pm$0.000 \\
ranolazine\_mpo & 0.783$\pm$0.029 & 0.759$\pm$0.010 & 0.762$\pm$0.007 & 0.719$\pm$0.023 & 0.515$\pm$0.007 \\
scaffold\_hop & 0.524$\pm$0.035 & 0.541$\pm$0.005 & 0.517$\pm$0.013 & 0.498$\pm$0.012 & 0.518$\pm$0.001 \\
sitagliptin\_mpo & 0.406$\pm$0.083 & 0.102$\pm$0.019 & 0.060$\pm$0.034 & 0.396$\pm$0.052 & 0.100$\pm$0.013 \\
thiothixene\_rediscovery & 0.374$\pm$0.027 & 0.411$\pm$0.006 & 0.444$\pm$0.029 & 0.322$\pm$0.018 & 0.356$\pm$0.000 \\
troglitazone\_rediscovery & 0.325$\pm$0.018 & \textbf{0.492$\pm$0.025} & 0.299$\pm$0.006 & 0.275$\pm$0.018 & 0.290$\pm$0.000 \\
valsartan\_smarts & 0.000$\pm$0.000 & 0.000$\pm$0.000 & 0.000$\pm$0.000 & 0.000$\pm$0.000 & 0.000$\pm$0.000 \\
zaleplon\_mpo & 0.333$\pm$0.026 & 0.171$\pm$0.021 & 0.376$\pm$0.019 & 0.349$\pm$0.042 & 0.262$\pm$0.004 \\
\midrule
Sum & 13.256 & 12.721 & 12.425 & 12.357 & 12.214 \\
Rank & 6 & 7 & 8 & 9 & 10 \\
\bottomrule
\end{tabularx}
\end{adjustbox}
\end{table*}

\newpage

\begin{table*}[h]
\centering
\caption{(Continued)
}
\label{tab:auctop1_2}
\begin{adjustbox}{width=\textwidth}
\scriptsize
\begin{tabularx}{\textwidth}{c | Y | Y | Y | Y | Y}
\toprule
Method & DST & MARS & LSTM HC SELFIES & MIMOSA & DoG-AE \\
Assembly & Fragments & Fragments & SELFIES & Fragments & Synthesis \\
\midrule
albuterol\_similarity & 0.671$\pm$0.021 & 0.668$\pm$0.121 & 0.726$\pm$0.029 & 0.649$\pm$0.023 & 0.621$\pm$0.045 \\
amlodipine\_mpo & 0.573$\pm$0.047 & 0.523$\pm$0.022 & 0.569$\pm$0.006 & 0.590$\pm$0.009 & 0.534$\pm$0.013 \\
celecoxib\_rediscovery & 0.422$\pm$0.005 & 0.428$\pm$0.049 & 0.425$\pm$0.015 & 0.420$\pm$0.017 & 0.401$\pm$0.024 \\
deco\_hop & 0.619$\pm$0.010 & 0.597$\pm$0.003 & 0.601$\pm$0.004 & 0.625$\pm$0.004 & 0.841$\pm$0.009 \\
drd2 & 0.886$\pm$0.021 & 0.938$\pm$0.014 & 0.847$\pm$0.036 & 0.879$\pm$0.024 & 0.985$\pm$0.003 \\
fexofenadine\_mpo & 0.741$\pm$0.005 & 0.729$\pm$0.007 & 0.716$\pm$0.006 & 0.721$\pm$0.013 & 0.716$\pm$0.041 \\
gsk3b & 0.737$\pm$0.036 & 0.628$\pm$0.055 & 0.537$\pm$0.040 & 0.639$\pm$0.046 & 0.754$\pm$0.118 \\
isomers\_c7h8n2o2 & 0.664$\pm$0.074 & 0.807$\pm$0.048 & 0.695$\pm$0.024 & 0.635$\pm$0.058 & 0.549$\pm$0.187 \\
isomers\_c9h10n2o2pf2cl & 0.551$\pm$0.040 & 0.640$\pm$0.023 & 0.476$\pm$0.039 & 0.345$\pm$0.045 & 0.134$\pm$0.072 \\
jnk3 & 0.600$\pm$0.062 & 0.548$\pm$0.088 & 0.303$\pm$0.053 & 0.401$\pm$0.071 & 0.539$\pm$0.133 \\
median1 & 0.256$\pm$0.017 & 0.226$\pm$0.012 & 0.268$\pm$0.014 & 0.270$\pm$0.005 & 0.200$\pm$0.009 \\
median2 & 0.194$\pm$0.021 & 0.196$\pm$0.009 & 0.228$\pm$0.006 & 0.227$\pm$0.005 & 0.198$\pm$0.008 \\
mestranol\_similarity & 0.491$\pm$0.049 & 0.430$\pm$0.024 & 0.492$\pm$0.014 & 0.509$\pm$0.033 & 0.429$\pm$0.027 \\
osimertinib\_mpo & 0.799$\pm$0.005 & 0.797$\pm$0.007 & 0.801$\pm$0.005 & 0.801$\pm$0.014 & 0.787$\pm$0.024 \\
perindopril\_mpo & 0.487$\pm$0.012 & 0.475$\pm$0.007 & 0.472$\pm$0.006 & 0.506$\pm$0.019 & 0.459$\pm$0.023 \\
qed & 0.941$\pm$0.000 & 0.940$\pm$0.001 & 0.942$\pm$0.000 & 0.942$\pm$0.000 & 0.938$\pm$0.001 \\
ranolazine\_mpo & 0.657$\pm$0.057 & 0.763$\pm$0.017 & 0.677$\pm$0.014 & 0.673$\pm$0.020 & 0.735$\pm$0.015 \\
scaffold\_hop & 0.507$\pm$0.004 & 0.482$\pm$0.009 & 0.495$\pm$0.007 & 0.517$\pm$0.017 & 0.519$\pm$0.020 \\
sitagliptin\_mpo & 0.159$\pm$0.074 & 0.040$\pm$0.013 & 0.203$\pm$0.025 & 0.136$\pm$0.029 & 0.037$\pm$0.031 \\
thiothixene\_rediscovery & 0.391$\pm$0.011 & 0.382$\pm$0.031 & 0.370$\pm$0.009 & 0.365$\pm$0.017 & 0.352$\pm$0.015 \\
troglitazone\_rediscovery & 0.295$\pm$0.019 & 0.274$\pm$0.019 & 0.283$\pm$0.004 & 0.314$\pm$0.008 & 0.344$\pm$0.052 \\
valsartan\_smarts & 0.000$\pm$0.000 & 0.000$\pm$0.000 & 0.000$\pm$0.000 & 0.000$\pm$0.000 & 0.000$\pm$0.000 \\
zaleplon\_mpo & 0.257$\pm$0.025 & 0.291$\pm$0.020 & 0.303$\pm$0.027 & 0.204$\pm$0.033 & 0.145$\pm$0.082 \\
\midrule
Sum & 11.911 & 11.814 & 11.441 & 11.378 & 11.227 \\
Rank & 11 & 12 & 13 & 14 & 15 \\

\hline
\hline

Method & VAE BO SELFIES & Screening & VAE BO SMILES & Pasithea & GFlowNet \\
Assembly & SELFIES & - & SMILES & SELFIES & Fragments \\
\midrule
albuterol\_similarity & 0.572$\pm$0.043 & 0.546$\pm$0.029 & 0.563$\pm$0.019 & 0.499$\pm$0.005 & 0.501$\pm$0.029 \\
amlodipine\_mpo & 0.580$\pm$0.004 & 0.580$\pm$0.014 & 0.602$\pm$0.032 & 0.582$\pm$2.676 & 0.467$\pm$0.006 \\
celecoxib\_rediscovery & 0.386$\pm$0.022 & 0.394$\pm$0.005 & 0.406$\pm$0.013 & 0.351$\pm$0.010 & 0.374$\pm$0.007 \\
deco\_hop & 0.590$\pm$0.002 & 0.611$\pm$0.002 & 0.608$\pm$0.003 & 0.603$\pm$0.012 & 0.590$\pm$0.001 \\
drd2 & 0.808$\pm$0.055 & 0.797$\pm$0.059 & 0.818$\pm$0.073 & 0.557$\pm$0.087 & 0.791$\pm$0.041 \\
fexofenadine\_mpo & 0.698$\pm$0.006 & 0.690$\pm$0.011 & 0.699$\pm$0.008 & 0.702$\pm$0.039 & 0.714$\pm$0.007 \\
gsk3b & 0.506$\pm$0.091 & 0.657$\pm$0.078 & 0.536$\pm$0.046 & 0.401$\pm$0.075 & 0.691$\pm$0.033 \\
isomers\_c7h8n2o2 & 0.497$\pm$0.052 & 0.395$\pm$0.079 & 0.332$\pm$0.052 & 0.792$\pm$0.057 & 0.539$\pm$0.068 \\
isomers\_c9h10n2o2pf2cl & 0.367$\pm$0.083 & 0.218$\pm$0.047 & 0.175$\pm$0.032 & 0.499$\pm$0.081 & 0.173$\pm$0.046 \\
jnk3 & 0.341$\pm$0.070 & 0.362$\pm$0.063 & 0.375$\pm$0.054 & 0.206$\pm$0.033 & 0.492$\pm$0.024 \\
median1 & 0.226$\pm$0.008 & 0.249$\pm$0.010 & 0.252$\pm$0.035 & 0.212$\pm$0.018 & 0.224$\pm$0.006 \\
median2 & 0.200$\pm$0.001 & 0.232$\pm$0.015 & 0.211$\pm$0.003 & 0.193$\pm$0.006 & 0.193$\pm$0.005 \\
mestranol\_similarity & 0.495$\pm$0.050 & 0.507$\pm$0.121 & 0.508$\pm$0.035 & 0.446$\pm$0.012 & 0.363$\pm$0.017 \\
osimertinib\_mpo & 0.790$\pm$0.003 & 0.784$\pm$0.005 & 0.792$\pm$0.004 & 0.787$\pm$0.008 & 0.801$\pm$0.008 \\
perindopril\_mpo & 0.458$\pm$0.015 & 0.478$\pm$0.018 & 0.469$\pm$0.019 & 0.445$\pm$0.015 & 0.455$\pm$0.008 \\
qed & 0.941$\pm$0.001 & 0.942$\pm$0.000 & 0.942$\pm$0.000 & 0.938$\pm$0.003 & 0.939$\pm$0.001 \\
ranolazine\_mpo & 0.534$\pm$0.046 & 0.485$\pm$0.026 & 0.563$\pm$0.049 & 0.437$\pm$0.050 & 0.679$\pm$0.004 \\
scaffold\_hop & 0.474$\pm$0.007 & 0.503$\pm$0.004 & 0.493$\pm$0.009 & 0.493$\pm$0.019 & 0.474$\pm$0.003 \\
sitagliptin\_mpo & 0.173$\pm$0.041 & 0.076$\pm$0.023 & 0.088$\pm$0.043 & 0.176$\pm$0.050 & 0.028$\pm$0.012 \\
thiothixene\_rediscovery & 0.329$\pm$0.007 & 0.350$\pm$0.007 & 0.355$\pm$0.017 & 0.330$\pm$0.015 & 0.312$\pm$0.011 \\
troglitazone\_rediscovery & 0.275$\pm$0.023 & 0.272$\pm$0.010 & 0.286$\pm$0.011 & 0.256$\pm$0.007 & 0.202$\pm$0.006 \\
valsartan\_smarts & 0.017$\pm$0.034 & 0.000$\pm$0.000 & 0.019$\pm$0.039 & 0.060$\pm$0.121 & 0.000$\pm$0.000 \\
zaleplon\_mpo & 0.322$\pm$0.033 & 0.222$\pm$0.058 & 0.094$\pm$0.028 & 0.185$\pm$0.033 & 0.066$\pm$0.042 \\
\midrule
Sum & 10.589 & 10.363 & 10.197 & 10.162 & 10.079 \\
Rank & 16 & 17 & 18 & 19 & 20 \\
\bottomrule
\end{tabularx}
\end{adjustbox}
\end{table*}

\newpage

\begin{table*}[h]
\centering
\caption{(Continued)
}
\label{tab:auctop1_3}
\begin{adjustbox}{width=\textwidth}
\scriptsize
\begin{tabularx}{\textwidth}{c | Y | Y | Y | Y | Y}
\toprule
Method & JT-VAE BO & GFlowNet-AL & GA+D & Graph MCTS & MolDQN \\
Assembly & Fragments & Fragments & SELFIES & Atoms & Atoms \\
\midrule
albuterol\_similarity & 0.541$\pm$0.051 & 0.440$\pm$0.020 & 0.528$\pm$0.029 & 0.625$\pm$0.028 & 0.348$\pm$0.022 \\
amlodipine\_mpo & 0.582$\pm$1.791 & 0.448$\pm$0.007 & 0.421$\pm$0.033 & 0.472$\pm$0.019 & 0.343$\pm$0.013 \\
celecoxib\_rediscovery & 0.385$\pm$0.025 & 0.289$\pm$0.005 & 0.241$\pm$0.023 & 0.297$\pm$0.009 & 0.114$\pm$0.016 \\
deco\_hop & 0.595$\pm$0.003 & 0.591$\pm$0.004 & 0.553$\pm$0.005 & 0.561$\pm$0.003 & 0.549$\pm$0.001 \\
drd2 & 0.741$\pm$0.185 & 0.716$\pm$0.073 & 0.425$\pm$0.207 & 0.476$\pm$0.111 & 0.030$\pm$0.003 \\
fexofenadine\_mpo & 0.695$\pm$0.012 & 0.713$\pm$0.004 & 0.607$\pm$0.008 & 0.596$\pm$0.011 & 0.498$\pm$0.015 \\
gsk3b & 0.482$\pm$0.054 & 0.640$\pm$0.031 & 0.363$\pm$0.022 & 0.354$\pm$0.032 & 0.286$\pm$0.012 \\
isomers\_c7h8n2o2 & 0.243$\pm$0.075 & 0.450$\pm$0.097 & 0.878$\pm$0.012 & 0.701$\pm$0.048 & 0.594$\pm$0.077 \\
isomers\_c9h10n2o2pf2cl & 0.273$\pm$0.121 & 0.131$\pm$0.024 & 0.681$\pm$0.022 & 0.601$\pm$0.066 & 0.481$\pm$0.043 \\
jnk3 & 0.353$\pm$0.063 & 0.431$\pm$0.035 & 0.234$\pm$0.021 & 0.144$\pm$0.031 & 0.134$\pm$0.013 \\
median1 & 0.209$\pm$0.017 & 0.223$\pm$0.001 & 0.201$\pm$0.007 & 0.234$\pm$0.014 & 0.144$\pm$0.013 \\
median2 & 0.191$\pm$0.003 & 0.182$\pm$0.004 & 0.128$\pm$0.005 & 0.141$\pm$0.003 & 0.094$\pm$0.003 \\
mestranol\_similarity & 0.448$\pm$0.055 & 0.327$\pm$0.016 & 0.396$\pm$0.019 & 0.307$\pm$0.007 & 0.209$\pm$0.007 \\
osimertinib\_mpo & 0.794$\pm$0.007 & 0.803$\pm$0.008 & 0.689$\pm$0.029 & 0.718$\pm$0.007 & 0.689$\pm$0.006 \\
perindopril\_mpo & 0.453$\pm$0.012 & 0.448$\pm$0.009 & 0.187$\pm$0.095 & 0.310$\pm$0.023 & 0.247$\pm$0.034 \\
qed & 0.940$\pm$0.000 & 0.930$\pm$0.004 & 0.877$\pm$0.016 & 0.913$\pm$0.009 & 0.788$\pm$0.030 \\
ranolazine\_mpo & 0.583$\pm$0.039 & 0.680$\pm$0.018 & 0.575$\pm$0.014 & 0.316$\pm$0.051 & 0.084$\pm$0.034 \\
scaffold\_hop & 0.487$\pm$0.006 & 0.472$\pm$0.004 & 0.417$\pm$0.009 & 0.421$\pm$0.004 & 0.411$\pm$0.006 \\
sitagliptin\_mpo & 0.134$\pm$0.070 & 0.020$\pm$0.011 & 0.311$\pm$0.023 & 0.138$\pm$0.047 & 0.010$\pm$0.008 \\
thiothixene\_rediscovery & 0.311$\pm$0.011 & 0.294$\pm$0.012 & 0.240$\pm$0.035 & 0.249$\pm$0.009 & 0.108$\pm$0.011 \\
troglitazone\_rediscovery & 0.257$\pm$0.003 & 0.201$\pm$0.008 & 0.160$\pm$0.013 & 0.245$\pm$0.015 & 0.135$\pm$0.007 \\
valsartan\_smarts & 0.000$\pm$0.000 & 0.000$\pm$0.000 & 0.000$\pm$0.000 & 0.000$\pm$0.000 & 0.000$\pm$0.000 \\
zaleplon\_mpo & 0.266$\pm$0.047 & 0.029$\pm$0.009 & 0.263$\pm$0.014 & 0.113$\pm$0.035 & 0.026$\pm$0.015 \\
\midrule
Sum & 9.973 & 9.470 & 9.387 & 8.944 & 6.332 \\
Rank & 21 & 22 & 23 & 24 & 25 \\
\bottomrule
\end{tabularx}
\end{adjustbox}
\end{table*}

\begin{table*}[h]
\centering
\caption{The mean and standard deviation of \textbf{AUC Top-100} from 5 independent runs. We ranked the methods by the summation of mean \textbf{AUC Top-100} of all tasks. 
(Continued)
}
\label{tab:auctop100_1}
\begin{adjustbox}{width=\textwidth}
\scriptsize
\begin{tabularx}{\textwidth}{c | Y | Y | Y | Y | Y}
\toprule
Method & REINVENT & Graph GA & STONED & REINVENT SELFIES & GP BO \\
Assembly & SMILES & Fragments & SELFIES & SELFIES & Fragments \\
\midrule
albuterol\_similarity & \textbf{0.842$\pm$0.013} & 0.759$\pm$0.014 & 0.727$\pm$0.070 & 0.781$\pm$0.033 & 0.839$\pm$0.019 \\
amlodipine\_mpo & 0.608$\pm$0.033 & \textbf{0.622$\pm$0.018} & 0.593$\pm$0.045 & 0.574$\pm$0.009 & 0.538$\pm$0.045 \\
celecoxib\_rediscovery & \textbf{0.646$\pm$0.053} & 0.558$\pm$0.075 & 0.366$\pm$0.035 & 0.515$\pm$0.044 & 0.637$\pm$0.041 \\
deco\_hop & 0.649$\pm$0.040 & 0.609$\pm$0.004 & 0.605$\pm$0.007 & 0.610$\pm$0.003 & 0.611$\pm$0.014 \\
drd2 & 0.908$\pm$0.007 & \textbf{0.924$\pm$0.020} & 0.881$\pm$0.026 & 0.898$\pm$0.008 & 0.870$\pm$0.031 \\
fexofenadine\_mpo & 0.752$\pm$0.005 & 0.731$\pm$0.012 & \textbf{0.777$\pm$0.013} & 0.705$\pm$0.002 & 0.685$\pm$0.005 \\
gsk3b & \textbf{0.823$\pm$0.042} & 0.737$\pm$0.072 & 0.621$\pm$0.045 & 0.711$\pm$0.043 & 0.808$\pm$0.046 \\
isomers\_c7h8n2o2 & 0.798$\pm$0.043 & 0.761$\pm$0.058 & 0.864$\pm$0.016 & 0.791$\pm$0.023 & 0.564$\pm$0.128 \\
isomers\_c9h10n2o2pf2cl & 0.590$\pm$0.050 & 0.628$\pm$0.048 & 0.765$\pm$0.039 & 0.656$\pm$0.045 & 0.399$\pm$0.184 \\
jnk3 & \textbf{0.742$\pm$0.025} & 0.488$\pm$0.126 & 0.481$\pm$0.092 & 0.567$\pm$0.057 & 0.524$\pm$0.149 \\
median1 & \textbf{0.325$\pm$0.009} & 0.264$\pm$0.019 & 0.244$\pm$0.013 & 0.299$\pm$0.012 & 0.275$\pm$0.012 \\
median2 & 0.258$\pm$0.006 & 0.251$\pm$0.011 & 0.236$\pm$0.031 & 0.232$\pm$0.005 & \textbf{0.275$\pm$0.007} \\
mestranol\_similarity & \textbf{0.586$\pm$0.046} & 0.523$\pm$0.019 & 0.577$\pm$0.094 & 0.578$\pm$0.026 & 0.572$\pm$0.086 \\
osimertinib\_mpo & \textbf{0.806$\pm$0.008} & 0.799$\pm$0.004 & 0.799$\pm$0.011 & 0.791$\pm$0.005 & 0.750$\pm$0.010 \\
perindopril\_mpo & 0.511$\pm$0.016 & 0.503$\pm$0.008 & 0.472$\pm$0.011 & 0.487$\pm$0.019 & 0.460$\pm$0.009 \\
qed & 0.931$\pm$0.000 & 0.930$\pm$0.000 & 0.930$\pm$0.000 & 0.929$\pm$0.000 & 0.919$\pm$0.002 \\
ranolazine\_mpo & 0.719$\pm$0.008 & 0.670$\pm$0.012 & \textbf{0.738$\pm$0.028} & 0.695$\pm$0.023 & 0.694$\pm$0.016 \\
scaffold\_hop & \textbf{0.537$\pm$0.015} & 0.502$\pm$0.005 & 0.512$\pm$0.031 & 0.502$\pm$0.011 & 0.527$\pm$0.015 \\
sitagliptin\_mpo & 0.006$\pm$0.000 & 0.330$\pm$0.074 & \textbf{0.351$\pm$0.078} & 0.118$\pm$0.105 & 0.117$\pm$0.036 \\
thiothixene\_rediscovery & 0.493$\pm$0.013 & 0.433$\pm$0.021 & 0.352$\pm$0.027 & 0.456$\pm$0.033 & \textbf{0.502$\pm$0.023} \\
troglitazone\_rediscovery & \textbf{0.411$\pm$0.029} & 0.358$\pm$0.014 & 0.307$\pm$0.018 & 0.314$\pm$0.013 & 0.379$\pm$0.013 \\
valsartan\_smarts & \textbf{0.168$\pm$0.336} & 0.000$\pm$0.000 & 0.000$\pm$0.000 & 0.000$\pm$0.000 & 0.000$\pm$0.000 \\
zaleplon\_mpo & \textbf{0.325$\pm$0.062} & 0.305$\pm$0.025 & 0.307$\pm$0.027 & 0.257$\pm$0.031 & 0.165$\pm$0.070 \\
\midrule
Sum & 13.445 & 12.696 & 12.518 & 12.475 & 12.122 \\
Rank & 1 & 2 & 3 & 4 & 5 \\

\hline
\hline

Method & SMILES GA & LSTM HC & SynNet & DST & MIMOSA \\
Assembly & SMILES & SMILES & Synthesis & Fragments & Fragments \\
\midrule
albuterol\_similarity & 0.643$\pm$0.068 & 0.602$\pm$0.014 & 0.494$\pm$0.026 & 0.539$\pm$0.012 & 0.566$\pm$0.014 \\
amlodipine\_mpo & 0.534$\pm$0.011 & 0.533$\pm$0.010 & 0.533$\pm$0.006 & 0.469$\pm$0.005 & 0.509$\pm$0.004 \\
celecoxib\_rediscovery & 0.331$\pm$0.027 & 0.448$\pm$0.012 & 0.374$\pm$0.023 & 0.333$\pm$0.005 & 0.353$\pm$0.003 \\
deco\_hop & 0.605$\pm$0.006 & \textbf{0.738$\pm$0.019} & 0.593$\pm$0.005 & 0.591$\pm$0.006 & 0.605$\pm$0.002 \\
drd2 & 0.875$\pm$0.022 & 0.788$\pm$0.017 & 0.897$\pm$0.015 & 0.738$\pm$0.025 & 0.709$\pm$0.021 \\
fexofenadine\_mpo & 0.700$\pm$0.014 & 0.680$\pm$0.003 & 0.720$\pm$0.011 & 0.690$\pm$0.004 & 0.672$\pm$0.009 \\
gsk3b & 0.586$\pm$0.043 & 0.670$\pm$0.011 & 0.655$\pm$0.039 & 0.598$\pm$0.036 & 0.475$\pm$0.040 \\
isomers\_c7h8n2o2 & \textbf{0.880$\pm$0.027} & 0.313$\pm$0.032 & 0.167$\pm$0.028 & 0.380$\pm$0.083 & 0.468$\pm$0.036 \\
isomers\_c9h10n2o2pf2cl & \textbf{0.823$\pm$0.073} & 0.186$\pm$0.015 & 0.053$\pm$0.022 & 0.307$\pm$0.084 & 0.259$\pm$0.046 \\
jnk3 & 0.288$\pm$0.022 & 0.489$\pm$0.025 & 0.466$\pm$0.038 & 0.489$\pm$0.059 & 0.302$\pm$0.055 \\
median1 & 0.185$\pm$0.012 & 0.213$\pm$0.007 & 0.187$\pm$0.005 & 0.193$\pm$0.006 & 0.212$\pm$0.004 \\
median2 & 0.191$\pm$0.005 & 0.217$\pm$0.004 & 0.205$\pm$0.003 & 0.166$\pm$0.016 & 0.195$\pm$0.004 \\
mestranol\_similarity & 0.449$\pm$0.028 & 0.428$\pm$0.018 & 0.352$\pm$0.018 & 0.400$\pm$0.016 & 0.391$\pm$0.013 \\
osimertinib\_mpo & 0.798$\pm$0.012 & 0.749$\pm$0.001 & 0.759$\pm$0.002 & 0.742$\pm$0.001 & 0.750$\pm$0.010 \\
perindopril\_mpo & 0.436$\pm$0.013 & 0.446$\pm$0.004 & \textbf{0.512$\pm$0.010} & 0.425$\pm$0.009 & 0.458$\pm$0.007 \\
qed & \textbf{0.932$\pm$0.001} & 0.923$\pm$0.001 & 0.930$\pm$0.001 & 0.925$\pm$0.001 & 0.925$\pm$0.000 \\
ranolazine\_mpo & 0.670$\pm$0.028 & 0.630$\pm$0.012 & 0.690$\pm$0.015 & 0.579$\pm$0.044 & 0.587$\pm$0.015 \\
scaffold\_hop & 0.487$\pm$0.010 & 0.491$\pm$0.004 & 0.478$\pm$0.007 & 0.480$\pm$0.003 & 0.488$\pm$0.009 \\
sitagliptin\_mpo & 0.307$\pm$0.058 & 0.020$\pm$0.006 & 0.007$\pm$0.004 & 0.017$\pm$0.005 & 0.052$\pm$0.012 \\
thiothixene\_rediscovery & 0.300$\pm$0.014 & 0.377$\pm$0.005 & 0.351$\pm$0.012 & 0.325$\pm$0.007 & 0.316$\pm$0.015 \\
troglitazone\_rediscovery & 0.256$\pm$0.024 & 0.301$\pm$0.008 & 0.254$\pm$0.007 & 0.250$\pm$0.020 & 0.273$\pm$0.008 \\
valsartan\_smarts & 0.000$\pm$0.000 & 0.000$\pm$0.000 & 0.000$\pm$0.000 & 0.000$\pm$0.000 & 0.000$\pm$0.000 \\
zaleplon\_mpo & 0.310$\pm$0.034 & 0.111$\pm$0.005 & 0.223$\pm$0.017 & 0.089$\pm$0.063 & 0.132$\pm$0.038 \\
\midrule
Sum & 11.598 & 10.365 & 9.914 & 9.737 & 9.708 \\
Rank & 6 & 7 & 8 & 9 & 10 \\
\bottomrule
\end{tabularx}
\end{adjustbox}
\end{table*}

\newpage

\begin{table*}[h]
\centering
\caption{(Continued)
}
\label{tab:auctop100_2}
\begin{adjustbox}{width=\textwidth}
\scriptsize
\begin{tabularx}{\textwidth}{c | Y | Y | Y | Y | Y}
\toprule
Method & DoG-Gen & MARS & LSTM HC SELFIES & GA+D & MolPal \\
Assembly & Synthesis & Fragments & SELFIES & SELFIES & - \\
\midrule
albuterol\_similarity & 0.578$\pm$0.011 & 0.478$\pm$0.121 & 0.572$\pm$0.027 & 0.448$\pm$0.018 & 0.528$\pm$0.002 \\
amlodipine\_mpo & 0.489$\pm$0.003 & 0.465$\pm$0.010 & 0.485$\pm$0.003 & 0.365$\pm$0.029 & 0.514$\pm$0.006 \\
celecoxib\_rediscovery & 0.387$\pm$0.006 & 0.317$\pm$0.056 & 0.324$\pm$0.004 & 0.200$\pm$0.024 & 0.349$\pm$0.002 \\
deco\_hop & 0.715$\pm$0.010 & 0.577$\pm$0.002 & 0.573$\pm$0.001 & 0.545$\pm$0.004 & 0.585$\pm$0.001 \\
drd2 & 0.740$\pm$0.003 & 0.752$\pm$0.019 & 0.510$\pm$0.035 & 0.314$\pm$0.190 & 0.403$\pm$0.009 \\
fexofenadine\_mpo & 0.640$\pm$0.001 & 0.669$\pm$0.003 & 0.650$\pm$0.004 & 0.553$\pm$0.007 & 0.639$\pm$0.002 \\
gsk3b & 0.629$\pm$0.018 & 0.463$\pm$0.042 & 0.292$\pm$0.003 & 0.309$\pm$0.016 & 0.319$\pm$0.007 \\
isomers\_c7h8n2o2 & 0.305$\pm$0.011 & 0.583$\pm$0.025 & 0.415$\pm$0.042 & 0.799$\pm$0.024 & 0.199$\pm$0.003 \\
isomers\_c9h10n2o2pf2cl & 0.095$\pm$0.004 & 0.471$\pm$0.015 & 0.208$\pm$0.007 & 0.608$\pm$0.024 & 0.071$\pm$0.001 \\
jnk3 & 0.436$\pm$0.022 & 0.386$\pm$0.081 & 0.136$\pm$0.003 & 0.195$\pm$0.020 & 0.200$\pm$0.004 \\
median1 & 0.181$\pm$0.000 & 0.169$\pm$0.015 & 0.200$\pm$0.003 & 0.152$\pm$0.006 & 0.202$\pm$0.001 \\
median2 & 0.188$\pm$0.001 & 0.159$\pm$0.012 & 0.178$\pm$0.004 & 0.111$\pm$0.005 & 0.191$\pm$0.000 \\
mestranol\_similarity & 0.369$\pm$0.004 & 0.323$\pm$0.033 & 0.381$\pm$0.006 & 0.333$\pm$0.014 & 0.433$\pm$0.002 \\
osimertinib\_mpo & 0.706$\pm$0.001 & 0.730$\pm$0.006 & 0.732$\pm$0.006 & 0.645$\pm$0.025 & 0.736$\pm$0.003 \\
perindopril\_mpo & 0.422$\pm$0.002 & 0.432$\pm$0.005 & 0.399$\pm$0.003 & 0.155$\pm$0.079 & 0.423$\pm$0.002 \\
qed & 0.912$\pm$0.000 & 0.886$\pm$0.012 & 0.920$\pm$0.001 & 0.821$\pm$0.013 & 0.930$\pm$0.000 \\
ranolazine\_mpo & 0.601$\pm$0.003 & 0.684$\pm$0.019 & 0.502$\pm$0.007 & 0.525$\pm$0.016 & 0.357$\pm$0.004 \\
scaffold\_hop & 0.483$\pm$0.004 & 0.450$\pm$0.004 & 0.445$\pm$0.001 & 0.406$\pm$0.008 & 0.461$\pm$0.000 \\
sitagliptin\_mpo & 0.015$\pm$0.005 & 0.004$\pm$0.000 & 0.040$\pm$0.002 & 0.232$\pm$0.021 & 0.014$\pm$0.000 \\
thiothixene\_rediscovery & 0.329$\pm$0.003 & 0.294$\pm$0.014 & 0.294$\pm$0.006 & 0.201$\pm$0.024 & 0.302$\pm$0.001 \\
troglitazone\_rediscovery & 0.331$\pm$0.016 & 0.228$\pm$0.013 & 0.225$\pm$0.002 & 0.139$\pm$0.012 & 0.245$\pm$0.000 \\
valsartan\_smarts & 0.000$\pm$0.000 & 0.000$\pm$0.000 & 0.000$\pm$0.000 & 0.000$\pm$0.000 & 0.000$\pm$0.000 \\
zaleplon\_mpo & 0.073$\pm$0.011 & 0.082$\pm$0.040 & 0.103$\pm$0.010 & 0.214$\pm$0.015 & 0.046$\pm$0.001 \\
\midrule
Sum & 9.635 & 9.612 & 8.595 & 8.280 & 8.156 \\
Rank & 11 & 12 & 13 & 14 & 15 \\

\hline
\hline

Method & GFlowNet & DoG-AE & GFlowNet-AL & Screening & VAE BO SMILES \\
Assembly & Fragments & Synthesis & Fragments & - & SMILES \\
\midrule
albuterol\_similarity & 0.374$\pm$0.009 & 0.423$\pm$0.020 & 0.324$\pm$0.002 & 0.410$\pm$0.003 & 0.412$\pm$0.003 \\
amlodipine\_mpo & 0.398$\pm$0.004 & 0.457$\pm$0.004 & 0.374$\pm$0.002 & 0.477$\pm$0.000 & 0.475$\pm$0.002 \\
celecoxib\_rediscovery & 0.275$\pm$0.006 & 0.282$\pm$0.019 & 0.213$\pm$0.002 & 0.289$\pm$0.002 & 0.291$\pm$0.001 \\
deco\_hop & 0.572$\pm$0.002 & 0.626$\pm$0.041 & 0.570$\pm$0.000 & 0.571$\pm$0.000 & 0.570$\pm$0.000 \\
drd2 & 0.279$\pm$0.065 & 0.543$\pm$0.069 & 0.165$\pm$0.010 & 0.186$\pm$0.005 & 0.187$\pm$0.014 \\
fexofenadine\_mpo & 0.653$\pm$0.004 & 0.618$\pm$0.007 & 0.645$\pm$0.002 & 0.613$\pm$0.002 & 0.616$\pm$0.002 \\
gsk3b & 0.585$\pm$0.022 & 0.356$\pm$0.074 & 0.504$\pm$0.011 & 0.235$\pm$0.008 & 0.214$\pm$0.007 \\
isomers\_c7h8n2o2 & 0.191$\pm$0.013 & 0.052$\pm$0.015 & 0.084$\pm$0.018 & 0.036$\pm$0.005 & 0.039$\pm$0.005 \\
isomers\_c9h10n2o2pf2cl & 0.047$\pm$0.012 & 0.012$\pm$0.004 & 0.021$\pm$0.002 & 0.027$\pm$0.002 & 0.025$\pm$0.001 \\
jnk3 & 0.367$\pm$0.022 & 0.245$\pm$0.065 & 0.272$\pm$0.016 & 0.126$\pm$0.005 & 0.123$\pm$0.003 \\
median1 & 0.165$\pm$0.004 & 0.134$\pm$0.006 & 0.145$\pm$0.001 & 0.161$\pm$0.002 & 0.160$\pm$0.001 \\
median2 & 0.164$\pm$0.001 & 0.156$\pm$0.006 & 0.156$\pm$0.001 & 0.170$\pm$0.001 & 0.169$\pm$0.000 \\
mestranol\_similarity & 0.273$\pm$0.006 & 0.304$\pm$0.013 & 0.246$\pm$0.002 & 0.328$\pm$0.005 & 0.323$\pm$0.001 \\
osimertinib\_mpo & 0.758$\pm$0.001 & 0.661$\pm$0.007 & 0.758$\pm$0.001 & 0.704$\pm$0.001 & 0.712$\pm$0.002 \\
perindopril\_mpo & 0.384$\pm$0.012 & 0.374$\pm$0.007 & 0.375$\pm$0.001 & 0.397$\pm$0.002 & 0.398$\pm$0.001 \\
qed & 0.861$\pm$0.007 & 0.877$\pm$0.004 & 0.820$\pm$0.007 & 0.922$\pm$0.000 & 0.922$\pm$0.000 \\
ranolazine\_mpo & 0.615$\pm$0.004 & 0.566$\pm$0.038 & 0.543$\pm$0.006 & 0.302$\pm$0.003 & 0.318$\pm$0.003 \\
scaffold\_hop & 0.445$\pm$0.001 & 0.453$\pm$0.011 & 0.442$\pm$0.000 & 0.443$\pm$0.000 & 0.441$\pm$0.001 \\
sitagliptin\_mpo & 0.001$\pm$0.000 & 0.001$\pm$0.000 & 0.001$\pm$0.000 & 0.006$\pm$0.000 & 0.006$\pm$0.000 \\
thiothixene\_rediscovery & 0.246$\pm$0.009 & 0.256$\pm$0.012 & 0.224$\pm$0.003 & 0.272$\pm$0.001 & 0.270$\pm$0.002 \\
troglitazone\_rediscovery & 0.170$\pm$0.001 & 0.207$\pm$0.007 & 0.167$\pm$0.000 & 0.218$\pm$0.002 & 0.220$\pm$0.001 \\
valsartan\_smarts & 0.000$\pm$0.000 & 0.000$\pm$0.000 & 0.000$\pm$0.000 & 0.000$\pm$0.000 & 0.000$\pm$0.000 \\
zaleplon\_mpo & 0.011$\pm$0.013 & 0.005$\pm$0.002 & 0.002$\pm$0.000 & 0.010$\pm$0.002 & 0.006$\pm$0.001 \\
\midrule
Sum & 7.844 & 7.620 & 7.060 & 6.915 & 6.909 \\
Rank & 16 & 17 & 18 & 19 & 20 \\
\bottomrule
\end{tabularx}
\end{adjustbox}
\end{table*}

\newpage

\begin{table*}[h]
\centering
\caption{(Continued)
}
\label{tab:auctop100_3}
\begin{adjustbox}{width=\textwidth}
\scriptsize
\begin{tabularx}{\textwidth}{c | Y | Y | Y | Y | Y}
\toprule
Method & VAE BO SELFIES & JT-VAE BO & Pasithea & Graph MCTS & MolDQN \\
Assembly & SELFIES & Fragments & SELFIES & Atoms & Atoms \\
\midrule
albuterol\_similarity & 0.413$\pm$0.003 & 0.412$\pm$0.018 & 0.365$\pm$0.004 & 0.497$\pm$0.014 & 0.273$\pm$0.008 \\
amlodipine\_mpo & 0.465$\pm$0.002 & 0.468$\pm$0.007 & 0.442$\pm$0.004 & 0.385$\pm$0.005 & 0.230$\pm$0.011 \\
celecoxib\_rediscovery & 0.260$\pm$0.004 & 0.240$\pm$0.012 & 0.242$\pm$0.005 & 0.204$\pm$0.007 & 0.080$\pm$0.002 \\
deco\_hop & 0.560$\pm$0.001 & 0.567$\pm$0.002 & 0.558$\pm$0.000 & 0.539$\pm$0.001 & 0.534$\pm$0.001 \\
drd2 & 0.174$\pm$0.007 & 0.170$\pm$0.046 & 0.060$\pm$0.010 & 0.121$\pm$0.008 & 0.018$\pm$0.000 \\
fexofenadine\_mpo & 0.619$\pm$0.006 & 0.616$\pm$0.009 & 0.583$\pm$0.011 & 0.522$\pm$0.005 & 0.431$\pm$0.010 \\
gsk3b & 0.206$\pm$0.006 & 0.201$\pm$0.025 & 0.141$\pm$0.029 & 0.183$\pm$0.008 & 0.176$\pm$0.008 \\
isomers\_c7h8n2o2 & 0.094$\pm$0.008 & 0.025$\pm$0.009 & 0.450$\pm$0.021 & 0.304$\pm$0.018 & 0.269$\pm$0.011 \\
isomers\_c9h10n2o2pf2cl & 0.063$\pm$0.004 & 0.021$\pm$0.006 & 0.193$\pm$0.093 & 0.203$\pm$0.044 & 0.134$\pm$0.013 \\
jnk3 & 0.113$\pm$0.002 & 0.119$\pm$0.007 & 0.076$\pm$0.008 & 0.066$\pm$0.005 & 0.075$\pm$0.003 \\
median1 & 0.159$\pm$0.003 & 0.144$\pm$0.006 & 0.133$\pm$0.007 & 0.143$\pm$0.003 & 0.094$\pm$0.003 \\
median2 & 0.159$\pm$0.001 & 0.159$\pm$0.003 & 0.153$\pm$0.002 & 0.117$\pm$0.000 & 0.072$\pm$0.002 \\
mestranol\_similarity & 0.316$\pm$0.001 & 0.299$\pm$0.010 & 0.274$\pm$0.007 & 0.229$\pm$0.006 & 0.150$\pm$0.008 \\
osimertinib\_mpo & 0.711$\pm$0.002 & 0.727$\pm$0.005 & 0.643$\pm$0.019 & 0.655$\pm$0.003 & 0.636$\pm$0.005 \\
perindopril\_mpo & 0.382$\pm$0.002 & 0.390$\pm$0.007 & 0.364$\pm$0.003 & 0.219$\pm$0.005 & 0.125$\pm$0.019 \\
qed & 0.914$\pm$0.001 & 0.912$\pm$0.004 & 0.896$\pm$0.004 & 0.832$\pm$0.006 & 0.630$\pm$0.006 \\
ranolazine\_mpo & 0.313$\pm$0.019 & 0.337$\pm$0.062 & 0.211$\pm$0.011 & 0.122$\pm$0.010 & 0.018$\pm$0.006 \\
scaffold\_hop & 0.427$\pm$0.002 & 0.443$\pm$0.005 & 0.424$\pm$0.001 & 0.392$\pm$0.002 & 0.388$\pm$0.003 \\
sitagliptin\_mpo & 0.021$\pm$0.003 & 0.010$\pm$0.005 & 0.026$\pm$0.004 & 0.013$\pm$0.002 & 0.000$\pm$0.000 \\
thiothixene\_rediscovery & 0.252$\pm$0.003 & 0.242$\pm$0.007 & 0.238$\pm$0.006 & 0.193$\pm$0.002 & 0.081$\pm$0.004 \\
troglitazone\_rediscovery & 0.207$\pm$0.002 & 0.206$\pm$0.004 & 0.200$\pm$0.002 & 0.194$\pm$0.005 & 0.101$\pm$0.002 \\
valsartan\_smarts & 0.000$\pm$0.000 & 0.000$\pm$0.000 & 0.000$\pm$0.000 & 0.000$\pm$0.000 & 0.000$\pm$0.000 \\
zaleplon\_mpo & 0.059$\pm$0.004 & 0.024$\pm$0.012 & 0.027$\pm$0.007 & 0.014$\pm$0.005 & 0.002$\pm$0.001 \\
\midrule
Sum & 6.899 & 6.740 & 6.712 & 6.156 & 4.528 \\
Rank & 21 & 22 & 23 & 24 & 25 \\

\bottomrule
\end{tabularx}
\end{adjustbox}
\end{table*}


\begin{table*}[h]
\centering
\caption{The mean and standard deviation of \textbf{Top-100} from 5 independent runs. We ranked the methods by the summation of mean \textbf{Top-100} of all tasks. 
(Continued)
}
\label{tab:top100_1}
\begin{adjustbox}{width=\textwidth}
\scriptsize
\begin{tabularx}{\textwidth}{c | Y | Y | Y | Y | Y}
\toprule
Method & REINVENT & REINVENT SELFIES & Graph GA & LSTM HC & GP BO \\
Assembly & SMILES & SELFIES & Fragments & SMILES & Fragments \\
\midrule
albuterol\_similarity & 0.991$\pm$0.223 & 0.948$\pm$0.228 & 0.951$\pm$0.190 & 0.953$\pm$0.198 & \textbf{0.995$\pm$0.181} \\
amlodipine\_mpo & 0.728$\pm$0.114 & 0.684$\pm$0.097 & \textbf{0.743$\pm$0.106} & 0.669$\pm$0.073 & 0.638$\pm$0.090 \\
celecoxib\_rediscovery & \textbf{0.821$\pm$0.199} & 0.684$\pm$0.169 & 0.692$\pm$0.168 & 0.694$\pm$0.143 & 0.802$\pm$0.177 \\
deco\_hop & 0.796$\pm$0.104 & 0.706$\pm$0.044 & 0.642$\pm$0.026 & \textbf{0.921$\pm$0.140} & 0.698$\pm$0.053 \\
drd2 & \textbf{0.999$\pm$0.250} & 0.999$\pm$0.262 & 0.999$\pm$0.209 & 0.999$\pm$0.312 & 0.998$\pm$0.273 \\
fexofenadine\_mpo & \textbf{0.892$\pm$0.110} & 0.818$\pm$0.080 & 0.817$\pm$0.089 & 0.763$\pm$0.072 & 0.774$\pm$0.088 \\
gsk3b & \textbf{0.965$\pm$0.243} & 0.934$\pm$0.262 & 0.919$\pm$0.226 & 0.942$\pm$0.249 & 0.957$\pm$0.218 \\
isomers\_c7h8n2o2 & 0.986$\pm$0.338 & 0.948$\pm$0.301 & 0.948$\pm$0.241 & 0.848$\pm$0.289 & 0.749$\pm$0.264 \\
isomers\_c9h10n2o2pf2cl & 0.820$\pm$0.336 & 0.865$\pm$0.313 & 0.837$\pm$0.227 & 0.610$\pm$0.189 & 0.538$\pm$0.257 \\
jnk3 & \textbf{0.943$\pm$0.295} & 0.782$\pm$0.268 & 0.796$\pm$0.258 & 0.851$\pm$0.255 & 0.675$\pm$0.244 \\
median1 & \textbf{0.382$\pm$0.080} & 0.339$\pm$0.070 & 0.310$\pm$0.056 & 0.315$\pm$0.064 & 0.316$\pm$0.052 \\
median2 & \textbf{0.313$\pm$0.055} & 0.300$\pm$0.056 & 0.300$\pm$0.048 & 0.290$\pm$0.046 & 0.313$\pm$0.046 \\
mestranol\_similarity & 0.733$\pm$0.177 & \textbf{0.755$\pm$0.181} & 0.680$\pm$0.127 & 0.652$\pm$0.118 & 0.733$\pm$0.168 \\
osimertinib\_mpo & \textbf{0.896$\pm$0.101} & 0.865$\pm$0.084 & 0.861$\pm$0.098 & 0.829$\pm$0.095 & 0.812$\pm$0.091 \\
perindopril\_mpo & \textbf{0.635$\pm$0.099} & 0.608$\pm$0.105 & 0.591$\pm$0.078 & 0.532$\pm$0.060 & 0.523$\pm$0.058 \\
qed & \textbf{0.948$\pm$0.030} & 0.947$\pm$0.031 & 0.946$\pm$0.028 & 0.947$\pm$0.029 & 0.943$\pm$0.026 \\
ranolazine\_mpo & 0.848$\pm$0.163 & 0.836$\pm$0.170 & 0.781$\pm$0.150 & 0.783$\pm$0.149 & 0.790$\pm$0.147 \\
scaffold\_hop & \textbf{0.708$\pm$0.089} & 0.608$\pm$0.064 & 0.550$\pm$0.040 & 0.573$\pm$0.046 & 0.593$\pm$0.056 \\
sitagliptin\_mpo & 0.010$\pm$0.003 & 0.269$\pm$0.172 & \textbf{0.578$\pm$0.197} & 0.088$\pm$0.023 & 0.195$\pm$0.080 \\
thiothixene\_rediscovery & \textbf{0.644$\pm$0.149} & 0.616$\pm$0.144 & 0.536$\pm$0.099 & 0.554$\pm$0.097 & 0.613$\pm$0.111 \\
troglitazone\_rediscovery & \textbf{0.570$\pm$0.140} & 0.469$\pm$0.101 & 0.464$\pm$0.083 & 0.465$\pm$0.084 & 0.482$\pm$0.082 \\
valsartan\_smarts & \textbf{0.194$\pm$0.363} & 0.000$\pm$0.000 & 0.000$\pm$0.000 & 0.000$\pm$0.000 & 0.000$\pm$0.000 \\
zaleplon\_mpo & \textbf{0.463$\pm$0.187} & 0.384$\pm$0.159 & 0.389$\pm$0.105 & 0.330$\pm$0.105 & 0.216$\pm$0.093 \\
\midrule
Sum & 16.297 & 15.377 & 15.342 & 14.621 & 14.365 \\
Rank & 1 & 2 & 3 & 4 & 5 \\

\hline
\hline

Method & STONED & DoG-Gen & SMILES GA & DST & MIMOSA \\
Assembly & SELFIES & Synthesis & SMILES & Fragments & Fragments \\
\midrule
albuterol\_similarity & 0.805$\pm$0.163 & 0.852$\pm$0.154 & 0.698$\pm$0.127 & 0.682$\pm$0.117 & 0.667$\pm$0.127 \\
amlodipine\_mpo & 0.631$\pm$0.080 & 0.583$\pm$0.070 & 0.558$\pm$0.048 & 0.482$\pm$0.031 & 0.550$\pm$0.049 \\
celecoxib\_rediscovery & 0.393$\pm$0.062 & 0.583$\pm$0.115 & 0.349$\pm$0.047 & 0.381$\pm$0.056 & 0.405$\pm$0.068 \\
deco\_hop & 0.626$\pm$0.023 & 0.910$\pm$0.140 & 0.624$\pm$0.022 & 0.613$\pm$0.022 & 0.637$\pm$0.029 \\
drd2 & 0.997$\pm$0.274 & 0.999$\pm$0.314 & 0.986$\pm$0.260 & 0.993$\pm$0.377 & 0.981$\pm$0.383 \\
fexofenadine\_mpo & 0.847$\pm$0.100 & 0.736$\pm$0.085 & 0.756$\pm$0.075 & 0.753$\pm$0.083 & 0.723$\pm$0.075 \\
gsk3b & 0.733$\pm$0.178 & 0.959$\pm$0.251 & 0.687$\pm$0.171 & 0.821$\pm$0.267 & 0.672$\pm$0.207 \\
isomers\_c7h8n2o2 & 0.993$\pm$0.280 & 0.809$\pm$0.285 & \textbf{0.995$\pm$0.264} & 0.713$\pm$0.291 & 0.732$\pm$0.303 \\
isomers\_c9h10n2o2pf2cl & 0.919$\pm$0.271 & 0.337$\pm$0.100 & \textbf{0.966$\pm$0.272} & 0.698$\pm$0.272 & 0.413$\pm$0.182 \\
jnk3 & 0.587$\pm$0.179 & 0.802$\pm$0.221 & 0.374$\pm$0.094 & 0.748$\pm$0.249 & 0.457$\pm$0.160 \\
median1 & 0.264$\pm$0.045 & 0.261$\pm$0.055 & 0.198$\pm$0.029 & 0.231$\pm$0.037 & 0.251$\pm$0.050 \\
median2 & 0.260$\pm$0.045 & 0.263$\pm$0.038 & 0.204$\pm$0.019 & 0.178$\pm$0.022 & 0.216$\pm$0.025 \\
mestranol\_similarity & 0.665$\pm$0.151 & 0.552$\pm$0.105 & 0.508$\pm$0.085 & 0.469$\pm$0.078 & 0.443$\pm$0.072 \\
osimertinib\_mpo & 0.847$\pm$0.104 & 0.826$\pm$0.123 & 0.834$\pm$0.095 & 0.802$\pm$0.095 & 0.804$\pm$0.092 \\
perindopril\_mpo & 0.514$\pm$0.055 & 0.546$\pm$0.074 & 0.454$\pm$0.041 & 0.453$\pm$0.044 & 0.530$\pm$0.065 \\
qed & 0.943$\pm$0.025 & 0.947$\pm$0.046 & 0.946$\pm$0.026 & 0.941$\pm$0.026 & 0.939$\pm$0.025 \\
ranolazine\_mpo & \textbf{0.855$\pm$0.190} & 0.782$\pm$0.170 & 0.766$\pm$0.155 & 0.730$\pm$0.223 & 0.757$\pm$0.222 \\
scaffold\_hop & 0.545$\pm$0.055 & 0.559$\pm$0.044 & 0.509$\pm$0.030 & 0.512$\pm$0.035 & 0.527$\pm$0.040 \\
sitagliptin\_mpo & 0.482$\pm$0.174 & 0.085$\pm$0.025 & 0.436$\pm$0.151 & 0.027$\pm$0.011 & 0.101$\pm$0.044 \\
thiothixene\_rediscovery & 0.382$\pm$0.053 & 0.463$\pm$0.079 & 0.316$\pm$0.033 & 0.374$\pm$0.053 & 0.348$\pm$0.048 \\
troglitazone\_rediscovery & 0.351$\pm$0.052 & 0.534$\pm$0.130 & 0.270$\pm$0.037 & 0.286$\pm$0.044 & 0.316$\pm$0.049 \\
valsartan\_smarts & 0.000$\pm$0.000 & 0.000$\pm$0.000 & 0.000$\pm$0.000 & 0.000$\pm$0.000 & 0.000$\pm$0.000 \\
zaleplon\_mpo & 0.369$\pm$0.108 & 0.253$\pm$0.090 & 0.377$\pm$0.114 & 0.156$\pm$0.095 & 0.239$\pm$0.108 \\
\midrule
Sum & 14.017 & 13.653 & 12.824 & 12.052 & 11.717 \\
Rank & 6 & 7 & 8 & 9 & 10 \\
\bottomrule
\end{tabularx}
\end{adjustbox}
\end{table*}

\newpage

\begin{table*}[h]
\centering
\caption{(Continued)
}
\label{tab:top100_2}
\begin{adjustbox}{width=\textwidth}
\scriptsize
\begin{tabularx}{\textwidth}{c | Y | Y | Y | Y | Y}
\toprule
Method & LSTM HC SELFIES & GA+D & MARS & SynNet & GFlowNet \\
Assembly & SELFIES & SELFIES & Fragments & Synthesis & Fragments \\
\midrule
albuterol\_similarity & 0.891$\pm$0.173 & 0.576$\pm$0.123 & 0.554$\pm$0.154 & 0.554$\pm$0.065 & 0.422$\pm$0.051 \\
amlodipine\_mpo & 0.553$\pm$0.044 & 0.513$\pm$0.150 & 0.496$\pm$0.063 & 0.559$\pm$0.046 & 0.439$\pm$0.064 \\
celecoxib\_rediscovery & 0.473$\pm$0.082 & 0.252$\pm$0.066 & 0.394$\pm$0.091 & 0.410$\pm$0.052 & 0.308$\pm$0.041 \\
deco\_hop & 0.610$\pm$0.019 & 0.580$\pm$0.024 & 0.587$\pm$0.013 & 0.603$\pm$0.012 & 0.584$\pm$0.009 \\
drd2 & 0.992$\pm$0.371 & 0.678$\pm$0.339 & 0.959$\pm$0.279 & 0.985$\pm$0.201 & 0.492$\pm$0.152 \\
fexofenadine\_mpo & 0.726$\pm$0.069 & 0.717$\pm$0.206 & 0.717$\pm$0.102 & 0.749$\pm$0.057 & 0.678$\pm$0.037 \\
gsk3b & 0.503$\pm$0.120 & 0.482$\pm$0.130 & 0.536$\pm$0.133 & 0.797$\pm$0.182 & 0.637$\pm$0.076 \\
isomers\_c7h8n2o2 & 0.765$\pm$0.259 & 0.993$\pm$0.302 & 0.845$\pm$0.309 & 0.212$\pm$0.067 & 0.341$\pm$0.105 \\
isomers\_c9h10n2o2pf2cl & 0.436$\pm$0.138 & 0.811$\pm$0.286 & 0.737$\pm$0.287 & 0.079$\pm$0.036 & 0.100$\pm$0.032 \\
jnk3 & 0.216$\pm$0.050 & 0.356$\pm$0.109 & 0.497$\pm$0.164 & 0.563$\pm$0.150 & 0.438$\pm$0.077 \\
median1 & 0.285$\pm$0.051 & 0.171$\pm$0.035 & 0.181$\pm$0.026 & 0.197$\pm$0.020 & 0.186$\pm$0.026 \\
median2 & 0.240$\pm$0.032 & 0.148$\pm$0.034 & 0.169$\pm$0.020 & 0.213$\pm$0.016 & 0.175$\pm$0.012 \\
mestranol\_similarity & 0.520$\pm$0.079 & 0.497$\pm$0.131 & 0.375$\pm$0.060 & 0.385$\pm$0.044 & 0.306$\pm$0.034 \\
osimertinib\_mpo & 0.804$\pm$0.092 & 0.768$\pm$0.169 & 0.776$\pm$0.128 & 0.789$\pm$0.079 & 0.779$\pm$0.041 \\
perindopril\_mpo & 0.469$\pm$0.043 & 0.293$\pm$0.137 & 0.463$\pm$0.058 & 0.546$\pm$0.053 & 0.424$\pm$0.054 \\
qed & 0.945$\pm$0.026 & 0.928$\pm$0.140 & 0.903$\pm$0.038 & 0.942$\pm$0.029 & 0.913$\pm$0.086 \\
ranolazine\_mpo & 0.724$\pm$0.183 & 0.763$\pm$0.264 & 0.720$\pm$0.115 & 0.744$\pm$0.106 & 0.648$\pm$0.090 \\
scaffold\_hop & 0.495$\pm$0.027 & 0.459$\pm$0.038 & 0.461$\pm$0.019 & 0.489$\pm$0.017 & 0.460$\pm$0.014 \\
sitagliptin\_mpo & 0.101$\pm$0.029 & 0.436$\pm$0.165 & 0.010$\pm$0.003 & 0.008$\pm$0.005 & 0.004$\pm$0.001 \\
thiothixene\_rediscovery & 0.399$\pm$0.054 & 0.271$\pm$0.073 & 0.378$\pm$0.060 & 0.379$\pm$0.035 & 0.268$\pm$0.027 \\
troglitazone\_rediscovery & 0.286$\pm$0.033 & 0.189$\pm$0.044 & 0.264$\pm$0.033 & 0.273$\pm$0.023 & 0.181$\pm$0.012 \\
valsartan\_smarts & 0.000$\pm$0.000 & 0.000$\pm$0.000 & 0.000$\pm$0.000 & 0.000$\pm$0.000 & 0.000$\pm$0.000 \\
zaleplon\_mpo & 0.213$\pm$0.070 & 0.336$\pm$0.124 & 0.101$\pm$0.057 & 0.280$\pm$0.076 & 0.029$\pm$0.019 \\
\midrule
Sum & 11.657 & 11.230 & 11.133 & 10.768 & 8.824 \\
Rank & 11 & 12 & 13 & 14 & 15 \\

\hline
\hline

Method & MolPal & DoG-AE & GFlowNet-AL & Screening & VAE BO SMILES \\
Assembly & - & Synthesis & Fragments & - & SMILES \\
\midrule
albuterol\_similarity & 0.545$\pm$0.049 & 0.434$\pm$0.034 & 0.356$\pm$0.032 & 0.448$\pm$0.035 & 0.452$\pm$0.037 \\
amlodipine\_mpo & 0.554$\pm$0.059 & 0.468$\pm$0.040 & 0.411$\pm$0.059 & 0.505$\pm$0.037 & 0.501$\pm$0.033 \\
celecoxib\_rediscovery & 0.364$\pm$0.041 & 0.288$\pm$0.028 & 0.239$\pm$0.025 & 0.317$\pm$0.029 & 0.324$\pm$0.033 \\
deco\_hop & 0.596$\pm$0.015 & 0.639$\pm$0.052 & 0.580$\pm$0.008 & 0.582$\pm$0.010 & 0.582$\pm$0.010 \\
drd2 & 0.476$\pm$0.135 & 0.589$\pm$0.146 & 0.253$\pm$0.065 & 0.308$\pm$0.084 & 0.319$\pm$0.091 \\
fexofenadine\_mpo & 0.665$\pm$0.067 & 0.634$\pm$0.058 & 0.672$\pm$0.036 & 0.649$\pm$0.056 & 0.649$\pm$0.048 \\
gsk3b & 0.369$\pm$0.083 & 0.377$\pm$0.098 & 0.555$\pm$0.066 & 0.312$\pm$0.064 & 0.284$\pm$0.060 \\
isomers\_c7h8n2o2 & 0.220$\pm$0.056 & 0.055$\pm$0.018 & 0.132$\pm$0.038 & 0.065$\pm$0.020 & 0.067$\pm$0.020 \\
isomers\_c9h10n2o2pf2cl & 0.078$\pm$0.019 & 0.013$\pm$0.005 & 0.033$\pm$0.009 & 0.045$\pm$0.013 & 0.039$\pm$0.010 \\
jnk3 & 0.233$\pm$0.053 & 0.263$\pm$0.084 & 0.324$\pm$0.059 & 0.167$\pm$0.034 & 0.161$\pm$0.034 \\
median1 & 0.210$\pm$0.025 & 0.138$\pm$0.014 & 0.164$\pm$0.019 & 0.181$\pm$0.019 & 0.180$\pm$0.020 \\
median2 & 0.197$\pm$0.016 & 0.159$\pm$0.010 & 0.167$\pm$0.011 & 0.183$\pm$0.013 & 0.181$\pm$0.012 \\
mestranol\_similarity & 0.451$\pm$0.053 & 0.313$\pm$0.028 & 0.272$\pm$0.027 & 0.368$\pm$0.036 & 0.356$\pm$0.034 \\
osimertinib\_mpo & 0.770$\pm$0.096 & 0.687$\pm$0.089 & 0.779$\pm$0.041 & 0.750$\pm$0.084 & 0.753$\pm$0.074 \\
perindopril\_mpo & 0.440$\pm$0.039 & 0.384$\pm$0.034 & 0.404$\pm$0.046 & 0.425$\pm$0.035 & 0.423$\pm$0.030 \\
qed & 0.944$\pm$0.025 & 0.890$\pm$0.036 & 0.884$\pm$0.081 & 0.939$\pm$0.023 & 0.939$\pm$0.025 \\
ranolazine\_mpo & 0.396$\pm$0.071 & 0.589$\pm$0.083 & 0.617$\pm$0.113 & 0.357$\pm$0.061 & 0.392$\pm$0.070 \\
scaffold\_hop & 0.470$\pm$0.017 & 0.460$\pm$0.019 & 0.454$\pm$0.012 & 0.458$\pm$0.015 & 0.457$\pm$0.014 \\
sitagliptin\_mpo & 0.018$\pm$0.004 & 0.002$\pm$0.000 & 0.001$\pm$0.000 & 0.010$\pm$0.003 & 0.010$\pm$0.002 \\
thiothixene\_rediscovery & 0.311$\pm$0.027 & 0.262$\pm$0.019 & 0.245$\pm$0.021 & 0.295$\pm$0.023 & 0.293$\pm$0.022 \\
troglitazone\_rediscovery & 0.253$\pm$0.019 & 0.211$\pm$0.013 & 0.177$\pm$0.011 & 0.235$\pm$0.016 & 0.236$\pm$0.016 \\
valsartan\_smarts & 0.000$\pm$0.000 & 0.000$\pm$0.000 & 0.000$\pm$0.000 & 0.000$\pm$0.000 & 0.000$\pm$0.000 \\
zaleplon\_mpo & 0.054$\pm$0.015 & 0.006$\pm$0.003 & 0.004$\pm$0.001 & 0.020$\pm$0.006 & 0.013$\pm$0.004 \\
\midrule
Sum & 8.625 & 7.872 & 7.735 & 7.630 & 7.623 \\
Rank & 16 & 17 & 18 & 19 & 20 \\
\bottomrule
\end{tabularx}
\end{adjustbox}
\end{table*}

\newpage

\begin{table*}[h]
\centering
\caption{(Continued)
}
\label{tab:top100_3}
\begin{adjustbox}{width=\textwidth}
\scriptsize
\begin{tabularx}{\textwidth}{c | Y | Y | Y | Y | Y}
\toprule
Method & VAE BO SELFIES & Pasithea & JT-VAE BO & Graph MCTS & MolDQN \\
Assembly & SELFIES & SELFIES & Fragments & Atoms & Atoms \\
\midrule
albuterol\_similarity & 0.447$\pm$0.036 & 0.379$\pm$0.019 & 0.432$\pm$0.036 & 0.543$\pm$0.050 & 0.329$\pm$0.034 \\
amlodipine\_mpo & 0.489$\pm$0.032 & 0.449$\pm$0.024 & 0.484$\pm$0.033 & 0.425$\pm$0.062 & 0.315$\pm$0.068 \\
celecoxib\_rediscovery & 0.289$\pm$0.026 & 0.247$\pm$0.015 & 0.249$\pm$0.021 & 0.232$\pm$0.029 & 0.093$\pm$0.009 \\
deco\_hop & 0.570$\pm$0.008 & 0.563$\pm$0.005 & 0.575$\pm$0.010 & 0.549$\pm$0.008 & 0.541$\pm$0.007 \\
drd2 & 0.293$\pm$0.082 & 0.065$\pm$0.015 & 0.196$\pm$0.068 & 0.180$\pm$0.047 & 0.023$\pm$0.004 \\
fexofenadine\_mpo & 0.643$\pm$0.046 & 0.596$\pm$0.037 & 0.633$\pm$0.045 & 0.562$\pm$0.067 & 0.482$\pm$0.049 \\
gsk3b & 0.261$\pm$0.051 & 0.151$\pm$0.036 & 0.223$\pm$0.046 & 0.231$\pm$0.042 & 0.217$\pm$0.040 \\
isomers\_c7h8n2o2 & 0.152$\pm$0.045 & 0.638$\pm$0.199 & 0.029$\pm$0.012 & 0.417$\pm$0.107 & 0.366$\pm$0.081 \\
isomers\_c9h10n2o2pf2cl & 0.106$\pm$0.030 & 0.291$\pm$0.135 & 0.026$\pm$0.010 & 0.298$\pm$0.089 & 0.260$\pm$0.064 \\
jnk3 & 0.146$\pm$0.029 & 0.080$\pm$0.013 & 0.139$\pm$0.029 & 0.083$\pm$0.017 & 0.093$\pm$0.020 \\
median1 & 0.174$\pm$0.018 & 0.138$\pm$0.012 & 0.150$\pm$0.014 & 0.164$\pm$0.021 & 0.138$\pm$0.021 \\
median2 & 0.168$\pm$0.009 & 0.156$\pm$0.006 & 0.165$\pm$0.009 & 0.127$\pm$0.010 & 0.084$\pm$0.008 \\
mestranol\_similarity & 0.348$\pm$0.032 & 0.279$\pm$0.016 & 0.312$\pm$0.025 & 0.261$\pm$0.034 & 0.213$\pm$0.038 \\
osimertinib\_mpo & 0.750$\pm$0.072 & 0.662$\pm$0.061 & 0.753$\pm$0.072 & 0.690$\pm$0.061 & 0.650$\pm$0.032 \\
perindopril\_mpo & 0.406$\pm$0.027 & 0.370$\pm$0.020 & 0.404$\pm$0.028 & 0.262$\pm$0.048 & 0.162$\pm$0.043 \\
qed & 0.935$\pm$0.025 & 0.906$\pm$0.020 & 0.929$\pm$0.027 & 0.875$\pm$0.053 & 0.802$\pm$0.113 \\
ranolazine\_mpo & 0.363$\pm$0.066 & 0.218$\pm$0.025 & 0.362$\pm$0.085 & 0.175$\pm$0.041 & 0.036$\pm$0.015 \\
scaffold\_hop & 0.440$\pm$0.011 & 0.431$\pm$0.007 & 0.454$\pm$0.015 & 0.407$\pm$0.013 & 0.398$\pm$0.012 \\
sitagliptin\_mpo & 0.038$\pm$0.011 & 0.049$\pm$0.016 & 0.014$\pm$0.008 & 0.026$\pm$0.008 & 0.001$\pm$0.000 \\
thiothixene\_rediscovery & 0.271$\pm$0.018 & 0.242$\pm$0.011 & 0.250$\pm$0.015 & 0.217$\pm$0.024 & 0.097$\pm$0.014 \\
troglitazone\_rediscovery & 0.220$\pm$0.013 & 0.204$\pm$0.008 & 0.213$\pm$0.012 & 0.210$\pm$0.019 & 0.121$\pm$0.015 \\
valsartan\_smarts & 0.000$\pm$0.000 & 0.000$\pm$0.000 & 0.000$\pm$0.000 & 0.000$\pm$0.000 & 0.000$\pm$0.000 \\
zaleplon\_mpo & 0.100$\pm$0.030 & 0.050$\pm$0.017 & 0.034$\pm$0.018 & 0.029$\pm$0.010 & 0.004$\pm$0.002 \\
\midrule
Sum & 7.622 & 7.173 & 7.037 & 6.975 & 5.435 \\
Rank & 21 & 22 & 23 & 24 & 25 \\
\bottomrule
\end{tabularx}
\end{adjustbox}
\end{table*}


\begin{table*}[h]
\centering
\caption{The mean and standard deviation of \textbf{Top-10} from 5 independent runs. We ranked the methods by the summation of mean \textbf{Top-10} of all tasks. 
(Continued)
}
\label{tab:top10_1}
\begin{adjustbox}{width=\textwidth}
\scriptsize
\begin{tabularx}{\textwidth}{c | Y | Y | Y | Y | Y}
\toprule
Method & REINVENT & Graph GA & REINVENT SELFIES & LSTM HC & GP BO \\
Assembly & SMILES & Fragments & SELFIES & SMILES & Fragments \\
\midrule
albuterol\_similarity & 0.998$\pm$0.188 & 0.997$\pm$0.180 & 0.960$\pm$0.194 & 0.998$\pm$0.199 & \textbf{1.0$\pm$0.156} \\
amlodipine\_mpo & 0.733$\pm$0.095 & \textbf{0.769$\pm$0.088} & 0.700$\pm$0.076 & 0.714$\pm$0.065 & 0.663$\pm$0.072 \\
celecoxib\_rediscovery & \textbf{0.861$\pm$0.190} & 0.756$\pm$0.179 & 0.722$\pm$0.155 & 0.785$\pm$0.160 & 0.859$\pm$0.181 \\
deco\_hop & 0.802$\pm$0.107 & 0.650$\pm$0.021 & 0.735$\pm$0.057 & \textbf{0.944$\pm$0.117} & 0.714$\pm$0.056 \\
drd2 & \textbf{0.999$\pm$0.178} & 0.999$\pm$0.120 & 0.999$\pm$0.178 & 0.999$\pm$0.203 & 0.999$\pm$0.200 \\
fexofenadine\_mpo & \textbf{0.903$\pm$0.080} & 0.830$\pm$0.059 & 0.835$\pm$0.051 & 0.794$\pm$0.044 & 0.793$\pm$0.056 \\
gsk3b & 0.968$\pm$0.195 & 0.937$\pm$0.191 & 0.956$\pm$0.219 & 0.984$\pm$0.171 & 0.974$\pm$0.188 \\
isomers\_c7h8n2o2 & 1.0$\pm$0.300 & 0.984$\pm$0.214 & 0.961$\pm$0.243 & 0.931$\pm$0.310 & 0.818$\pm$0.237 \\
isomers\_c9h10n2o2pf2cl & 0.851$\pm$0.318 & 0.891$\pm$0.207 & 0.903$\pm$0.269 & 0.764$\pm$0.249 & 0.565$\pm$0.238 \\
jnk3 & \textbf{0.948$\pm$0.262} & 0.812$\pm$0.259 & 0.821$\pm$0.247 & 0.935$\pm$0.218 & 0.689$\pm$0.233 \\
median1 & \textbf{0.399$\pm$0.069} & 0.330$\pm$0.050 & 0.396$\pm$0.072 & 0.350$\pm$0.059 & 0.333$\pm$0.045 \\
median2 & 0.325$\pm$0.049 & 0.315$\pm$0.043 & 0.309$\pm$0.048 & 0.317$\pm$0.046 & \textbf{0.329$\pm$0.039} \\
mestranol\_similarity & 0.742$\pm$0.154 & 0.736$\pm$0.122 & 0.761$\pm$0.156 & \textbf{0.792$\pm$0.130} & 0.768$\pm$0.161 \\
osimertinib\_mpo & \textbf{0.905$\pm$0.046} & 0.872$\pm$0.040 & 0.873$\pm$0.034 & 0.847$\pm$0.033 & 0.828$\pm$0.031 \\
perindopril\_mpo & \textbf{0.642$\pm$0.078} & 0.613$\pm$0.059 & 0.609$\pm$0.081 & 0.553$\pm$0.042 & 0.548$\pm$0.041 \\
qed & \textbf{0.948$\pm$0.007} & 0.947$\pm$0.006 & 0.948$\pm$0.007 & 0.948$\pm$0.005 & 0.947$\pm$0.006 \\
ranolazine\_mpo & 0.857$\pm$0.109 & 0.801$\pm$0.106 & 0.846$\pm$0.121 & 0.807$\pm$0.101 & 0.807$\pm$0.114 \\
scaffold\_hop & \textbf{0.714$\pm$0.089} & 0.558$\pm$0.034 & 0.615$\pm$0.058 & 0.647$\pm$0.058 & 0.610$\pm$0.054 \\
sitagliptin\_mpo & 0.034$\pm$0.011 & \textbf{0.657$\pm$0.211} & 0.362$\pm$0.185 & 0.186$\pm$0.055 & 0.267$\pm$0.106 \\
thiothixene\_rediscovery & \textbf{0.663$\pm$0.138} & 0.574$\pm$0.095 & 0.637$\pm$0.135 & 0.645$\pm$0.104 & 0.651$\pm$0.106 \\
troglitazone\_rediscovery & 0.587$\pm$0.133 & 0.494$\pm$0.081 & 0.496$\pm$0.098 & 0.539$\pm$0.100 & 0.514$\pm$0.081 \\
valsartan\_smarts & \textbf{0.196$\pm$0.376} & 0.000$\pm$0.000 & 0.000$\pm$0.000 & 0.000$\pm$0.000 & 0.000$\pm$0.000 \\
zaleplon\_mpo & \textbf{0.475$\pm$0.172} & 0.412$\pm$0.096 & 0.433$\pm$0.141 & 0.390$\pm$0.124 & 0.252$\pm$0.093 \\
\midrule
Sum & 16.564 & 15.946 & 15.889 & 15.880 & 14.940 \\
Rank & 1 & 2 & 3 & 4 & 5 \\

\hline
\hline

Method & DoG-Gen & STONED & SMILES GA & LSTM HC SELFIES & DST \\
Assembly & Synthesis & SELFIES & SMILES & SELFIES & Fragments \\
\midrule
albuterol\_similarity & 0.925$\pm$0.150 & 0.805$\pm$0.146 & 0.703$\pm$0.109 & 0.971$\pm$0.181 & 0.748$\pm$0.115 \\
amlodipine\_mpo & 0.605$\pm$0.036 & 0.635$\pm$0.062 & 0.563$\pm$0.022 & 0.579$\pm$0.023 & 0.525$\pm$0.015 \\
celecoxib\_rediscovery & 0.682$\pm$0.117 & 0.398$\pm$0.053 & 0.356$\pm$0.035 & 0.535$\pm$0.083 & 0.422$\pm$0.045 \\
deco\_hop & 0.925$\pm$0.122 & 0.627$\pm$0.017 & 0.624$\pm$0.017 & 0.626$\pm$0.016 & 0.627$\pm$0.020 \\
drd2 & 0.999$\pm$0.116 & 0.997$\pm$0.228 & 0.986$\pm$0.208 & 0.999$\pm$0.304 & 0.998$\pm$0.298 \\
fexofenadine\_mpo & 0.769$\pm$0.043 & 0.851$\pm$0.065 & 0.764$\pm$0.044 & 0.753$\pm$0.039 & 0.767$\pm$0.047 \\
gsk3b & \textbf{0.989$\pm$0.141} & 0.756$\pm$0.148 & 0.709$\pm$0.138 & 0.601$\pm$0.107 & 0.843$\pm$0.220 \\
isomers\_c7h8n2o2 & 0.923$\pm$0.316 & 1.0$\pm$0.255 & \textbf{1.0$\pm$0.229} & 0.879$\pm$0.264 & 0.804$\pm$0.299 \\
isomers\_c9h10n2o2pf2cl & 0.483$\pm$0.147 & 0.935$\pm$0.254 & \textbf{0.976$\pm$0.243} & 0.597$\pm$0.185 & 0.810$\pm$0.300 \\
jnk3 & 0.886$\pm$0.179 & 0.613$\pm$0.164 & 0.393$\pm$0.084 & 0.303$\pm$0.053 & 0.781$\pm$0.223 \\
median1 & 0.296$\pm$0.052 & 0.282$\pm$0.039 & 0.199$\pm$0.020 & 0.338$\pm$0.054 & 0.262$\pm$0.031 \\
median2 & 0.287$\pm$0.039 & 0.264$\pm$0.041 & 0.207$\pm$0.013 & 0.261$\pm$0.032 & 0.194$\pm$0.023 \\
mestranol\_similarity & 0.609$\pm$0.102 & 0.671$\pm$0.141 & 0.513$\pm$0.067 & 0.585$\pm$0.075 & 0.506$\pm$0.069 \\
osimertinib\_mpo & 0.843$\pm$0.043 & 0.848$\pm$0.037 & 0.834$\pm$0.032 & 0.822$\pm$0.029 & 0.817$\pm$0.029 \\
perindopril\_mpo & 0.575$\pm$0.052 & 0.521$\pm$0.034 & 0.456$\pm$0.021 & 0.502$\pm$0.028 & 0.480$\pm$0.027 \\
qed & 0.948$\pm$0.013 & 0.947$\pm$0.005 & 0.948$\pm$0.005 & 0.947$\pm$0.005 & 0.946$\pm$0.006 \\
ranolazine\_mpo & 0.807$\pm$0.090 & \textbf{0.859$\pm$0.149} & 0.775$\pm$0.118 & 0.769$\pm$0.138 & 0.745$\pm$0.184 \\
scaffold\_hop & 0.590$\pm$0.039 & 0.546$\pm$0.050 & 0.512$\pm$0.023 & 0.519$\pm$0.024 & 0.519$\pm$0.026 \\
sitagliptin\_mpo & 0.181$\pm$0.057 & 0.517$\pm$0.173 & 0.480$\pm$0.150 & 0.230$\pm$0.063 & 0.111$\pm$0.054 \\
thiothixene\_rediscovery & 0.506$\pm$0.080 & 0.388$\pm$0.042 & 0.326$\pm$0.025 & 0.439$\pm$0.054 & 0.406$\pm$0.046 \\
troglitazone\_rediscovery & \textbf{0.619$\pm$0.134} & 0.359$\pm$0.043 & 0.272$\pm$0.030 & 0.315$\pm$0.031 & 0.308$\pm$0.038 \\
valsartan\_smarts & 0.000$\pm$0.000 & 0.000$\pm$0.000 & 0.000$\pm$0.000 & 0.000$\pm$0.000 & 0.000$\pm$0.000 \\
zaleplon\_mpo & 0.314$\pm$0.111 & 0.373$\pm$0.100 & 0.389$\pm$0.107 & 0.310$\pm$0.093 & 0.259$\pm$0.105 \\
\midrule
Sum & 14.772 & 14.201 & 12.997 & 12.894 & 12.889 \\
Rank & 6 & 7 & 8 & 9 & 10 \\
\bottomrule
\end{tabularx}
\end{adjustbox}
\end{table*}

\newpage

\begin{table*}[h]
\centering
\caption{(Continued)
}
\label{tab:top10_2}
\begin{adjustbox}{width=\textwidth}
\scriptsize
\begin{tabularx}{\textwidth}{c | Y | Y | Y | Y | Y}
\toprule
Method & SynNet & MIMOSA & MARS & GA+D & MolPal \\
Assembly & Synthesis & Fragments & Fragments & SELFIES & - \\
\midrule
albuterol\_similarity & 0.646$\pm$0.075 & 0.702$\pm$0.112 & 0.652$\pm$0.155 & 0.623$\pm$0.118 & 0.625$\pm$0.046 \\
amlodipine\_mpo & 0.585$\pm$0.023 & 0.564$\pm$0.021 & 0.526$\pm$0.030 & 0.525$\pm$0.136 & 0.614$\pm$0.042 \\
celecoxib\_rediscovery & 0.478$\pm$0.056 & 0.428$\pm$0.050 & 0.448$\pm$0.091 & 0.269$\pm$0.063 & 0.426$\pm$0.033 \\
deco\_hop & 0.624$\pm$0.016 & 0.641$\pm$0.023 & 0.596$\pm$0.007 & 0.583$\pm$0.023 & 0.662$\pm$0.031 \\
drd2 & 0.998$\pm$0.109 & 0.990$\pm$0.301 & 0.988$\pm$0.190 & 0.772$\pm$0.365 & 0.872$\pm$0.200 \\
fexofenadine\_mpo & 0.785$\pm$0.033 & 0.737$\pm$0.038 & 0.741$\pm$0.044 & 0.729$\pm$0.187 & 0.696$\pm$0.018 \\
gsk3b & 0.901$\pm$0.164 & 0.700$\pm$0.156 & 0.607$\pm$0.105 & 0.511$\pm$0.128 & 0.619$\pm$0.118 \\
isomers\_c7h8n2o2 & 0.529$\pm$0.135 & 0.798$\pm$0.294 & 0.949$\pm$0.303 & 1.0$\pm$0.274 & 0.523$\pm$0.115 \\
isomers\_c9h10n2o2pf2cl & 0.332$\pm$0.126 & 0.444$\pm$0.179 & 0.820$\pm$0.304 & 0.820$\pm$0.265 & 0.177$\pm$0.040 \\
jnk3 & 0.715$\pm$0.148 & 0.483$\pm$0.140 & 0.587$\pm$0.166 & 0.378$\pm$0.111 & 0.404$\pm$0.077 \\
median1 & 0.228$\pm$0.019 & 0.275$\pm$0.044 & 0.216$\pm$0.018 & 0.199$\pm$0.036 & 0.257$\pm$0.024 \\
median2 & 0.244$\pm$0.017 & 0.229$\pm$0.019 & 0.190$\pm$0.017 & 0.156$\pm$0.031 & 0.237$\pm$0.017 \\
mestranol\_similarity & 0.427$\pm$0.040 & 0.470$\pm$0.051 & 0.444$\pm$0.053 & 0.527$\pm$0.129 & 0.585$\pm$0.061 \\
osimertinib\_mpo & 0.810$\pm$0.027 & 0.813$\pm$0.030 & 0.797$\pm$0.049 & 0.777$\pm$0.143 & 0.794$\pm$0.029 \\
perindopril\_mpo & 0.589$\pm$0.040 & 0.548$\pm$0.050 & 0.480$\pm$0.025 & 0.324$\pm$0.144 & 0.480$\pm$0.024 \\
qed & 0.947$\pm$0.003 & 0.945$\pm$0.005 & 0.938$\pm$0.012 & 0.941$\pm$0.118 & 0.947$\pm$0.004 \\
ranolazine\_mpo & 0.771$\pm$0.055 & 0.767$\pm$0.176 & 0.759$\pm$0.068 & 0.771$\pm$0.252 & 0.494$\pm$0.064 \\
scaffold\_hop & 0.515$\pm$0.019 & 0.534$\pm$0.034 & 0.476$\pm$0.009 & 0.465$\pm$0.038 & 0.501$\pm$0.015 \\
sitagliptin\_mpo & 0.029$\pm$0.017 & 0.179$\pm$0.078 & 0.034$\pm$0.011 & 0.469$\pm$0.173 & 0.051$\pm$0.012 \\
thiothixene\_rediscovery & 0.433$\pm$0.042 & 0.367$\pm$0.036 & 0.426$\pm$0.067 & 0.294$\pm$0.072 & 0.347$\pm$0.023 \\
troglitazone\_rediscovery & 0.303$\pm$0.022 & 0.332$\pm$0.041 & 0.296$\pm$0.033 & 0.198$\pm$0.041 & 0.273$\pm$0.013 \\
valsartan\_smarts & 0.000$\pm$0.000 & 0.000$\pm$0.000 & 0.000$\pm$0.000 & 0.000$\pm$0.000 & 0.000$\pm$0.000 \\
zaleplon\_mpo & 0.381$\pm$0.078 & 0.274$\pm$0.111 & 0.213$\pm$0.074 & 0.353$\pm$0.123 & 0.191$\pm$0.049 \\
\midrule
Sum & 12.279 & 12.233 & 12.193 & 11.696 & 10.786 \\
Rank & 11 & 12 & 13 & 14 & 15 \\

\hline
\hline

Method & GFlowNet & DoG-AE & VAE BO SELFIES & Screening & VAE BO SMILES \\
Assembly & Fragments & Synthesis & SELFIES & - & SMILES \\
\midrule
albuterol\_similarity & 0.502$\pm$0.054 & 0.543$\pm$0.043 & 0.528$\pm$0.038 & 0.526$\pm$0.033 & 0.530$\pm$0.035 \\
amlodipine\_mpo & 0.465$\pm$0.024 & 0.513$\pm$0.012 & 0.531$\pm$0.015 & 0.563$\pm$0.024 & 0.559$\pm$0.021 \\
celecoxib\_rediscovery & 0.359$\pm$0.036 & 0.360$\pm$0.018 & 0.352$\pm$0.024 & 0.372$\pm$0.022 & 0.382$\pm$0.027 \\
deco\_hop & 0.594$\pm$0.007 & 0.789$\pm$0.084 & 0.587$\pm$0.004 & 0.600$\pm$0.007 & 0.604$\pm$0.007 \\
drd2 & 0.836$\pm$0.208 & 0.978$\pm$0.122 & 0.772$\pm$0.197 & 0.741$\pm$0.189 & 0.773$\pm$0.193 \\
fexofenadine\_mpo & 0.711$\pm$0.019 & 0.686$\pm$0.023 & 0.682$\pm$0.013 & 0.686$\pm$0.019 & 0.692$\pm$0.016 \\
gsk3b & 0.691$\pm$0.056 & 0.624$\pm$0.114 & 0.420$\pm$0.078 & 0.560$\pm$0.099 & 0.473$\pm$0.080 \\
isomers\_c7h8n2o2 & 0.530$\pm$0.141 & 0.251$\pm$0.088 & 0.423$\pm$0.115 & 0.254$\pm$0.079 & 0.226$\pm$0.064 \\
isomers\_c9h10n2o2pf2cl & 0.199$\pm$0.063 & 0.052$\pm$0.018 & 0.286$\pm$0.086 & 0.153$\pm$0.047 & 0.118$\pm$0.030 \\
jnk3 & 0.499$\pm$0.062 & 0.492$\pm$0.156 & 0.262$\pm$0.054 & 0.309$\pm$0.056 & 0.302$\pm$0.065 \\
median1 & 0.216$\pm$0.022 & 0.174$\pm$0.012 & 0.211$\pm$0.016 & 0.222$\pm$0.018 & 0.222$\pm$0.020 \\
median2 & 0.188$\pm$0.008 & 0.184$\pm$0.009 & 0.192$\pm$0.006 & 0.212$\pm$0.012 & 0.207$\pm$0.010 \\
mestranol\_similarity & 0.347$\pm$0.027 & 0.378$\pm$0.023 & 0.414$\pm$0.029 & 0.447$\pm$0.040 & 0.427$\pm$0.031 \\
osimertinib\_mpo & 0.798$\pm$0.009 & 0.759$\pm$0.028 & 0.780$\pm$0.013 & 0.783$\pm$0.019 & 0.784$\pm$0.014 \\
perindopril\_mpo & 0.457$\pm$0.025 & 0.437$\pm$0.018 & 0.445$\pm$0.011 & 0.464$\pm$0.018 & 0.458$\pm$0.015 \\
qed & 0.939$\pm$0.027 & 0.933$\pm$0.010 & 0.945$\pm$0.007 & 0.946$\pm$0.004 & 0.946$\pm$0.007 \\
ranolazine\_mpo & 0.674$\pm$0.046 & 0.700$\pm$0.037 & 0.488$\pm$0.061 & 0.456$\pm$0.052 & 0.523$\pm$0.066 \\
scaffold\_hop & 0.475$\pm$0.010 & 0.495$\pm$0.016 & 0.464$\pm$0.007 & 0.485$\pm$0.010 & 0.483$\pm$0.011 \\
sitagliptin\_mpo & 0.017$\pm$0.006 & 0.010$\pm$0.005 & 0.140$\pm$0.044 & 0.040$\pm$0.012 & 0.034$\pm$0.011 \\
thiothixene\_rediscovery & 0.309$\pm$0.026 & 0.320$\pm$0.020 & 0.314$\pm$0.016 & 0.336$\pm$0.019 & 0.336$\pm$0.022 \\
troglitazone\_rediscovery & 0.196$\pm$0.009 & 0.264$\pm$0.020 & 0.253$\pm$0.009 & 0.264$\pm$0.013 & 0.270$\pm$0.013 \\
valsartan\_smarts & 0.000$\pm$0.000 & 0.000$\pm$0.000 & 0.006$\pm$0.007 & 0.000$\pm$0.000 & 0.006$\pm$0.008 \\
zaleplon\_mpo & 0.070$\pm$0.042 & 0.054$\pm$0.032 & 0.280$\pm$0.075 & 0.124$\pm$0.039 & 0.071$\pm$0.024 \\
\midrule
Sum & 10.084 & 10.007 & 9.788 & 9.553 & 9.435 \\
Rank & 16 & 17 & 18 & 19 & 20 \\
\bottomrule
\end{tabularx}
\end{adjustbox}
\end{table*}

\newpage

\begin{table*}[h]
\centering
\caption{(Continued)
}
\label{tab:top10_3}
\begin{adjustbox}{width=\textwidth}
\scriptsize
\begin{tabularx}{\textwidth}{c | Y | Y | Y | Y | Y}
\toprule
Method & GFlowNet-AL & Pasithea & JT-VAE BO & Graph MCTS & MolDQN \\
Assembly & Fragments & SELFIES & Fragments & Atoms & Atoms \\
\midrule
albuterol\_similarity & 0.420$\pm$0.027 & 0.460$\pm$0.020 & 0.499$\pm$0.039 & 0.626$\pm$0.041 & 0.362$\pm$0.034 \\
amlodipine\_mpo & 0.443$\pm$0.020 & 0.508$\pm$0.007 & 0.526$\pm$0.014 & 0.462$\pm$0.017 & 0.354$\pm$0.035 \\
celecoxib\_rediscovery & 0.285$\pm$0.023 & 0.317$\pm$0.014 & 0.305$\pm$0.016 & 0.296$\pm$0.038 & 0.111$\pm$0.008 \\
deco\_hop & 0.590$\pm$0.005 & 0.583$\pm$0.003 & 0.591$\pm$0.006 & 0.563$\pm$0.007 & 0.552$\pm$0.006 \\
drd2 & 0.637$\pm$0.168 & 0.275$\pm$0.060 & 0.557$\pm$0.177 & 0.401$\pm$0.118 & 0.032$\pm$0.005 \\
fexofenadine\_mpo & 0.706$\pm$0.015 & 0.665$\pm$0.017 & 0.675$\pm$0.015 & 0.594$\pm$0.028 & 0.516$\pm$0.038 \\
gsk3b & 0.623$\pm$0.040 & 0.293$\pm$0.047 & 0.379$\pm$0.074 & 0.333$\pm$0.053 & 0.285$\pm$0.046 \\
isomers\_c7h8n2o2 & 0.322$\pm$0.090 & 0.824$\pm$0.233 & 0.113$\pm$0.026 & 0.623$\pm$0.124 & 0.523$\pm$0.088 \\
isomers\_c9h10n2o2pf2cl & 0.090$\pm$0.025 & 0.448$\pm$0.200 & 0.108$\pm$0.046 & 0.563$\pm$0.138 & 0.504$\pm$0.119 \\
jnk3 & 0.403$\pm$0.052 & 0.158$\pm$0.021 & 0.257$\pm$0.048 & 0.134$\pm$0.031 & 0.130$\pm$0.025 \\
median1 & 0.203$\pm$0.015 & 0.182$\pm$0.013 & 0.183$\pm$0.010 & 0.212$\pm$0.021 & 0.168$\pm$0.023 \\
median2 & 0.182$\pm$0.008 & 0.181$\pm$0.005 & 0.183$\pm$0.005 & 0.140$\pm$0.008 & 0.100$\pm$0.007 \\
mestranol\_similarity & 0.318$\pm$0.020 & 0.365$\pm$0.021 & 0.365$\pm$0.022 & 0.308$\pm$0.031 & 0.265$\pm$0.038 \\
osimertinib\_mpo & 0.800$\pm$0.009 & 0.756$\pm$0.013 & 0.785$\pm$0.016 & 0.722$\pm$0.017 & 0.685$\pm$0.017 \\
perindopril\_mpo & 0.437$\pm$0.017 & 0.424$\pm$0.010 & 0.438$\pm$0.014 & 0.311$\pm$0.038 & 0.253$\pm$0.066 \\
qed & 0.932$\pm$0.034 & 0.938$\pm$0.006 & 0.943$\pm$0.008 & 0.916$\pm$0.025 & 0.846$\pm$0.081 \\
ranolazine\_mpo & 0.666$\pm$0.046 & 0.354$\pm$0.025 & 0.524$\pm$0.074 & 0.303$\pm$0.069 & 0.104$\pm$0.046 \\
scaffold\_hop & 0.469$\pm$0.008 & 0.462$\pm$0.006 & 0.479$\pm$0.012 & 0.426$\pm$0.013 & 0.414$\pm$0.013 \\
sitagliptin\_mpo & 0.009$\pm$0.003 & 0.137$\pm$0.044 & 0.063$\pm$0.037 & 0.106$\pm$0.034 & 0.005$\pm$0.003 \\
thiothixene\_rediscovery & 0.286$\pm$0.018 & 0.291$\pm$0.010 & 0.287$\pm$0.012 & 0.249$\pm$0.020 & 0.115$\pm$0.015 \\
troglitazone\_rediscovery & 0.193$\pm$0.008 & 0.242$\pm$0.005 & 0.241$\pm$0.007 & 0.240$\pm$0.016 & 0.141$\pm$0.014 \\
valsartan\_smarts & 0.000$\pm$0.000 & 0.006$\pm$0.013 & 0.000$\pm$0.000 & 0.000$\pm$0.000 & 0.000$\pm$0.000 \\
zaleplon\_mpo & 0.020$\pm$0.006 & 0.140$\pm$0.043 & 0.161$\pm$0.061 & 0.096$\pm$0.034 & 0.017$\pm$0.009 \\
\midrule
Sum & 9.044 & 9.020 & 8.671 & 8.635 & 6.495 \\
Rank & 21 & 22 & 23 & 24 & 25 \\
\bottomrule
\end{tabularx}
\end{adjustbox}
\end{table*}


\begin{table*}[h]
\centering
\caption{The mean and standard deviation of \textbf{Top-1} from 5 independent runs. We ranked the methods by the summation of mean \textbf{Top-1} of all tasks. 
(Continued)
}
\label{tab:top1_1}
\begin{adjustbox}{width=\textwidth}
\scriptsize
\begin{tabularx}{\textwidth}{c | Y | Y | Y | Y | Y}
\toprule
Method & REINVENT & LSTM HC & Graph GA & REINVENT SELFIES & DoG-Gen \\
Assembly & SMILES & SMILES & Fragments & SELFIES & Synthesis \\
\midrule
albuterol\_similarity & 1.0$\pm$0.166 & 1.0$\pm$0.188 & \textbf{1.0$\pm$0.168} & 0.960$\pm$0.162 & 0.978$\pm$0.150 \\
amlodipine\_mpo & 0.735$\pm$0.086 & 0.739$\pm$0.063 & \textbf{0.783$\pm$0.078} & 0.706$\pm$0.068 & 0.621$\pm$0.034 \\
celecoxib\_rediscovery & \textbf{0.959$\pm$0.226} & 0.850$\pm$0.188 & 0.810$\pm$0.199 & 0.750$\pm$0.146 & 0.760$\pm$0.127 \\
deco\_hop & 0.805$\pm$0.109 & \textbf{0.955$\pm$0.083} & 0.654$\pm$0.019 & 0.736$\pm$0.069 & 0.932$\pm$0.076 \\
drd2 & 0.999$\pm$0.108 & \textbf{0.999$\pm$0.149} & 0.999$\pm$0.008 & 0.999$\pm$0.062 & 0.999$\pm$0.003 \\
fexofenadine\_mpo & \textbf{0.910$\pm$0.073} & 0.818$\pm$0.047 & 0.845$\pm$0.053 & 0.842$\pm$0.044 & 0.808$\pm$0.036 \\
gsk3b & 0.972$\pm$0.160 & \textbf{1.0$\pm$0.119} & 0.946$\pm$0.156 & 0.964$\pm$0.187 & 1.0$\pm$0.076 \\
isomers\_c7h8n2o2 & 1.0$\pm$0.260 & 0.971$\pm$0.285 & \textbf{1.0$\pm$0.196} & 0.961$\pm$0.172 & 0.990$\pm$0.324 \\
isomers\_c9h10n2o2pf2cl & 0.855$\pm$0.290 & 0.832$\pm$0.267 & 0.905$\pm$0.190 & 0.913$\pm$0.214 & 0.624$\pm$0.148 \\
jnk3 & 0.954$\pm$0.233 & \textbf{0.968$\pm$0.196} & 0.818$\pm$0.257 & 0.838$\pm$0.227 & 0.948$\pm$0.146 \\
median1 & \textbf{0.399$\pm$0.058} & 0.388$\pm$0.064 & 0.350$\pm$0.050 & 0.399$\pm$0.063 & 0.322$\pm$0.053 \\
median2 & 0.332$\pm$0.045 & \textbf{0.339$\pm$0.049} & 0.324$\pm$0.040 & 0.313$\pm$0.040 & 0.297$\pm$0.040 \\
mestranol\_similarity & 0.748$\pm$0.140 & \textbf{0.894$\pm$0.154} & 0.761$\pm$0.118 & 0.761$\pm$0.134 & 0.657$\pm$0.106 \\
osimertinib\_mpo & \textbf{0.909$\pm$0.040} & 0.859$\pm$0.023 & 0.880$\pm$0.029 & 0.878$\pm$0.028 & 0.850$\pm$0.028 \\
perindopril\_mpo & \textbf{0.644$\pm$0.071} & 0.568$\pm$0.037 & 0.625$\pm$0.054 & 0.610$\pm$0.070 & 0.587$\pm$0.044 \\
qed & \textbf{0.948$\pm$0.000} & 0.948$\pm$0.002 & 0.948$\pm$0.001 & 0.948$\pm$0.002 & 0.948$\pm$0.007 \\
ranolazine\_mpo & \textbf{0.865$\pm$0.068} & 0.824$\pm$0.073 & 0.810$\pm$0.072 & 0.851$\pm$0.095 & 0.823$\pm$0.057 \\
scaffold\_hop & 0.716$\pm$0.088 & \textbf{0.797$\pm$0.136} & 0.561$\pm$0.031 & 0.617$\pm$0.052 & 0.621$\pm$0.040 \\
sitagliptin\_mpo & 0.080$\pm$0.034 & 0.262$\pm$0.079 & \textbf{0.689$\pm$0.214} & 0.409$\pm$0.170 & 0.252$\pm$0.099 \\
thiothixene\_rediscovery & 0.665$\pm$0.128 & \textbf{0.734$\pm$0.116} & 0.601$\pm$0.092 & 0.642$\pm$0.127 & 0.553$\pm$0.087 \\
troglitazone\_rediscovery & 0.593$\pm$0.127 & 0.587$\pm$0.115 & 0.505$\pm$0.079 & 0.509$\pm$0.094 & \textbf{0.707$\pm$0.124} \\
valsartan\_smarts & \textbf{0.197$\pm$0.382} & 0.000$\pm$0.000 & 0.000$\pm$0.000 & 0.000$\pm$0.000 & 0.000$\pm$0.000 \\
zaleplon\_mpo & \textbf{0.478$\pm$0.150} & 0.413$\pm$0.126 & 0.421$\pm$0.086 & 0.441$\pm$0.109 & 0.343$\pm$0.111 \\
\midrule
Sum & 16.772 & 16.754 & 16.244 & 16.059 & 15.633 \\
Rank & 1 & 2 & 3 & 4 & 5 \\

\hline
\hline

Method & GP BO & STONED & LSTM HC SELFIES & DST & SMILES GA \\
Assembly & Fragments & SELFIES & SELFIES & Fragments & SMILES \\
\midrule
albuterol\_similarity & 1.0$\pm$0.140 & 0.805$\pm$0.136 & 1.0$\pm$0.185 & 0.792$\pm$0.113 & 0.715$\pm$0.095 \\
amlodipine\_mpo & 0.681$\pm$0.067 & 0.638$\pm$0.054 & 0.600$\pm$0.012 & 0.582$\pm$0.054 & 0.570$\pm$0.006 \\
celecoxib\_rediscovery & 0.946$\pm$0.206 & 0.401$\pm$0.051 & 0.585$\pm$0.090 & 0.459$\pm$0.039 & 0.358$\pm$0.031 \\
deco\_hop & 0.727$\pm$0.067 & 0.627$\pm$0.015 & 0.637$\pm$0.018 & 0.635$\pm$0.019 & 0.624$\pm$0.014 \\
drd2 & 0.999$\pm$0.130 & 0.997$\pm$0.182 & 0.999$\pm$0.237 & 0.999$\pm$0.209 & 0.986$\pm$0.161 \\
fexofenadine\_mpo & 0.805$\pm$0.053 & 0.851$\pm$0.058 & 0.769$\pm$0.039 & 0.778$\pm$0.041 & 0.771$\pm$0.041 \\
gsk3b & 0.986$\pm$0.164 & 0.766$\pm$0.106 & 0.65$\pm$0.074 & 0.861$\pm$0.160 & 0.722$\pm$0.090 \\
isomers\_c7h8n2o2 & 0.858$\pm$0.216 & 1.0$\pm$0.234 & 0.937$\pm$0.242 & 0.836$\pm$0.235 & 1.0$\pm$0.204 \\
isomers\_c9h10n2o2pf2cl & 0.583$\pm$0.219 & 0.935$\pm$0.230 & 0.713$\pm$0.210 & 0.861$\pm$0.281 & \textbf{0.976$\pm$0.217} \\
jnk3 & 0.698$\pm$0.221 & 0.62$\pm$0.150 & 0.428$\pm$0.101 & 0.789$\pm$0.200 & 0.414$\pm$0.080 \\
median1 & 0.345$\pm$0.044 & 0.295$\pm$0.036 & 0.362$\pm$0.058 & 0.281$\pm$0.036 & 0.207$\pm$0.014 \\
median2 & 0.337$\pm$0.033 & 0.265$\pm$0.038 & 0.274$\pm$0.031 & 0.201$\pm$0.024 & 0.210$\pm$0.009 \\
mestranol\_similarity & 0.796$\pm$0.153 & 0.671$\pm$0.132 & 0.646$\pm$0.079 & 0.529$\pm$0.070 & 0.515$\pm$0.057 \\
osimertinib\_mpo & 0.837$\pm$0.020 & 0.848$\pm$0.024 & 0.832$\pm$0.018 & 0.827$\pm$0.018 & 0.835$\pm$0.019 \\
perindopril\_mpo & 0.562$\pm$0.036 & 0.522$\pm$0.027 & 0.521$\pm$0.028 & 0.502$\pm$0.026 & 0.459$\pm$0.014 \\
qed & 0.947$\pm$0.002 & 0.947$\pm$0.001 & 0.948$\pm$0.001 & 0.947$\pm$0.003 & 0.948$\pm$0.002 \\
ranolazine\_mpo & 0.817$\pm$0.080 & 0.862$\pm$0.113 & 0.795$\pm$0.099 & 0.752$\pm$0.163 & 0.780$\pm$0.082 \\
scaffold\_hop & 0.619$\pm$0.055 & 0.548$\pm$0.047 & 0.543$\pm$0.029 & 0.521$\pm$0.019 & 0.512$\pm$0.020 \\
sitagliptin\_mpo & 0.318$\pm$0.117 & 0.526$\pm$0.169 & 0.349$\pm$0.089 & 0.205$\pm$0.106 & 0.504$\pm$0.145 \\
thiothixene\_rediscovery & 0.663$\pm$0.097 & 0.390$\pm$0.036 & 0.468$\pm$0.057 & 0.427$\pm$0.042 & 0.329$\pm$0.021 \\
troglitazone\_rediscovery & 0.544$\pm$0.083 & 0.360$\pm$0.039 & 0.344$\pm$0.035 & 0.317$\pm$0.034 & 0.282$\pm$0.023 \\
valsartan\_smarts & 0.000$\pm$0.000 & 0.000$\pm$0.000 & 0.000$\pm$0.000 & 0.000$\pm$0.000 & 0.000$\pm$0.000 \\
zaleplon\_mpo & 0.269$\pm$0.084 & 0.373$\pm$0.088 & 0.360$\pm$0.093 & 0.344$\pm$0.119 & 0.396$\pm$0.097 \\
\midrule
Sum & 15.345 & 14.257 & 13.770 & 13.455 & 13.123 \\
Rank & 6 & 7 & 8 & 9 & 10 \\

\bottomrule
\end{tabularx}
\end{adjustbox}
\end{table*}

\newpage

\begin{table*}[h]
\centering
\caption{(Continued)
}
\label{tab:top1_2}
\begin{adjustbox}{width=\textwidth}
\scriptsize
\begin{tabularx}{\textwidth}{c | Y | Y | Y | Y | Y}
\toprule
Method & SynNet & MARS & MolPal & MIMOSA & GA+D \\
Assembly & Synthesis & Fragments & - & Fragments & SELFIES \\
\midrule
albuterol\_similarity & 0.697$\pm$0.083 & 0.710$\pm$0.149 & 0.714$\pm$0.054 & 0.720$\pm$0.099 & 0.633$\pm$0.109 \\
amlodipine\_mpo & 0.596$\pm$0.020 & 0.546$\pm$0.034 & 0.651$\pm$0.043 & 0.594$\pm$0.009 & 0.527$\pm$0.124 \\
celecoxib\_rediscovery & 0.525$\pm$0.062 & 0.486$\pm$0.082 & 0.511$\pm$0.041 & 0.438$\pm$0.032 & 0.289$\pm$0.060 \\
deco\_hop & 0.639$\pm$0.019 & 0.603$\pm$0.005 & 0.860$\pm$0.102 & 0.642$\pm$0.018 & 0.584$\pm$0.023 \\
drd2 & 0.999$\pm$0.084 & 0.994$\pm$0.141 & 0.964$\pm$0.165 & 0.993$\pm$0.203 & 0.836$\pm$0.374 \\
fexofenadine\_mpo & 0.797$\pm$0.031 & 0.755$\pm$0.034 & 0.709$\pm$0.006 & 0.743$\pm$0.030 & 0.737$\pm$0.174 \\
gsk3b & 0.932$\pm$0.146 & 0.683$\pm$0.109 & 0.82$\pm$0.128 & 0.718$\pm$0.097 & 0.534$\pm$0.127 \\
isomers\_c7h8n2o2 & 0.685$\pm$0.147 & 0.961$\pm$0.260 & 0.882$\pm$0.163 & 0.804$\pm$0.233 & 1.0$\pm$0.254 \\
isomers\_c9h10n2o2pf2cl & 0.507$\pm$0.173 & 0.864$\pm$0.302 & 0.391$\pm$0.091 & 0.465$\pm$0.164 & 0.820$\pm$0.246 \\
jnk3 & 0.797$\pm$0.141 & 0.646$\pm$0.160 & 0.608$\pm$0.117 & 0.497$\pm$0.120 & 0.392$\pm$0.111 \\
median1 & 0.244$\pm$0.019 & 0.233$\pm$0.017 & 0.309$\pm$0.028 & 0.296$\pm$0.039 & 0.219$\pm$0.037 \\
median2 & 0.259$\pm$0.016 & 0.203$\pm$0.015 & 0.273$\pm$0.021 & 0.238$\pm$0.016 & 0.161$\pm$0.028 \\
mestranol\_similarity & 0.447$\pm$0.040 & 0.481$\pm$0.047 & 0.733$\pm$0.081 & 0.523$\pm$0.049 & 0.543$\pm$0.128 \\
osimertinib\_mpo & 0.821$\pm$0.016 & 0.809$\pm$0.021 & 0.816$\pm$0.020 & 0.817$\pm$0.022 & 0.784$\pm$0.129 \\
perindopril\_mpo & 0.610$\pm$0.039 & 0.488$\pm$0.016 & 0.504$\pm$0.020 & 0.557$\pm$0.047 & 0.337$\pm$0.147 \\
qed & 0.948$\pm$0.001 & 0.946$\pm$0.001 & 0.948$\pm$0.002 & 0.947$\pm$0.002 & 0.945$\pm$0.104 \\
ranolazine\_mpo & 0.783$\pm$0.038 & 0.776$\pm$0.050 & 0.556$\pm$0.064 & 0.773$\pm$0.139 & 0.775$\pm$0.244 \\
scaffold\_hop & 0.531$\pm$0.022 & 0.489$\pm$0.012 & 0.525$\pm$0.016 & 0.534$\pm$0.026 & 0.467$\pm$0.038 \\
sitagliptin\_mpo & 0.067$\pm$0.040 & 0.083$\pm$0.037 & 0.117$\pm$0.030 & 0.209$\pm$0.085 & 0.482$\pm$0.175 \\
thiothixene\_rediscovery & 0.481$\pm$0.057 & 0.463$\pm$0.077 & 0.361$\pm$0.016 & 0.378$\pm$0.029 & 0.307$\pm$0.068 \\
troglitazone\_rediscovery & 0.326$\pm$0.022 & 0.328$\pm$0.040 & 0.296$\pm$0.013 & 0.341$\pm$0.036 & 0.201$\pm$0.039 \\
valsartan\_smarts & 0.000$\pm$0.000 & 0.000$\pm$0.000 & 0.000$\pm$0.000 & 0.000$\pm$0.000 & 0.000$\pm$0.000 \\
zaleplon\_mpo & 0.402$\pm$0.059 & 0.296$\pm$0.023 & 0.286$\pm$0.064 & 0.287$\pm$0.103 & 0.359$\pm$0.119 \\
\midrule
Sum & 13.105 & 12.853 & 12.844 & 12.524 & 11.942 \\
Rank & 11 & 12 & 13 & 14 & 15 \\

\hline
\hline

Method & VAE BO SELFIES & DoG-AE & Screening & GFlowNet & VAE BO SMILES \\
Assembly & SELFIES & Synthesis & - & Fragments & SMILES \\
\midrule
albuterol\_similarity & 0.594$\pm$0.063 & 0.633$\pm$0.054 & 0.603$\pm$0.056 & 0.550$\pm$0.069 & 0.593$\pm$0.048 \\
amlodipine\_mpo & 0.593$\pm$0.022 & 0.539$\pm$0.017 & 0.613$\pm$0.039 & 0.482$\pm$0.016 & 0.611$\pm$0.036 \\
celecoxib\_rediscovery & 0.391$\pm$0.027 & 0.406$\pm$0.027 & 0.419$\pm$0.023 & 0.409$\pm$0.042 & 0.425$\pm$0.026 \\
deco\_hop & 0.602$\pm$0.006 & 0.862$\pm$0.060 & 0.616$\pm$0.003 & 0.600$\pm$0.007 & 0.666$\pm$0.027 \\
drd2 & 0.940$\pm$0.183 & 0.999$\pm$0.059 & 0.949$\pm$0.206 & 0.951$\pm$0.185 & 0.899$\pm$0.164 \\
fexofenadine\_mpo & 0.707$\pm$0.011 & 0.723$\pm$0.045 & 0.706$\pm$0.021 & 0.727$\pm$0.017 & 0.719$\pm$0.016 \\
gsk3b & 0.564$\pm$0.128 & 0.778$\pm$0.143 & 0.836$\pm$0.185 & 0.726$\pm$0.058 & 0.606$\pm$0.100 \\
isomers\_c7h8n2o2 & 0.605$\pm$0.143 & 0.563$\pm$0.200 & 0.488$\pm$0.154 & 0.693$\pm$0.158 & 0.418$\pm$0.109 \\
isomers\_c9h10n2o2pf2cl & 0.461$\pm$0.162 & 0.140$\pm$0.078 & 0.273$\pm$0.075 & 0.290$\pm$0.094 & 0.209$\pm$0.067 \\
jnk3 & 0.414$\pm$0.117 & 0.554$\pm$0.143 & 0.456$\pm$0.100 & 0.54$\pm$0.047 & 0.432$\pm$0.098 \\
median1 & 0.231$\pm$0.017 & 0.203$\pm$0.014 & 0.271$\pm$0.029 & 0.237$\pm$0.019 & 0.267$\pm$0.043 \\
median2 & 0.206$\pm$0.006 & 0.201$\pm$0.010 & 0.244$\pm$0.021 & 0.198$\pm$0.009 & 0.222$\pm$0.011 \\
mestranol\_similarity & 0.507$\pm$0.059 & 0.436$\pm$0.036 & 0.552$\pm$0.143 & 0.388$\pm$0.038 & 0.523$\pm$0.049 \\
osimertinib\_mpo & 0.802$\pm$0.010 & 0.793$\pm$0.026 & 0.801$\pm$0.016 & 0.817$\pm$0.016 & 0.801$\pm$0.010 \\
perindopril\_mpo & 0.482$\pm$0.024 & 0.464$\pm$0.026 & 0.500$\pm$0.028 & 0.478$\pm$0.021 & 0.484$\pm$0.028 \\
qed & 0.947$\pm$0.003 & 0.944$\pm$0.004 & 0.947$\pm$0.001 & 0.945$\pm$0.005 & 0.947$\pm$0.003 \\
ranolazine\_mpo & 0.564$\pm$0.065 & 0.744$\pm$0.025 & 0.532$\pm$0.059 & 0.701$\pm$0.030 & 0.598$\pm$0.076 \\
scaffold\_hop & 0.487$\pm$0.013 & 0.526$\pm$0.024 & 0.509$\pm$0.006 & 0.488$\pm$0.010 & 0.504$\pm$0.015 \\
sitagliptin\_mpo & 0.244$\pm$0.083 & 0.039$\pm$0.033 & 0.142$\pm$0.060 & 0.045$\pm$0.020 & 0.114$\pm$0.068 \\
thiothixene\_rediscovery & 0.343$\pm$0.016 & 0.358$\pm$0.021 & 0.362$\pm$0.017 & 0.342$\pm$0.030 & 0.370$\pm$0.028 \\
troglitazone\_rediscovery & 0.287$\pm$0.032 & 0.349$\pm$0.056 & 0.294$\pm$0.018 & 0.211$\pm$0.013 & 0.306$\pm$0.024 \\
valsartan\_smarts & 0.064$\pm$0.072 & 0.000$\pm$0.000 & 0.000$\pm$0.000 & 0.000$\pm$0.000 & 0.064$\pm$0.077 \\
zaleplon\_mpo & 0.379$\pm$0.091 & 0.156$\pm$0.093 & 0.280$\pm$0.101 & 0.118$\pm$0.061 & 0.139$\pm$0.046 \\
\midrule
Sum & 11.423 & 11.418 & 11.403 & 10.945 & 10.926 \\
Rank & 16 & 17 & 18 & 19 & 20 \\
\bottomrule
\end{tabularx}
\end{adjustbox}
\end{table*}

\newpage

\begin{table*}[h]
\centering
\caption{(Continued)
}
\label{tab:top1_3}
\begin{adjustbox}{width=\textwidth}
\scriptsize
\begin{tabularx}{\textwidth}{c | Y | Y | Y | Y | Y}
\toprule
Method & Pasithea & JT-VAE BO & GFlowNet-AL & Graph MCTS & MolDQN \\
Assembly & SELFIES & Fragments & Fragments & Atoms & Atoms \\
\midrule
albuterol\_similarity & 0.507$\pm$0.018 & 0.550$\pm$0.057 & 0.472$\pm$0.032 & 0.686$\pm$0.052 & 0.383$\pm$0.038 \\
amlodipine\_mpo & 0.585$\pm$0.0 & 0.585$\pm$0.0 & 0.466$\pm$0.016 & 0.483$\pm$0.024 & 0.383$\pm$0.033 \\
celecoxib\_rediscovery & 0.355$\pm$0.015 & 0.390$\pm$0.031 & 0.332$\pm$0.030 & 0.329$\pm$0.037 & 0.128$\pm$0.019 \\
deco\_hop & 0.608$\pm$0.013 & 0.600$\pm$0.006 & 0.596$\pm$0.006 & 0.569$\pm$0.008 & 0.554$\pm$0.006 \\
drd2 & 0.592$\pm$0.122 & 0.778$\pm$0.215 & 0.863$\pm$0.198 & 0.586$\pm$0.197 & 0.049$\pm$0.012 \\
fexofenadine\_mpo & 0.707$\pm$0.041 & 0.702$\pm$0.016 & 0.732$\pm$0.015 & 0.611$\pm$0.024 & 0.532$\pm$0.039 \\
gsk3b & 0.414$\pm$0.084 & 0.511$\pm$0.086 & 0.675$\pm$0.052 & 0.404$\pm$0.067 & 0.344$\pm$0.061 \\
isomers\_c7h8n2o2 & 0.902$\pm$0.231 & 0.264$\pm$0.099 & 0.561$\pm$0.163 & 0.783$\pm$0.144 & 0.652$\pm$0.126 \\
isomers\_c9h10n2o2pf2cl & 0.607$\pm$0.186 & 0.307$\pm$0.147 & 0.182$\pm$0.061 & 0.704$\pm$0.150 & 0.583$\pm$0.122 \\
jnk3 & 0.210$\pm$0.035 & 0.404$\pm$0.104 & 0.463$\pm$0.060 & 0.178$\pm$0.051 & 0.152$\pm$0.029 \\
median1 & 0.216$\pm$0.021 & 0.212$\pm$0.019 & 0.229$\pm$0.012 & 0.242$\pm$0.023 & 0.188$\pm$0.028 \\
median2 & 0.194$\pm$0.006 & 0.192$\pm$0.003 & 0.191$\pm$0.009 & 0.148$\pm$0.010 & 0.108$\pm$0.009 \\
mestranol\_similarity & 0.449$\pm$0.015 & 0.454$\pm$0.060 & 0.351$\pm$0.024 & 0.330$\pm$0.030 & 0.294$\pm$0.041 \\
osimertinib\_mpo & 0.792$\pm$0.009 & 0.800$\pm$0.011 & 0.812$\pm$0.010 & 0.738$\pm$0.018 & 0.699$\pm$0.018 \\
perindopril\_mpo & 0.447$\pm$0.016 & 0.463$\pm$0.019 & 0.464$\pm$0.020 & 0.334$\pm$0.038 & 0.282$\pm$0.062 \\
qed & 0.943$\pm$0.005 & 0.946$\pm$0.003 & 0.944$\pm$0.015 & 0.928$\pm$0.019 & 0.871$\pm$0.067 \\
ranolazine\_mpo & 0.443$\pm$0.054 & 0.587$\pm$0.041 & 0.705$\pm$0.034 & 0.369$\pm$0.096 & 0.171$\pm$0.077 \\
scaffold\_hop & 0.503$\pm$0.022 & 0.496$\pm$0.013 & 0.479$\pm$0.009 & 0.434$\pm$0.014 & 0.421$\pm$0.015 \\
sitagliptin\_mpo & 0.230$\pm$0.085 & 0.169$\pm$0.096 & 0.028$\pm$0.017 & 0.210$\pm$0.088 & 0.015$\pm$0.009 \\
thiothixene\_rediscovery & 0.333$\pm$0.016 & 0.315$\pm$0.014 & 0.319$\pm$0.020 & 0.265$\pm$0.022 & 0.129$\pm$0.018 \\
troglitazone\_rediscovery & 0.258$\pm$0.007 & 0.259$\pm$0.003 & 0.206$\pm$0.011 & 0.267$\pm$0.027 & 0.153$\pm$0.016 \\
valsartan\_smarts & 0.064$\pm$0.126 & 0.000$\pm$0.000 & 0.000$\pm$0.000 & 0.000$\pm$0.000 & 0.000$\pm$0.000 \\
zaleplon\_mpo & 0.243$\pm$0.084 & 0.302$\pm$0.089 & 0.048$\pm$0.020 & 0.166$\pm$0.065 & 0.042$\pm$0.024 \\
\midrule
Sum & 10.611 & 10.296 & 10.130 & 9.778 & 7.143 \\
Rank & 21 & 22 & 23 & 24 & 25 \\
\bottomrule
\end{tabularx}
\end{adjustbox}
\end{table*}


\newpage

\subsection{Distribution of ZINC 250k's properties}
\label{sec:dist}

\begin{figure*}[h]
  \centering
  \includegraphics[width=0.6\textwidth]{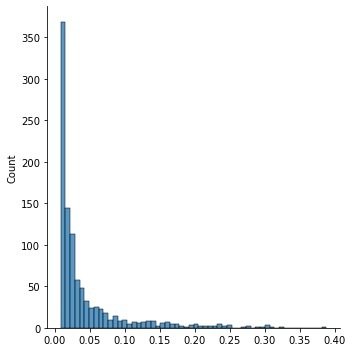}
  \caption{The distribution of Zaleplon MPO values in the ZINC 250k database. We show the values of top 1000 molecules only and all remaining are less than 0.02.}
  \label{fig:dist_zaleplon_mpo}
\end{figure*}

\end{document}